
\input harvmac
\def\npb#1(#2)#3{{ Nucl. Phys. }{B#1} (#2) #3}
\def\plb#1(#2)#3{{ Phys. Lett. }{#1B} (#2) #3}
\def\pla#1(#2)#3{{ Phys. Lett. }{#1A} (#2) #3}
\def\prl#1(#2)#3{{ Phys. Rev. Lett. }{#1} (#2) #3}
\def\mpla#1(#2)#3{{ Mod. Phys. Lett. }{A#1} (#2) #3}
\def\ijmpa#1(#2)#3{{ Int. J. Mod. Phys. }{A#1} (#2) #3}
\def\cmp#1(#2)#3{{ Commun. Math. Phys. }{#1} (#2) #3}
\def\cqg#1(#2)#3{{ Class. Quantum Grav. }{#1} (#2) #3}
\def\jmp#1(#2)#3{{ J. Math. Phys. }{#1} (#2) #3}
\def\anp#1(#2)#3{{ Ann. Phys. }{#1} (#2) #3}
\def\prd#1(#2)#3{{ Phys. Rev.} {D\bf{#1}} (#2) #3}

\def\p{\partial}

\def\Z{{\bf Z}}

\def\N{{\bf N}}
\def\t#1{{\theta_{#1}}}

\def\inbar{\,\vrule height1.5ex width.4pt depth0pt}
\def\IQ{\relax\,\hbox{$\inbar\kern-.3em{\rm Q}$}}
\def\IB{\relax{\rm I\kern-.18em B}}
\def\IC{\relax\hbox{$\inbar\kern-.3em{\rm C}$}}
\def\IP{\relax{\rm I\kern-.18em P}}
\def\IR{\relax{\rm I\kern-.18em R}}
\def\ZZ{\relax\ifmmode\mathchoice
{\hbox{Z\kern-.4em Z}}{\hbox{Z\kern-.4em Z}}
{\lower.9pt\hbox{Z\kern-.4em Z}}
{\lower1.2pt\hbox{Z\kern-.4em Z}}\else{Z\kern-.4em Z}\fi}

\def\n*#1{\nu^{* (#1)}}
\def\X{{\tilde X}}
\def\IP{{\bf P}}

\def\n#1{\nu_{#1}^*}
\def\a#1{a_{#1}}
\def\z#1{z_{#1}}
\def\IP{{\bf P}}

\def\n#1{\nu_{#1}^*}
\def\a#1{a_{#1}}
\def\z#1{z_{#1}}
\def\y#1{y_{#1}}
\def\x#1{x_{#1}}

\def\({ \left(  }
\def\){ \right) }
\def\t#1{\theta_{#1}}

\def\f#1{\phi_{#1}}
\def\ph{\phantom- }
\def\ie{\hbox{\it i.e.}}
\def\-{\phantom{-}}
\noblackbox

\catcode`@=12
\baselineskip16pt
\noblackbox
\newif\ifdraft

\noblackbox
\newif\ifhypertex
\ifx\hyperdef\UnDeFiNeD
    \hypertexfalse
    \message{[HYPERTEX MODE OFF}
    \def\hyperref#1#2#3#4{#4}
    \def\hyperdef#1#2#3#4{#4}
    \def\hypernoname{}
    \def\e@tf@ur#1{}
    \def\hth/#1#2#3#4#5#6#7{{\tt hep-th/#1#2#3#4#5#6#7}}
    
\else
    \hypertextrue
    \message{[HYPERTEX MODE ON}
  \def\hth/#1#2#3#4#5#6#7{
  {\tt hep-th/#1#2#3#4#5#6#7}}

\fi

\catcode`\@=11
\newif\iffigureexists
\newif\ifepsfloaded
\def\epsfcheck{
\ifdraft
\input epsf\epsfloadedtrue
\else
  \openin 1 epsf
  \ifeof 1 \epsfloadedfalse \else \epsfloadedtrue \fi
  \closein 1
  \ifepsfloaded
    \input epsf
  \else
\immediate\write20{NO EPSF FILE --- FIGURES WILL BE IGNORED}
  \fi
\fi
\def\epsfcheck{}}
\def\checkex#1{
\ifdraft
\figureexistsfalse\immediate%
\write20{Draftmode: figure #1 not included}
\else\relax
    \ifepsfloaded \openin 1 #1
        \ifeof 1
           \figureexistsfalse
  \immediate\write20{FIGURE FILE #1 NOT FOUND}
        \else \figureexiststrue
        \fi \closein 1
    \else \figureexistsfalse
    \fi
\fi}
\def\missbox#1#2{$\vcenter{\hrule
\hbox{\vrule height#1\kern1.truein
\raise.5truein\hbox{#2} \kern1.truein \vrule} \hrule}$}
\def\lfig#1{
\let\labelflag=#1%
\def\numb@rone{#1}%
\ifx\labelflag\UnDeFiNeD%
{\xdef#1{\the\figno}%
\writedef{#1\leftbracket{\the\figno}}%
\global\advance\figno by1%
}\fi{\hyperref{}{figure}{{\numb@rone}}{Fig.{\numb@rone}}}}
\def\figinsert#1#2#3#4{
\epsfcheck\checkex{#4}%
\def\figsize{#3}%
\let\flag=#1\ifx\flag\UnDeFiNeD
{\xdef#1{\the\figno}%
\writedef{#1\leftbracket{\the\figno}}%
\global\advance\figno by1%
}\fi
\goodbreak\midinsert%
\iffigureexists
\centerline{\epsfysize\figsize\epsfbox{#4}}%
\else%
\vskip.05truein
  \ifepsfloaded
  \ifdraft
  \centerline{\missbox\figsize{Draftmode: #4 not included}}%
  \else
  \centerline{\missbox\figsize{#4 not found}}
  \fi
  \else
  \centerline{\missbox\figsize{epsf.tex not found}}
  \fi
\vskip.05truein
\fi%
{\smallskip%
\leftskip 4pc \rightskip 4pc%
\noindent\ninepoint\sl \baselineskip=11pt%
{\bf{\hyperdef\hypernoname{figure}{{#1}}{Fig.{#1}}}:~}#2%
\smallskip}\bigskip\endinsert%
}
%


\nopagenumbers\abstractfont\hsize=\hstitle
\null
\rightline{\vbox{\baselineskip12pt\hbox{CERN-TH-7528/95}
                                  \hbox{IASSNS-HEP-94/106}
                                  \hbox{OSU-M-94-3}
                                  \hbox{hep-th/9506091}}}%
\vfill
\centerline{\titlefont Mirror Symmetry and the Moduli Space for }
\vskip5pt
\centerline{\titlefont Generic Hypersurfaces in Toric Varieties}
\abstractfont\vfill\pageno=0

\vskip-0.3cm
\centerline{Per Berglund}                                \vskip-.2ex
 \centerline{\it School of Natural Sciences}                  \vskip-.4ex
 \centerline{\it Institute for Advanced Study}         \vskip-.4ex
 \centerline{\it Princeton, NJ 08540, USA}       \vskip-.4ex
\vskip .05in
\centerline{Sheldon Katz}                                \vskip-.2ex
 \centerline{\it Department of Mathematics} \vskip-.4ex
 \centerline{\it Oklahoma State University}                   \vskip-.4ex
 \centerline{\it Stillwater, OK 74078, USA} \vskip0ex
\vskip -.05in
\centerline{and}
\vskip -.05in
\centerline{Albrecht Klemm\footnote{$^{}$}
      {Email: berglund@guinness.ias.edu, katz@math.okstate.edu,
klemm@nxth04.cern.ch}}                                \vskip-.2ex
 \centerline{\it Theory Division, CERN} \vskip-.4ex
 \centerline{\it CH-1211 Geneva  23, Switzerland}\vskip-.4ex
\vfill
\vskip-0.3cm
\vbox{\narrower\baselineskip=12pt\noindent
The moduli dependence of $(2,2)$ superstring compactifications
based on Calabi--Yau hypersurfaces in weighted projective space has so
far only been investigated for Fermat-type polynomial constraints. These
correspond to Landau-Ginzburg orbifolds with $c=9$ whose potential is a sum
of $A$-type singularities. Here we consider the generalization to arbitrary
quasi-homogeneous singularities at $c=9$.
We use mirror symmetry to derive the dependence
of the models on the  complexified K\"ahler moduli and check the
expansions of some topological correlation functions against explicit
genus zero and genus one instanton calculations. As an important application
we give examples of how non-algebraic (``twisted'') deformations can be
mapped to algebraic ones, hence allowing us to study the full moduli
space. We also study how moduli spaces can be nested in each other,
thus enabling a (singular) transition from one theory to
another. Following the recent work of Greene, Morrison and Strominger
we show that this corresponds to black hole condensation in type II string
theories compactified on Calabi-Yau manifolds.}

\Date{\vbox{\line{CERN-TH-7528/95\hfill}
            \line{6/95 \hfill}}}

\vfill\eject
\baselineskip=14pt plus 1 pt minus 1 pt
\newsec{Motivation and Outline of the Strategy}

The study of the moduli dependence of two-dimensional conformal
field theories is essential for understanding
the symmetries and the vacuum structure of the
critical string.
For conformal field theories with extended $N=2$  superconformal
symmetry remarkable progress in this question was
made after realizing the similarity of this problem
with the geometrical problem of the variation of the complex
structure~\ref\fs{S.~Ferrara and A.~Strominger, in the Proceedings of the
Strings '89 workshop, College Stattion, 1989\semi
A.~Strominger, \cmp{133}(1990)163\semi
P.~Candelas and X.~de~la~Ossa, \npb{355}(1991)455.}.
This implies that the (topological) correlation functions are
related to solutions of Fuchsian differential equations
and have natural modular properties with respect to the four-dimensional
spacetime moduli~\ref\dkl{L.~Dixon, V.~Kaplunovsky and J.~Louis,
\npb{329}(1990)27.}.
Closed string compactifications with $N=1$ spacetime supersymmetry
actually require an extension of the conformal symmetry to a
global $N=2$ superconformal symmetry for the right moving modes
on the worldsheet; heterotic compactifications
with $E_8\times E_6$ gauge group are based on a left-right symmetric
$(N,\bar N)=(2,2)$ superconformal internal sectors with $c=\bar c=9$.
The truly marginal operators, which preserve the $(2,2)$ structure,
 in these phenomenologically
motivated string compactifications come in two equivalent
types related to the left-right chiral $(c,c)$ states and the left
anti-chiral, right chiral $(a,c)$ states of the $(2,2)$
theory \ref\lvw{W. Lerche, C. Vafa and N. Warner,
\npb324(1989)427.}~\foot{These marginal operators are in one to one
correspondence with the ${\bf 27}$ and $\bar{\bf 27}$ of $E_6$. There is
yet a third class,
namely the marginal operators which correspond to
${\bf 1}$. They could potentially  enlarge the set of moduli fields.
They correspond to deformations of the tangent bundle of the
Calabi--Yau manifold
in question, $X$, and are given by $H^1(X,End T_X)$. We will focus
our attention to the traditional space of $(2,2)$ preserving deformations.
However, see~\ref\dk{J.~Distler and S.~Kachru,
{\sl Duality of (0,2) String Vacua}, hep-th/9501111.} for some recent work.}.
The two types of states form two rings whose structure
constants depend only on one type of moduli respectively; our convention
will be
to identify the $(c,c)$ ring with the complex structure deformations (also
known as the B-model in the language of topological field theory
\ref\Witten{ E. Witten,
{\sl Mirror Manifolds And Topological Field Theory}, in
{\it Essays on Mirror Manifolds}, (ed. S.-T. Yau),
Int. Press, Hong Kong, 1992 p.~120--158, hep-th/9112056.})
and the $(a,c)$ ring with
deformations of the complexified K\"ahler structure (the A-model in the
topological sigma model) of the target space $X$,
a Calabi--Yau threefold.

Unlike the dependence of the theory on the complexified
K\"ahler structure, which contains
the information about the holomorphic instantons on $X$,
the geometrical problem of complex structure deformations is a
well studied subject in classical geometry~\ref\kod{E.g. K.~Kodaira,
{\sl Complex Manifolds and Deformations of Complex Structure} (Springer Verlag
1985).}.
This fact and the equivalent structure of the two rings from
the point of view of the $(2,2)$ theory, which is the origin of mirror
symmetry\lref\ld{L.~Dixon: in {\it
     Superstrings, Unified Theories and Cosmology 1987},
     eds.~G.~Furlan et al.\ (World Scientific, Singapore, 1988)
     p.~67--127.}~\refs{\lvw,\ld}, has motivated the
key trick to solve both sectors; to find a
geometrical object $X^*$ yielding the identical $(2,2)$ theory in such a
way that the deformations of the
complexified K\"ahler structure on $X$ ($X^*$) can be identified
with the complex structure deformations on $X^*$ ($X$).
If this holds for $X$ and $X^*$,
the two manifolds form a so called mirror pair.

Before going on let us briefly recall the phenomenological implications
of the above. The physical (normalized) Yukawa couplings can
in principle be computed
since we know the metric on the moduli space for {\it both} the complex
structure {\it and} the K\"ahler structure deformations; the latter thanks
to mirror symmetry. In addition, as was first pointed out by Bershadsky et.
al.~\ref\bcov{M.~Bershadsky, S.~Cecotti, H.~Ooguri and C.~Vafa
(with an appendix by S.~Katz), \npb405(1993)279, hep-th/9302103.} and recently
amplified by Kaplunovsky and Louis~\ref\kl{V.~Kaplunovsky and J.~Louis,
{\sl On Gauge Couplings in String Theory}, hep-th/9502077.},
threshold corrections to gauge couplings in the four-dimensional effective
field theory can be inferred from the detailed knowledge of the singularity
structure of the moduli space.

Recently it has become clear that the moduli spaces of
Calabi-Yau manifolds might play an essential r\^ ole already
when writing down a consistent four dimensional $N=2$ supergravity
theory, independently of whether one believes that it arises
as a low energy effective theory from string compactification
or not. In the work of  Seiberg and Witten
\ref\sw{Seiberg and Witten, \npb426(1994)19-52,
Erratum-ibid.\npb430(1994)485-486}
the moduli space of four dimensional pure $SU(2)$
Super-Yang-Mills theory with global $N=2$ supersymmetry is
governed by rigid special geometry, and due to the consistency
condition of the positive kinetic terms uniquely it can be identified with
the moduli space of a torus\foot{The positivity condition is
solved in this approach by the second Riemann inequality for
the period lattice. In general, from lattices of dimension six on,
not every such such structure, which defines an abelian variety,
comes from a geometric curve; this is known as
Schottky problem. Families of curves however that
generalize\sw~for the $SU(N)$ series have been identified
\lref\klty{A. Klemm, W. Lerche, S. Theisen and S. Yankielowicz
Phys. Lett. B 344 (1995) 169-175, HEP-TH/9411048
and {\sl On The Monodromies  Of $N=2$ Supersymmetric  Yang-Mills Theory},
HEP-TH/9412158, P. C. Argyres and A. E. Faraggi
{\sl The Vacuum Structure and Spectrum of $N=2$ Supersymmetric
$SU(n)$ Gauge Theory} HEP-TH/9411057}\lref\klt{A. Klemm, W. Lerche,
S. Theisen, {\sl Nonperturbative Effective Actions of $N=2$
Supersymmetric Gauge Theories}. CERN-TH-95-104
HEP-TH/9505150}~\refs{\klty,\klt}.}. Similarly, the moduli space
of $N=2$ supergravity is known to exhibit non-rigid special
geometry \ref\cdfv{A. Ceresole, R. D'Auria, S. Ferrara, A. Van Proeyen,
{\sl On Electromagnetic Duality In Locally Supersymmetric $N=2$
Yang-Mills Theory}. CERN-TH-7510-94, HEP-TH/9412200 and
{\sl Duality Transformations in Supersymmetric Yang-Mills-Theories
Coupled to Supergravity} CERN-TH-7547-94, HEP-TH/9502072}
and a natural geometrical object associated to
this structure is a Calabi-Yau threefold.
Indeed, in a recent paper, Kachru and Vafa~\ref\kv{S. Kachru and C. Vafa,
{\sl Exact Results for N=2 Compactifications of Heterotic Strings},
hep-th/9505105.} give examples of heterotic stringy
realizations of~\sw.

The classification of $N=2$ SCFT with
$c < 3$ follows an A-D-E scheme which has,
via the Landau-Ginzburg (LG) approach \ref\lg{E. Martinec,
\plb217(1989)431, C. Vafa and N. P. Warner, \plb218(1989)51.},
a beautiful relation to the singularities of modality
zero \ref\arnold{
V. I. Arnol'd, V. A. Vasil'ev, V. V. Goryunov and O. V. Lyashko,
{\sl Singularities Local and Global Theory} in {\it Dynamical
Systems}, Enc. Math. Sc. Vol 6, Ed. V. I. Arnold, (1991) Springer,
Heidelberg.}. Superstring compactifications with $N=1$ spacetime
supersymmetry can be constructed by taking
suitable tensor products of these models such that $c=9$, adding free theories
for the uncompactified spacetime degrees of freedom including the gauge
degrees of freedom in the left-moving sector, and implementing
a generalized GSO-projection \ref\gepner{D. Gepner, \plb199(1987)380,
\npb 296 (1988) 757,\npb311(1988)191;\hfil\break C. Vafa, \mpla4(1989)1169.}.

The program for solving the full moduli dependence of the $(2,2)$
theory and checking the instanton predictions
has so far been studied only for theories based on tensor products
of the $A$-series\lref\cdgp{
P. Candelas, X. De la Ossa, P. Green and L. Parkes,
\npb359(1991)21.}
\lref\dave{D. Morrison, {\sl Picard-Fuchs Equations and
     Mirror Maps for Hypersurfaces}, in {\it Essays on Mirror
Manifolds}, ed. S.-T. Yau, (Int. Press, Hong Kong, 1992),
 hep-th/9111025.}
\lref\onemod{
A. Klemm and S. Theisen, \npb389(1993)153, hep-th/9205041;
Theor. Math. Phys. {\bf 95} (1993) 583, hep-th/9210142;
\hfill\break
A. Font, \npb391(1993)358, hep-th/9203084.}
\lref\cdfkm{P. Candelas, X. de la Ossa, A. Font,
S. Katz and D. Morrison, \npb416(1994)481, hep-th/9308083; \hfill\break
P. Candelas, A. Font, S. Katz and D. Morrison, \npb429(1994)626,
hep-th/9403187.}\lref\hkty{S. Hosono, A. Klemm, S. Theisen and S.-T. Yau:
{\sl Mirror Symmetry, Mirror Map and Applications to Calabi--Yau
Hypersurfaces}, HUTMP-93/0801, LMU-TPW-93-22,
to be published in Commun. Math. Phys.,
hep-th/9308122.}~\refs{\cdgp,\dave,\onemod,\cdfkm,\hkty}.
(For a rather different approach than the one pursued here,
using the linear sigma model,
see~\ref\pm{ D.R.~Morrison and M.R.~Plesser,
{\sl Summing the Instantons: Quantum Cohomology and Mirror Symmetry in Toric
Varieties}, hep-th/9412236.}, following the original mirror symmetry
construction by Greene and Plesser~\ref\gp{B. Greene and R. Plesser,
\npb338(1990)15.}.)
Geometrically they can be identified with hypersurfaces of
Fermat-type  in weighted projective spaces
$X=\{\vec x\in \IP^4(\vec w)|\sum_{i=1}^5 x_i^{n_i}=0\}$.
These cases are however only a very tiny subset of all transversal
quasi-homogeneous singularities or Landau-Ginzburg potentials
with singularity index $\beta=3/2$, which correspond to
rational $N=2$ SCFT with $c=9$. Theories of this general type have been
classified in \ref\ks{ A. Klemm and R. Schimmrigk,
\npb411(1994)559, hep-th/9204060;\hfil\break
M. Kreuzer and H. Skarke, \npb388(1993)113, hep-th/9205004.}.
Here we develop the methods to treat these generic
quasi-homogeneous potentials involving arbitrary
combinations of $A$-$D$-$E$ invariants as well as new
types of singularities.

Already the first step, to find the geometrical object
$X^*$, is more  tricky for the general case.
For models with $A$, $D_k,\,k\le 3$ and/or $E$  invariants
one has always the simplification, that the complex structure moduli
space of $X^*$ is the restriction of the moduli space of
$X$ to the invariant subsector with respect to a discrete group $H$; in other
words $X^*\simeq X/H$ (modulo desingularizations).
Therefore in these cases one in effect only needs to
consider the restricted complex structure deformation of
the original manifold $X$ itself. The general case,
however, requires the construction of a new variety.
The following two approaches to that problem will become
relevant for our calculation.

{\sl i)} For models which admit a description in terms of
Fermat-type polynomials Batyrev has suggested
a method of constructing the pair $X$ and $X^*$ as hypersurfaces
in toric varieties defined by a pair of reflexive
simplices\lref\batyrevI{V. Batyrev, Journal Alg. Geom. 3 (1994) 493,
alg-geom/9310003.}
\lref\batyrevII{V. Batyrev, Duke Math. Journ.,
69 (1993) 349}~\refs{\batyrevI,\batyrevII}, see also
\ref\roan{S.-S.~Roan: Int.\ J.\ Math 2 (1991) 439}.
It was later noticed\lref\bkii{P.Berglund and S. Katz,
{\sl Mirror Symmetry Constructions: A Review}
to appear in {\it Essays of Mirror Manifolds II} (Ed. S. T. Yau),
hep-th/9406008.}\lref\CdK{P.~Candelas, X.~de~la~Ossa and S.~Katz:
{\sl Mirror Symmetry for Calabi--Yau Hypersurfaces in Weighted
$\IP^4$ and an Extension of
Landau-Ginzburg Theory}, IASSNS-HEP-94/100,NEIP-94-009, OSU-M-93-3,
UTTG-25-93, hep-th/9412117.}~\refs{\hkty,\bkii,\CdK} that this method applies
also to general quasi-homogeneous hypersurfaces and can be
used to construct all the mirror manifolds for the hypersurfaces
in $\IP^4(\vec w)$ which where classified in \ks.

{\sl ii)}
An alternative approach\lref\gptwo{B. Greene and R. Plesser, in the Proceedings
of PASCOS '91, Boston 1991, hep-th/9110014.}
\lref\bh{P. Berglund and T. H\"ubsch, \npb393(1993)377,
hep-th/9201014.}~\refs{\gptwo,\bh}
starts from the following symmetry consideration.
The LG-theory ${\cal P}$ is defined by a transversal
potential $p(x_1,\ldots,x_5)$ quasi-homogeneous of degree $d$
with respect to the weights $w_i$,
\ie\ $p(\lambda^{w_1} x_1,\ldots,\lambda^{w_5} x_5)=
\lambda^d p(x_1,\ldots,x_5)$. The potential has an invariance group
${\cal G}({\cal P})$
whose elements $\gamma$ act on the coordinates by phase multiplication
$x_i\rightarrow x_i \exp(\gamma_i)$.
The GSO-projection onto integral charge states is implemented in the
internal sector by orbifoldization with respect to a subgroup
$\ZZ_d\in {\cal G}({\cal P})$ acting by
$x_i\rightarrow x_i \exp(2 \pi i{w_i\over d})$ \gepner.
The string compactification based on the internal
sector ${\cal P}/\ZZ_d$ has a simple geometrical interpretation.
Namely the compact part of the target space is given by
$X=\{\vec x\in \IP^4(\vec w)|p(x_1,\ldots,x_5)=0\}$.
Any orbifold with respect to a group $H$ with $\ZZ_d\subset H\subset
{\cal G}({\cal P})$
and $\prod_{i=1}^5 \exp(\kappa_i)=1$ for all $\kappa\in H$, leads
likewise to a consistent string compactification. The compact
part of the target space is now $X/(H/\ZZ_d)$.
It now follows from general arguments
\ref\dvvv{R. Dijkgraaf, E. Verlinde, H. Verlinde and C. Vafa,
\cmp123(1989)485.} that the orbifold ${\cal O}$ with respect to an
abelian group $H$, will have a dual symmetry group ${\cal G}_q({\cal O})$
called the quantum symmetry, which is isomorphic to $H$ and
manifest in the operator algebra of the twisted states
of ${\cal O}$. In the case of the above orbifold
${\cal O}={\cal P}/\ZZ_d$ all $(a,c)$ states belong
to the twisted states and ${\cal G}_q({\cal O})\cong \ZZ_d$.
On the other hand the operator algebra of
the invariant $(c,c)$ states is determined by
${\cal G}_g({\cal O})={\cal G}({\cal P})/\ZZ_d$, the so called geometrical
symmetry group. To exchange the r\^ole of the
$(c,c)$ and the $(a,c)$ ring and to construct a mirror pair
one therefore tries to construct two orbifolds
${\cal O}$ and ${\cal O}^*$ in which the geometrical
symmetry and the quantum symmetry are exchanged, \ie\
${\cal G}_g({\cal O})\cong{\cal G}_q({\cal O}^*)$
and ${\cal G}_q({\cal O})\cong{\cal G}_g({\cal O}^*)$.
As was recognized in~\gptwo~$N=2$ models based on tensor products of
minimal models always have a symmetry group $H$
with $\ZZ_d \subset H\subset {\cal G}({\cal P})$ such that
${\tilde {\ZZ_d}}={\cal G}({\cal P})/H$ and the quantum symmetry and the
geometrical symmetry are in fact exchanged for the pair
${\cal O}={\cal P}/\ZZ_d$, ${\cal O}^*={\cal P}/H$.

For general LG-models such  an $H$ need not exist.
In \bh~the authors present a generalization\foot{
Contrary to the original mirror symmetry construction~\gp,
there is no supporting argument at the
level of the underlying conformal field theory, since the exact $N=2$
SCFT theory is not known. However, the calculation here,
the affirmative check of the elliptic
genus~\ref\bhenn{P.~Berglund and M.~Henningson, \npb433(1995)311,
hep-th/9401029; see also {\sl On the Elliptic Genus and Mirror Symmetry}
to appear in {\it Essays of Mirror Symmetry II}, eds. B.~Greene and
S.T.~Yau, hep-th/9406045.} as well as the chiral
ring~\ref\max{M.~Kreuzer, \plb328(1994)312, hep-th/9402114.} suggests that
this is true.  Moreover, Morrison and Plesser have a promising approach to
providing such an argument.\ref\mp{D.R.~Morrison and M.R.~Plesser, work in
progress.}}
of the argument of
\gptwo~and consider a LG-potentials $p(x_1,\ldots,x_r)$,
which is transversal for a polynomial configuration with $r$ monomials,
\ie\ $p=\sum_{i=1}^r X^{v^{(i)}}$ with $v^{(i)}$ vectors of
exponents and $X^{v^{(i)}}:= x_1^{v^{(i)}_1}\cdots x_r^{v^{(i)}_r}$,
$v^{(i)}_j\in \N_0$. Then
one can consider the ``transposed'' polynomial
$\hat p=\sum_{i=1}^r Y^{\hat v^{(i)} }$, where the new exponent vectors
$\hat v^{(i)}$ are defined by $\hat v^{(i)}_r=v^{(r)}_i$.
It follows from the transversality condition \ks~that
all monomials $X^{v^{(i)}}$ in $p$ as well as the transposed
monomials in $\hat p$ have the form $x_i^{n_i} x_j$, with $i$, $j$
not necessarily different.
Using this one can see that
${\cal G}({\cal P})\cong {\cal G}(\hat {\cal P})$ and it was argued
in \bh~that there exists a group $H$ with $\ZZ_{\hat d}\subset H
\subset {\cal G}({\cal P})$
such that ${\cal G}_q ({\cal P}/\ZZ_d)
\cong {\cal G}_g( \hat {\cal P}/H)$ and ${\cal G}_g ({\cal P}/\ZZ_d)
\cong {\cal G}_q( \hat {\cal P}/H)$.  In fact the transposition rule also holds
for many non-transverse polynomials \CdK, where it was also shown to be
consistent with approach {\sl i)}.

In section~2 we will briefly review the construction of Calabi--Yau
hypersurfaces
in toric varieties, heavily relying on the methods introduced in~\hkty.
In particular we will introduce the Batyrev-Cox homogeneous coordinate
ring~\ref\bc{V.~Batyrev and D.~Cox, Duke Math. Journ. 74
(1994) 293, alg-geom/9306011.} which will simplify the construction of the
period vector as well as the Picard-Fuchs equation associated to it.
To generalize this approach to cases which have no ordinary
LG-description, \ie\ where no description as hypersurface
in $\IP^4(\vec w)$ or orbifolds thereof is available, we
develop methods to derive the Picard-Fuchs equations
directly from the combinatorial data of the dual polyhedron
$\Delta^*$.\foot{As usual, the Mori cone of the associated toric
variety enters in a crucial way.  We observe that the Mori cone of the
Calabi--Yau hypersurface may be strictly
smaller than the Mori cone of the toric variety.}

In section~3.1 we then consider generalized
hypersurfaces in $\IP^4(\vec w)$ with two K\"ahler moduli.
We use the constructions $(i),(ii)$ to derive
the Picard-Fuchs equations for the period
integrals in order to study the complex structure
deformations of $X^*$. It is convenient to follow first
$(ii)$ and use $\hat {\cal P}/H$ as a representation of $X^*$. Then one can
apply the standard Dwork-Katz-Griffiths\lref\Katz{N.\ Katz, Publ.\
Math. I.H.E.S. 35 (1968) 71.}\lref\dwork{B.~Dwork, Ann. of Math. (2)
{\bf 80} (1964) 227.}      \lref\griffiths{P. Griffiths,
Ann. of Math. 90 (1969) 460.}~\refs{\dwork,\Katz,\griffiths}
 reduction formulas
adapted to weighted projective spaces for the derivation of
the differential equations.
{}From this information we calculate the number of holomorphic
instantons of genus zero and genus one on $X$.
A detailed check of these predictions will
be presented in the section~4, where we also note that
the discrepancy between the Mori cones
affects the nature of the large complex structure limit, but
does not affect the validity of the instanton expansions.

The construction of \bh~allows one to obtain $X^*$
for the majority of the LG-potentials in \ks. It fails however
for LG-potentials for which a transversal configuration
requires more than $r$ monomials; in that case, Batyrev's approach will
apply, as shown in~\refs{\hkty,\CdK}.
We consider therefore in section~3.2 such a case.
It turns out that if we restrict to a suitable non-transversal
configuration involving $r$ terms (where $r=4+1$ in our case since
the hypersurfaces are embedded in a toric variety of dimension four)
and consider the transposed
polynomial $\hat P/H$ as before, we get an orbifold of a
non-transversal hypersurface $X^*$ in $\IP^4(\vec w)$,
which can be conveniently used to derive the set of differential
equations of Fuchsian type.
The latter reproduce the topological couplings and at least to lowest order
the genus zero instantons of $X$, see section~4.
This approach can be justified by translating the operator identities,
which hold modulo the ideal of $\hat P$ (equations of motions),
to the Laurent-monomials and derivatives of the Laurent-polynomial
and using the partial differentiation rule of section~(3.1).

In section~5 we apply the techniques developed in the earlier sections
to study some new phenomena. Given a toric variety based on a polyhedron
$\Delta$, we construct new reflexive polyhedra. In 5.1 we give examples of
reflexive polyhedra, which are
not associated to hypersurfaces in $\IP^4(\vec w)$ and apply the
methods outlined in section~2 to get the Picard-Fuchs equations also
in this case.
In section~5.2, we the study the phenomenon
of embedding the moduli space of one Calabi--Yau space into the
moduli space of another Calabi--Yau space and develop a quite general
strategy for constructing an algebraic realization of the
space of deformations (section~5.3).
The latter approach removes, what is
sometimes described as the ``twisted sector problem'' in the physics
literature.
Finally, in section~6 we will discuss the general validity of
the computation and give an algorithm for which,
in principle, a model with any number of moduli can be solved.
We also briefly comment on the connection with the recent
developments in type II string theory compactified on Calabi-Yau manifolds.

\newsec{General construction of Picard--Fuchs equations for Calabi--Yau
hypersurfaces in toric varieties}

Let us start by giving a brief review of the existing methods by which the
the Picard--Fuchs equations are obtained. For more details, see~\hkty.

Given a weighted projective space $\IP^4(\vec w)$ we construct the
Newton polyhedron, $\Delta$, as the convex hull (shifted by $(-1,-1,-1,-1,-1)$)
of the most general polynomial $p$ of degree $d=\sum_{i=1}^5w_i$,
\eqn\newtonpoly{\Delta={\rm Conv}\left(\{n\in \ZZ^5|
\sum_{i=1}^5 n_i w_i=0,\, n_i\ge -1\,\forall i\}\right),\,}
which lies in a hyperplane in $\IR^5$ passing through the origin.
For any set of weights which admits a transverse polynomial, it has been
shown in \CdK\ that the Newton polyhedron is reflexive, yielding a toric
variety birational to $\IP^4(\vec w)$.  (Reflexivity in these cases
has been checked independently by the third author.) The polar polyhedron,
$\Delta^*$, is given by
\eqn\dualpoly{\Delta^*={\rm Conv}\left(\{m\in\Lambda^*|<n,m>\geq -1,\quad
\forall n\in\Delta\}\right)}
where $\Lambda^*$ is the  dual lattice to
$\Lambda=\{n\in \ZZ^5| \sum_{i=1}^5 n_i w_i=0\}$. We can identify five
vertices, $v^{*(i)}\,,i=1,\ldots,5$ satisfying the linear
relation~\bkii\
\eqn\verrel{
\sum_{i=1}^5 k_i\cdot v^{*(i)}=0~.}
When $k_1=1$ we have
\eqn\vstari{
v^{*(1)}=(-k_2,-k_3,-k_4,-k_5)\,,v^{*(i+1)}=e_i\,,\quad
i=1,\ldots,4~,}
where the $e_i$ are the standard basis elements in $\ZZ^4$.

We are interested in studying hyperplane sections, $X^*$, of the toric variety
$\IP_{\Delta^*}$, given by
\eqn\Lpol{
P=\sum_{i\in \Delta^*\cap\Lambda^*} a_i \phi_i=
\sum_{i\in \Delta^*\cap\Lambda^*}\prod_j
Y_j^{\nu^{*(i)}_j}}
The claim~\batyrevI\ is that hypersurfaces $X\in\IP^4(\vec w)$ and
$X^*\in \IP_{\Delta^*}$
are mirror partners.
An alternative way is to specify
an \'etale map $Y(y)$ to the homogeneous coordinates $(y_1:\ldots :y_5)$
of a suitable four dimensional weighted projective space $\hat\IP^4(\vec w)$
(this map is actually only \'etale on the respective tori),
which identifies~\Lpol\
in $\IP_{\Delta^*}$ with the hypersurface defined by an
ordinary polynomial constraint $\hat p=(y_1 y_2 y_3 y_4 y_5 ) P(Y(y))=0$
in this $\hat\IP^4(\vec w)$.
Note that in general $\hat\IP^4(\vec w)\neq \IP^4(\vec w)$
In fact if we set $a_i=0,\,i\neq 1,\ldots,5$ then~\Lpol\ reduces to the
transposed polynomial $\hat p_0(y_i)$, of $p_0$; here $p_0(x_i)$ defines
a (possibly non-transverse) hypersurface in $\IP^4(\vec w)$, see {\sl ii)}
in the previous section.
Note that the map $Y(y)$ is in general not one-to-one, but rather there is an
automorphism which is isomorphic to $H$ the group which we need to
divide by in order to make $\{\hat p(y_i)=0\}/H$ the mirror of  a
hypersurface in $\IP^4(\vec w)$ described by $p=0$.
(For a more detailed account of the
correspondence between the toric construction and the transposition scheme,
see~\refs{\bkii,\CdK}.)

Yet a third way of identifying the hyperplane section in $\IP_{\Delta^*}$
is to use the homogeneous coordinate ring defined by Batyrev and Cox~\bc.
To each of the vertices, $\nu^{(i)}$ in $\Delta$ we associate a coordinate
$x_i$. As for a homogeneous coordinate in a weighted projective space,
each $x_i$ has a weight, or rather a multiple weight, given by positive
linear relations among the $\nu^{(i)}$. However, for our purposes it is
enough to note that to each vertex in the polar polytope, $\Delta^*$,
 corresponds a monomial
$\phi_j$in terms of the $x_i$;
\eqn\bcmonomial{
\phi_j=\sum_{i\in \Lambda \cap \Delta}<(v^{(i)},1),(v^{(j)*},1)>~.}
Thus, $P=\sum_{j\in \Lambda^*\cap\Delta^*}a_i\phi_i$ defines a hyperplanes
section in $\IP_{\Delta^*}$. Note that if we were to restrict to a particular
set of five vertices in $\Delta$ one can show $P$ reduces to $\hat p$
defined above through the \'etale map~\refs{\bkii,\CdK}.
In the examples we will mostly be using the second and third representation,
but that is only as a matter of convenience; all results can be derived using
Batyrev's original set of coordinates~\Lpol.

The next step is to find the generators of the Mori cone as they are
relevant in studying the large complex structure limit;
recall that the Mori cone is dual to the K\"ahler cone.
Given $\Delta^*$
we have to specify a particular triangulation; more
precisely a star subdivision of $\Delta^*$ from the interior point
$\nu^{*(0)}$. In general this triangulation is not unique. In particular there
may exist more than one subdivision which admits a K\"ahler resolution, \ie\
there is more than one Calabi--Yau phase~\ref\agm{P.S.~Aspinwall, B.R.~Greene
and D.R.~Morrison,\npb416(1994)414, hep-th/9309097.};
see section~3 for examples.

Once a particular subdivision is picked
an algorithm for constructing the generators
of the Mori cone is as follows:\foot{This algorithm is
equivalent to the one described by Oda and Park
\ref\OP{T. Oda and H.~S. Park, T\^ohoku Math.\ J.\ 43 (1991) 375.},
but is simpler to apply.}
\vskip .15in
{\it
\item {i)} Extend $\nu^{*(i)}$ to $\bar \nu^{*(i)}=(1,\nu^{*(i)})$

\item {ii)} Consider every pair $(S_k,S_l)$ of  four-dimensional
     simplices in the star subdivision of $\Delta^*$
     which have a common three-dimensional simplex $s_i=S_k\cap S_l$.

\item {iii)} Find for all such pairs the unique linear relation
      $\sum_{i=1}^6 l^{(k,l)}_i \bar \nu^{*(i)}=0$ among the six points
      $\nu^{*(i)}$ of $S_l\cup S_k$ in which the $l^{(k,l)}_i$ are minimal
      integers and the coefficients of the two points in
      $(S_k\cup S_l)\setminus (S_l\cap S_k)$ are non-negative.

\item {iv)}  Find the minimal integer vectors $l^{(i)}$ by which
      every $l^{(k,l)}$ can be expressed as positive integer
      linear combination. These are the generators of the
      Mori cone.}
\vskip .3in

Next we derive the Picard-Fuchs equation for $X^*$ from
the residue expression for the periods. There are two residue
forms for the periods; the Laurent polynomial
\Lpol\
or
the transposed polynomial from the \'etale map,
both of which we will use for the
derivation.
The first one reads \batyrevII
\eqn\batvariant{\Pi_i(a_0,\ldots,a_p)=
\int_{\Gamma_i}{\omega\over P},
\quad i=1,\dots , 2(h^{2,1}+1),}
where $\Gamma_i\in H^4({\bf T}\setminus X)$ ($\bf T$ is the algebraic
torus associated to the toric variety~\batyrevII), and
$\omega={d Y_1\over Y_1} \wedge {d Y_2\over Y_2}
\wedge {d Y_3\over Y_3} \wedge {d Y_4\over Y_4}$. Alternatively, we can write
the following expression for the periods in terms of the transposed
polynomial ${\hat p}=0$, \griffiths
\eqn\periods{\hat \Pi_i(a_0,\ldots,a_p)=
             \int_{\gamma}\int_{\hat \Gamma_i}{\hat \omega\over \hat p\,},
             \quad i=1,\dots , 2(h^{2,1}+1)\,.}
Here
$\hat \omega=\sum_{i=1}^5(-1)^i w_i y_i dy_1
\wedge\dots\wedge
{\widehat{dy_i}}
\wedge\dots\wedge d y_{5}$;
$\hat \Gamma_i$ is an element of $H_3(\hat X,\ZZ)$ and $\gamma$
a small curve around  ${\hat p}=0$ in the $4$-dimensional embedding space.
Note that by performing a change of variables as dictated by the \'etale
map one can show that~\batvariant\ is equivalent to~\periods.
Symmetry considerations now make the derivation of the Picard-Fuchs
equations a short argument.
The linear relations
between the points ${\bar \nu^{*\,(i)}}=(\nu^{*\,(i)},1)$ in the extended dual
polyhedron $\bar \Delta^*$ as expressed
by the $l^{(i)}$ translates into relations among the $\phi_i$,
\eqn\relations{
{\cal L}_k(\phi_i)~=~0~.}
However, from the definition of $\Pi_i$, the \relations\ are equivalent to
\eqn\diffrel{
{\cal L}_k(\del_{a_i})\Pi_j~=~0~.}
Finally, due to the $(\IC^*)^5$-invariance of $\Pi_j(a_i)$ we choose the
following combinations as the coordinates relevant in the large complex
structure limit~\batyrevII\
\eqn\goodvar{z_k=(-1)^{l_0^{(k)}} \prod_i a_i^{l^{(k)}_i}~.}
They will
lead directly to the large K\"ahler structure limit at $z_k=0$ at
which the monodromy is maximal unipotent
\ref\morrison{D. Morrison, Amer. Math. Soc. 6 (1993) 223;
{\sl Compactifications of Moduli spaces inspired by mirror symmetry}
DUK-M-93-06, alg-geom/9304007.}.
Thus, using $\t{i}=z_i {d\over z_i}$, eq. \diffrel\ is readily
transformed to\foot{For historical reasons
we rescale also the periods and use $\Pi_i={1\over a_0} \hat \Pi_i$. This
brings the system of differential equations in the form, first
discussed by Gelfand-Kapranov and Zelevinskii \ref\gkzI{
I. M. Gel'fand, A. V. Zelevinkii and M. M. Kapranov, Functional Anal. Appl.
{\bf 23} 2 (1989) 12, English trans. 94.}.}
\eqn\gkzth{
{\cal L}_k(\theta_i,z_i)\hat\Pi_j~=~0~.}

However, the above system~\gkzth\ has in general more than $2(h^{2,1}+1)$
solutions as can be seen by studying the number of solutions to the indicial
problem. This can be resurrected in the following two ways.
On one hand, one can try to factorize the set of differential operators
${\cal L}_k$ in order to reproduce operators of lower order; for examples
see~\hkty\ and section~3.1.
Alternatively, we can derive the differential operators by the more
standard manipulations of the residuum expressions~\batvariant\ and~\periods.
In particular, there exist  relations of order $n$ based on the use of the
ideal $\del_{y_i} \hat p$,\foot{There are in general
many ways of obtaining relations such as the one below. From a technical
point of view, it is preferable to use the ideal in such away that the
use of the partial integration rule becomes trivial.},
\eqn\ideal{
\Xi_k(\del_{y_i} \hat p,\phi_j,y_i)=0~.}
This
implies  $\int_{\gamma}\int_{\Gamma_i}{\Xi \hat \omega
\over {\hat p}^{n+1}}=0$. We can use the partial integration rule
${m(y) \p_{y_i} \hat p\over {\hat p}^{n+1}}={1\over n}{\p_{y_i} m(y)
\over {\hat p}^{n}}$ under the integral sign~\periods,
which follows from the fact that
$\p_{y_i}\left(m(y)\over {\hat p}^{n} \right) \hat \omega$
is exact, provided the integral over both sides of the
above equation makes sense as periods, which is the case
if $m(y) \p_{y_i} \hat p$ is homogeneous of degree $d\,n$, see
e.g.\lref\lsw{W.~Lerche, D.~J.~Smit and N.~P.~Warner,
\npb372(1992)87, hep-th/9108013.}~\refs{\lsw,\dave}.
In analogy to~\relations\ and~\diffrel\ we get a differential equation
satisfied by the period vector,
\eqn\diffideal{
\hat{\cal L}_k(\theta_i,z_i)\hat\Pi_j~=~0~.}

It is clear that~\ideal\
can be used analogously in
\batvariant.
To be more precise
$\Theta_i=Y_i {d\over d Y_i}$ and the  partial integration rule, which
due to the measure $\omega$ of \batvariant~now
reads\foot{The indexing of the $d_i$ is such that
$d_i \prod_j Y_j^{\nu^{*(k)}_j}=
(1-\langle \nu^{(i)},\nu^{*(k)}\rangle)\prod_j Y_j^{\nu^{*(k)}_j}$.}
${m(Y) d_i P\over {P}^n}={1\over n-1}
{(d_i+n-2) m(Y)\over {P}^{n-1}}$.
Also this integration rule is valid only if the integral
\batvariant~over both sides makes sense as periods,
which is the case if the points associated to
$m(Y) d_i P$ are in $(n-1)\Delta^*$ and the ones
associated to $(d_i+n-2)m(Y)$ are in $(n-2)\Delta^*$, where
$k\Delta^*$ denotes the polyhedron $\Delta^*$ scaled by $k$.

It should be obvious from the first and the last
derivation of the Picard-Fuchs equation, that the \'etale map
and the transposed polynomial are auxiliary constructions.
Their virtue is to
introduce a suitable grading which facilitate the calculations.
All relevant data however can be obtained directly
from the polyhedron $\Delta^*$ and \batvariant.

Finally, from the sets of operators $({\cal L}_k,\hat{\cal L}_k)$ one then has
to select the set of operators $(L_k)$ which will reproduce the
relevant ring structure. Unfortunately, we do not know of a general
recipe for how this is done, and at present we have to resort to a case by case
study, see section~3.

With the relevant set of operators at hand $(L_k)$ we are now ready to compute
the Yukawa couplings, the mirror map and then the instanton expansion.
Schematically the procedure is as follows. (For more details,
see\lref\hktyII{S. Hosono, A. Klemm, S. Theisen and S.-T. Yau,
\npb433(1995)501, hep-th/9406055.}~\refs{\hkty,\hktyII}.)
One first finds solutions, $w_j$, of the Picard-Fuchs system $(L_k\hat\Pi=0)$
with maximally unipotent monodromy at $z_i=0$. Then the flat coordinates
are given by $t_j=w_j/w_0$ where $w_0$ is the power series solution at $z_i=0$.
 From the $(L_k)$ one can  derive linear relations among the Yukawa
couplings $K_{z_i z_j z_k}$ and their derivatives.
Thus rather than using the explicit
solutions, and a knowledge of an integral symplectic basis, we derive
the Yukawa couplings (up to an overall normalization) directly from the
differential operators. This also gives us the discriminant locus, $\Delta$,
\ie\ the codimension one set where the Calabi--Yau hypersurface $X^*$
is singular.

{}From the Picard-Fuchs operators, we can also determine the intersection
numbers up to a normalization as the
coefficients of the unique
degree three element in
$$\IC[\theta_i]/\{\lim_{z_i\rightarrow 0}
L_k\}~.$$
The normalization may be fixed  by the intersection of
\eqn\normwcp{
K_{J_1J_1J_1}:= \int_{X} J_1\wedge J_1 \wedge J_1=
\left(d\over \prod_{i=1}^5 w_i\right) n_0^3,}
where $n_0$ is the least common multiple of the orders
of all fixpoints in $X$.
Alternatively, by considering the
restriction to a one-parameter subspace of the  moduli space of complex
structure deformations
spanned by the deformation corresponding to the interior point in~$\Delta^*$,
 one can  directly compute~\normwcp~\ref\bki{P.~Berglund and S.~Katz,
\npb420(1994)289, hep-th/9311014.}.

With both the Yukawa couplings and the flat coordinates at hand it is then
straightforward to map our results to that of the Yukawa coupling as a
function of the K\"ahler moduli in $X$, the hypersurface in $\IP^4(\vec w)$.
The details of this are standard and can be found in~\refs{\hkty,\hktyII}.
Here we only record the result;
\eqn\yukzexp{
K_{\tilde t_i\tilde t_j\tilde t_k}(\tilde t)~=~
{1\over w_0^2} \sum_{l,m,n} {\del x_l\over \del\tilde t_i}
{\del x_m\over \del\tilde t_j}{\del x_n\over \del\tilde t_k}K_{z_lz_mz_n}~,}
where the $\tilde t_i$ corresponds to an integral basis of $H^{1,1}(X,\ZZ)$
and are related to the flat coordinates by an integral
 similarity transformation. Finally, by  expanding~\yukzexp\ in terms of the
variables $q_i=exp(2\pi i\tilde t_i)$, we can read off the invariants of
the rational curves $N(\{n_l\})$,
\eqn\instexp{
K_{\tilde t_i\tilde t_j\tilde t_k}(\tilde t)~=~
K^0_{ijk}~+~\sum_{n_i}{N(\{n_l\})n_in_jn_k\over 1-\prod_lq_l^{n_l}}
\prod_lq_l^{n_l}~.}

It is well-known that there are no corrections to the Yukawa couplings at
higher genus.
Still it is interesting to note that invariants of elliptic curves (genus one)
can be studied by means of the following index
$F_1^{top}$~\bcov,
\eqn\onelooptoplim{F_1^{top}=\log \left[\left(1\over
\omega_0\right)^{5-\chi/12}{\partial(z_i)\over
\partial(t_i)} f(z)\right]+ {\rm const.}}
where $f$ is a holomorphic function with singularities only at singular
points on $X$ and so is related to the discriminant locus. Thus,
we make the ansatz
\eqn\ansatz{
f(z)=(\prod_j\Delta_j^{r_j}) \prod_iz_i^{s_i},}
with the $s_i$ fixed
by using the following asymptotic relation (valid at the large radius limit)
\eqn\oneloopasym{
\lim_{t,\bar t \rightarrow \infty} F_1=
- {2 \pi i \over 12} \sum_i
(t_i + \bar t_i) \int c_2 J_i~.}
In~\ansatz\ the product over the $j$ is taken over the components $\Delta_j$
of the discriminant locus. The $r_j$ are determined
by knowing some of the lowest order invariants of the elliptic curves;
see the examples for more details.

\newsec{Examples}

\subsec{Non-Fermat hypersurfaces in $\IP^4$ with two moduli}
The minimal K\"ahler structure  moduli system of transversal non-Fermat
Calabi--Yau  hypersurfaces in weighted $\IP^4$ is two dimensional.
In Table~3.1 the complete set of these models~\ks\ is listed.
We treat the first two examples in some detail and
summarize the results of the remaining cases in Appendix A.

{\vbox{\ninepoint{
$$\vbox{
{\offinterlineskip\tabskip=0pt
\halign{
\strut\vrule#&~#~&\hskip-6pt\vrule#&\hfil$#$~&\vrule#&\hfil$#$~&\vrule#
&\hfil$#$~&\vrule#&\hfil$#$~&\vrule#&\hfil $#$~&\vrule#\cr
\noalign{\hrule}
&X &&X_7(1,1,1,2,2)
&&X_7(1,1,1,1,3)
&&X_8(1,1,1,2,3)
&&X_9(1,1,2,2,3)
&&X_{14}(1,1,2,3,7)& \cr
\noalign{\hrule}
&$(h^{1,1},h^{2,1})$
&& (2,95)
&& (2,122)
&& (2,106)
&& (2,86)
&& (2,132)& \cr
&$C(\Delta){,}C(\Delta^*{)}$
&& (9,6)
&& (8,6)
&& (8,6)
&& (9,6)
&& (7,6)  &\cr
&$\nu^{*(6)}$
&& (0,0,-1,-1)
&& (0,0,0,-1)
&& (0,0,0,-1)
&& (0,-1,-1,-1)
&& (0,0,-1,-2) & \cr
\noalign{\hrule}
&$ \hat X$
&& X_{21}(2,2,3,7,7)
&& X_{14}(1,2,2,2,7)
&& X_{8}(1,1,1,1,4)
&& X_{12}(1,1,3,3,4)
&& X_{28}(2,2,3,7,14)& \cr
\noalign{\hrule}
&$(\hat h^{1,1},\hat h^{2,1})$
&& (11,50)
&& (2,122)
&& (1,149)
&& (5,89)
&& (9,83)& \cr
&$H\,:\,\, \hat h^{(1)}$
&& {1\over 21}(2,1,18,7,14)
&& {1\over 7}(1,0,6,0,0)
&& {1\over 8}(7,0,0,1,0)
&& {1\over 9}(1,8,0,0,0)
&& {1\over 14}(1,13,0,0,0)
& \cr
&$\phantom{H}\,:\,\, \hat h^{(2)}$
&&
&& {1\over 7}(1,6,0,0,0)
&& {1\over 8}(7,0,1,0,0)
&& {1\over 4}(1,0,3,0,0)
&& {1\over 2}(1,0,0,0,1)
& \cr
\noalign{\hrule}}}}
$$
\vskip-10pt
\noindent{\bf Table 3.1:} Non-Fermat Calabi--Yau hypersurfaces in
$\IP^4(\vec w)$ with $h^{1,1}=2$. The table displays
the Hodge numbers and the number of corners ($C(\Delta),C(\delta^*)$) of
the Newton
polyhedron $\Delta$ and its dual
$\Delta^*$ respectively.
For the five corners of $\Delta^*$ besides
$\nu^{*(6)}$, see~\vstari\ and the discussion below.  Note that there are no
non-toric
states in $X$ because all non-interior points in $\Delta^*$ are
corners or interior points of codimension one faces,
which means that we can describe all
complex- and K\"ahler-structure perturbations of $X$ algebraically.
The mirror manifold is $X^*={\hat X}/H$, were $\hat X$ is
the transposed hypersurface with Hodge numbers $\hat h^{1,1},\hat h^{2,1}$,
which is transversal
in these cases and hence appears in \ks. The vectors $h^{(k)}$ specify the
generators of $H$,
which act by $x_i\rightarrow x_i \exp 2 \pi i h_i^{(k)}$ on the
homogeneous coordinates of $\hat X$.
}
\vskip10pt}}

The first model is defined by the zero locus $p=0$ of a quasihomogeneous
polynomial of degree $d=7$ in the weighted projective space
$\IP^4(1,1,1,2,2)$. By Bertini's Theorem  (see e.g. Remark III.10.9.2. in
\ref\hartshorne{R. Hartshorne,
{\it Algebraic Geometry}, Grad. Texts in Mathematics {\bf 52},
(Springer Verlag, Heidelberg 1977).}) transversality
can fail for a generic member of $F$ only at the base locus, \ie\
the locus where every $p$ vanishes identically.
This base locus is the hyperplane $H=\{x\in \IP^4(1,1,1,2,2)|x_1=x_2=x_3=0\}$.
However the possible singular locus $S=\{x\in H|dp=0\}$,
where transversality could fail, is empty for a generic member of $X$.
A transversal member of $X$ is given for instance by
\eqn\transconfig{X_7(1,1,1,2,2)=\{x\in \IP^4(1,1,1,2,2)|\,
p_0=x_1^7{+}x_2^7 {+} x_3^7 {+} x_1 x_4^3 {+} x_2 x_5^3=0\}\, .}
Note that $p_0$ is not a sum of $A$-$D$-$E$ singularities, but since
the transversal configuration involves only five terms the
construction of ref. \bh~will apply.
All complex structure deformations are algebraic and can be described
by $95$ monomial perturbations of $p_0$ with elements of
${\cal R}=\IC[x_1,\ldots,x_5]/ \{\partial_{x_1}p_0,\ldots,\partial_{x_5}p_0\}$.
The canonical resolution $\hat X$
of the hypersurface $X$ has two elements in $H^2(\hat{X})$; one corresponds
to the divisor associated to the generating element of
${\rm Pic}(X)$ and a second one stems from the exceptional divisor,
which is introduced by the resolution of the $Z_2$-singularity.

Returning to $\IP^4(1,1,1,2,2)$, we see that $\Delta$
has nine corners whose components read
\eqn\cornersdelta{\matrix{
\nu^{(1)}=(-1,-1,-1,2),&
\nu^{(2)}=(-1,-1,-1,-1),&
\nu^{(3)}=(0,-1,-1,2)\cr
\nu^{(4)}=(6,-1,-1,-1),&
\nu^{(5)}=(-1,0,-1,2),&
\nu^{(6)}=(-1,6,-1,-1)\cr
\nu^{(7)}=(-1,-1,2,-1),&
\nu^{(8)}=(0,-1,2,-1),&
\nu^{(9)}=(-1,0,2,-1)}}
in a convenient basis for the sublattice $\Lambda\in \ZZ^5$ within the
hyperplane: $e_1=(-1,1,0,0,0)$, $e_2=(-1,0,1,0,0)$, $e_3=(-2,0,0,1,0)$ and
$e_4=(-2,0,0,0,1)$.
Beside the corners, $\Delta$ contains 1 lattice point in the interior,
$20$ lattice points on codimension~1 faces, $54$ lattice points on
codimension~2 and
$36$ lattice points on codimension~3 faces. One can pick 95 monomials
corresponding to 4 of the corners, the internal point and the 90 points on
codimension~2 and 3 faces as representatives of ${\cal R}$.

Since we know that the weights $w_i$ at hand admits a transverse polynomial
the polyhedron $\Delta$ is reflexive~\CdK. Its dual $\Delta^*$
has six corners, whose components in the basis of the dual lattice
$\Lambda^*$  are given below
\eqn\cornersdeltadual{\matrix{
\nu^{*(1)}=(-1,-1,-2,-2),&
\nu^{*(2)}=( 1, 0, 0, 0),&
\nu^{*(3)}=( 0, 1, 0, 0),\cr
\nu^{*(4)}=( 0, 0, 1, 0),&
\nu^{*(5)}=( 0, 0, 0, 1),&
\nu^{*(6)}=(0,0,-1,-1).}}
Beside these corners, the point $\nu^{*(0)}=(0,0,0,0)$ is the only integral
point in $\Delta^*$. In all cases where $w_1=1$ we can chose the lattice
such that $\nu^{*(1)}=(-w_2,-w_3,-w_4,-w_5)$ and $\nu^{*(i)}$ is as above for
$i=2,\ldots,5$.  In all cases considered here, there is only the additional
corner $\nu^{*(6)}$.  It is always the case that $\nu^{*(0)}$ is an additional
integral point of $\Delta^*$.  It is in fact the only additional integral
point except in the case of $\IP^4(1,1,2,3,7)$. In that case $\Delta$ contains
the point $(0,0,0,-1)$; but this plays essentially no role since it is the
interior point of a codimension~1 face.
Hence it is sufficient to only list $\nu^{*(6)}$ for the
other cases, see Table~3.1.

Using~\cornersdeltadual\ and~\Lpol\ the Laurent polynomial
of $\Delta^*$ is given by
\eqn\Laurantpolynomial{
P=\sum_i a_i \phi_i=a_0 + a_1\,{1\over Y_1 Y_2 Y_3^2 Y_4^2} + a_2\,Y_1 +
a_3\,Y_2 + a_4\, Y_3+
a_5\, Y_4  + a_6\,{1\over Y_3 Y_4}.}

The \'etale map is given by
\eqn\etale{Y_1={y_2^6\over y_1 y_3 y_4 },\,\,
           Y_2={y_3^6\over y_1 y_2 y_4 y_5},\,\,
           Y_3={y_4^2\over y_1 y_2 y_3 y_5},\,\,
           Y_4={y_5^2\over y_1 y_2 y_3 y_4}}
which leads
to a polynomial constraint
\eqn\mirrorp{\hat p=
\sum_{i=0}^6 a_i \phi_i\equiv
a_1 y_1^7 y_4 +
a_2 y_2^7 y_5 +
a_3 y_3^7 +
a_4 y_4^3 +
a_5 y_5^3 +
a_0 y_1 y_2 y_3 y_4 y_5 +
a_6 (y_1 y_2 y_3 )^3,}
which is quasihomogeneous of degree $\hat d=21$ with respect to the
weights of $\IP^4(2,2,3,7,7)$.
In fact $\hat p_0=  y_1^7 y_4 +y_2^7 y_5 + y_3^7 +  y_4^3 + y_5^3$
is the transposed polynomial of $p_0$ in \transconfig~and the
symmetry, which is identified
in \etale~corresponds exactly to the symmetry group
$H\sim (\Z_{21}:2,1,18,7,14)$, which has to be modded out from the
configuration
$X_{21}(2,2,3,7,7)$ to obtain the mirror \bh~of the
configuration $X_{7}(1,1,1,2,2)$.
Note also that the family $X_{21}(2,2,3,7,7)$ admits 50 independent complex
structure
perturbations (which are all algebraic) but the terms of \mirrorp~are
the only invariant terms under the $H$ symmetry group.

The boundary of  $\Delta^*$ in \cornersdeltadual~consists of 9
three dimensional simplices and
joining each of these simplices with the origin $\nu^{*(0)}$
we get a unique star subdivision of $\Delta^*$ into 9
four dimensional simplices.
Application of the algorithm for constructing the generators of the Mori cone
described in the previous section leads to
\eqn\genmori{l^{(1)}=(-3,0,0,0,1,1,1),\,\,\,\,
l^{(2)}=(-1,1,1,1,0,0,-2).}
Note that the Mori cone of
$\IP_{\Delta^*}$ coincides with the Mori cone of $X$
in this example, as well as in all examples treated so
far~\refs{\hkty,\hktyII}; see however
sections~3.2, 4 and Appendix~A for models where this is not true.
 From the general discussion in section~2 this allows us to determine the
relevant coordinates in the large complex structure limit, see~\goodvar, as
$z_1=-{a_4 a_5 a_6\over a_0^3}$ and
$z_2=-{a_1 a_2  a_3 \over a_0 a_6^2}$.

{}From the relations $\phi_0^3-\phi_4\phi_5\phi_6=0$ and
$\phi_6^2 \phi_0-\phi_1 \phi_2\phi_3=0$, where the $\phi_i$ are defined
by~\Laurantpolynomial~or equivalently by \mirrorp, we get two
third order differential operators satisfied by all periods
$\p_{a_0}^3-\p_{a_4}\p_{a_5}\p_{a_6}=0$ and $\p_{a_6}^2 \p_{a_0}-\p_{a_1}
\p_{a_2}\p_{a_3}=0$. This is readily transformed in the good variables \goodvar
\eqn\gkz{
\eqalign{{\cal L}_1&=\t1^2 (2 \t2-\t1) +
( 3 \t1+ \t2-2) (3 \t1+\t2-1)(3 \t1+\t2)z_1\cr
{\cal L}_2&= \t2^3-(3\t1+\t2)(2 \t2- \t1-2)(2\t2-\t1-1)z_2.}}
A short consideration of the indicial problem reveals that
it has nine solutions. The six periods however are solutions of a system,
which consists of a third and a second order differential operator.
As in many examples in \hktyII~the second
order differential
can be obtained in this case rather simply by factoring
$7 {\cal L}_2- 27 {\cal L}_1=(3 \t1+\t2) L_2$ with
\eqn\lII{L_2=9\t1^2-21 \t1 \t2 + 7 \t2^2 - 27 z_1
\prod_{i=1}^2 (3\t1+\t2+i)-7 z_2 \prod_{i=0}^1(2 \t2 -\t1+i).}
As Picard-Fuchs system we may choose this operator and say $L_1={\cal L}_2$.

Instead of using the factorization we may derive e.g. the
second order differential operator from the vanishing of
\eqn\trivialidentity{\eqalign{\Xi=&
a_1 a_2 \phi_1 \phi_2{-}
3 a_1 a_6\phi_1 \phi_6{-}
3 a_2 a_6 \phi_2 \phi_6{-}
3 a_0 a_6 \phi_0 \phi_6{-}
27 z_1 a_0^2 \phi_0^2{-}
7 z_2 a_6^2 \phi_6^2{-}\cr
&{9 a_4 a_6\over a_0} (y_1y_2y_3y_4)^2 \p_{y_5}\hat p{+}
3 a_6 \left(
{a_2\over a_0} y_1^2 y_2^9 y_3^2+ y_4 (y_1 y_2 y_3)^3\right)
\p_{y_4} \hat p {-}
{a_1 a_2\over a_0} (y_1 y_2)^6\p_{y_3} \hat p}}
which implies  $\int_{\gamma}\int_{\Gamma_i}{\Xi \hat \omega
\over {\hat p}^3}=0$.
After the partial integration we get from the last three
terms of $\Xi$ the contribution
${3a_6\over 2}  \int_{\gamma}\int_{\Gamma_i}{\phi_6 \omega\over {\hat p}^2}$.
Replacing all $\phi_i$ by derivatives with respect to the $a_i$
and transforming to the $z_i$ variables yields \lII
It is clear that \trivialidentity~can be used analogously in
\batvariant, to obtain the second order differential relation.
To be more precise $\Xi$ transforms into the variables $Y_i$
as
\eqn\trivialidentityII{\eqalign{\Xi=&
a_1 a_2 \phi_1 \phi_2{-}
3 a_1 a_6\phi_1 \phi_6{-}
3 a_2 a_6 \phi_2 \phi_6{-}
3 a_0 a_6 \phi_0 \phi_6{-}
27 z_1 a_0^2 \phi_0^2{-}
7 z_2 a_6^2 \phi_6^2{-}\cr
&{9 a_4 a_6\over a_0} {1\over Y_4} d_1 P+
3 a_6 \left({a_2\over a_0} {Y_1\over Y_3 Y_4}+ {1\over Y_3 Y_4}\right)
d_7 P -
{a_1 a_2\over a_0} {1\over Y_2 Y_3^2Y_4^2  }d_6 P=0\, ,}}
where
$d_1=(1- \Theta_1-\Theta_2-\Theta_3+2\Theta_4)$,
$d_7=(1-\Theta_1-\Theta_2+2\Theta_3-\Theta_4)$,
$d_6=(1-\Theta_1+6\Theta_2-\Theta_3-\Theta_4)$ and
$\Theta_i=Y_i {d\over d Y_i}$. Using the partial integration rule
yields again \lII.

{}From the Picard-Fuchs operators, we first determine the intersection
numbers up to a normalization which we get from \normwcp. Hence
$K_{J_1J_1J_1}=14$, $K_{J_1J_1J_2}=7$, $K_{J_1J_2J_2}=3$ and $K_{J_2J_2J_2}=0$.
Secondly,
we derive the general
discriminant locus
\eqn\disI{\Delta=(1{-}27 \z1)^3{-}
z_2(8{-}675 z_1{+}71442z_1^2{-}16 z_2{+}1372z_1z_2
{-}453789z_1^2z_2{+}823543z_1^2 z_2^2)}
and the Yukawa couplings in our normalization
\eqn\yuk{\eqalign{
K_{111}&={( 14{-} 112 z_2  {+} 324z_1  {+} 729z_1^2
{-} 2213z_1z_2  {-} 1323z_1^2z_2 {+} 224z_2^2 {+} 4116z_1z_2^2)
\over z_1^3 \Delta}\cr
K_{112}&={ ( 7-135 z_1  {-} 1458z_1^2 {-} 56z_2 {+}1284z_1z_2  {+}
 3969z_1^2z_2  {+} 112z_2^2 {-} 2744z_1z_2^2)\over z_1^2 z_2 \Delta} \cr
K_{122}&={( 3- 162 z_1  {+} 2187z_1^2 {-} 26z_2 {+} 1629z_1z_2  {-}
11907z_1^2z_2 {+} 56z_2^2 {-} 3773z_1z_2^2)\over z_1 z_2^2 \Delta}\cr
K_{222}&={ z_1 (-11 + 1161 z_1 +35721 z_1^2 + 28 z_2 -
3087 z_1 z_2)\over z_2^3 \Delta}.}}
Finally, the number of rational curves is obtained by an expansion of~\yuk\
around $z_i=0$; the result is recorded in Table~3.2.

To obtain the invariants for the elliptic curves we first note that
$\int_X c_2 J_1=68$ and $\int_X c_2J_2=36$. Using the ansatz~\onelooptoplim\
and the asymptotic relation~\oneloopasym\
one obtains $s_1=-20/3,s_2=-4$. The invariants of the
elliptic curves still contain  $r_0$, e.g. $n^e_{0,1}=-(1/3) (2+ 12 r_0)$.
We will see in section~4.1 that this invariant vanishes, which fixes
$r_0=-(1/6)$. This allows us to obtain the
other invariants of the elliptic curves, see Table~3.2. In fact, we will
see later that $n^e_{0,1}=0$ for all of our two parameter models, which
will allow us to determine $r_0$ in each case.  We always obtain
$r_0=-1/6$, which supports the
conjecture \ref\am{P.S.~Aspinwall and D.R.~Morrison:
 Phys. Lett. {\bf 334B} (1994) 79, hep-th/9406032.},
\hktyII~that the exponent is universally
$-(1/6)$ for the component of the discriminant parameterizing nodal
hypersurfaces.  An intriguing possible explanation of this phenomenon
based on consideration of black hole states of the type II string
has been given in \ref\vafa{C.~Vafa, {\sl A Stringy Test of the Fate of the
Conifold}, HUTP-95/A014, hep-th/9505023.}.
These instanton predictions will be discussed in section four. The
successful check provides a very detailed verification that
the configuration $X_{21}(2,2,3,7,7)$ modded out by
$H\simeq(\ZZ_{21}:2,1,18,7,14)$ is in fact the correct mirror configuration.

{\vbox{\ninepoint{
$$
\vbox{\offinterlineskip\tabskip=0pt
\halign{\strut\vrule#
&\hfil~$#$
&\vrule#&~
\hfil ~$#$~
&\hfil ~$#$~
&\hfil $#$~
&\hfil $#$~
&\hfil $#$~
&\hfil $#$~
&\hfil $#$~
&\vrule#\cr
\noalign{\hrule}
&n_{i,j}&& j=0& j=1 &  j=2  &  j=3  &  j=4 & j=5 & j=6& \cr
\noalign{\hrule}
&n^r_{0,j} && 0& -2&  0&\forall j>1&&&&\cr
&n^e_{0,j} && 0&\forall j&&&&&&\cr
\noalign{\hrule}
&n^r_{1,j}&& 177&178&3&5&7&9&11&\cr
&n^e_{1,j} && 0&\forall j&&&&&&\cr
\noalign{\hrule}
&n^r_{2,j}&&177&20291&-177&-708&-1068&-1448&1880&\cr
&n^e_{2,j}&& 0&  0&    0&   0&     0&   9&68&\cr
\noalign{\hrule}
&n^r_{3,j}&& 186&317172& 332040& 44790 & 75225&110271 &157734 &\cr
&n^e_{3,j}&& 3&  4&    181&   534&     885&   -177&-11161&\cr
\noalign{\hrule}
&n^r_{4,j}&& 177&2998628& 73458379& 794368&-4468169 &
-7157586&-11253268 &\cr
&n^e_{4,j}&& 0&  -356&    316802&   -60844&     -121684&
-81636&857218&\cr
\noalign{\hrule}
&n^r_{5,j}&& 177&21195310& 3048964748& 3122149716 & 243105088&
396368217 & 676476353 &\cr
&n^e_{5,j}&& 0&  -40582& 21251999&  26695536& 16380749&
23269402&-21423697&\cr
\noalign{\hrule}}
\hrule}$$
\vskip-10pt
\noindent
{\bf Table 3.2} The invariants of rational and elliptic curves of degree
$(i,j)$, $n^r_{i,j}$ and $n_{i,j}^e$ respectively, for $X_7(1,1,1,2,2)$.
(The computation was done using the Mathematica code, INSTANTON~\hktyII.)}
\vskip10pt}}

The mirror configuration for the second example $X_7(1,1,1,1,3)$
is obtained by the resolved quotient of the family
$X_{14}(1,2,2,2,7)$, a manifold with $h^{2,1}=2$ and $h^{1,1}=122$. These
are in fact also the Hodge numbers of the mirror of the
Fermat configuration $X_{14}(1,2,2,2,7)$ itself, which suggests, that
$X_7(1,1,1,1,3)$ and $X_{14}(1,2,2,2,7)$ (as well as its mirror
pair) are isomorphic. This has been checked to be the case in \CdK.
Let us look into this in more detail and check
first the homotopy type of the mirror pair.
We get the following generators of the Mori cone
\eqn\genmoriII{l^{(1)}=(-2,0,0,0,0,1, 1),\quad
               l^{(2)}=(-1,1,1,1,1,0,-3)}
Applying similar methods as described above we derive
a third and a second order Picard-Fuchs equation
\eqn\pfII{
\eqalign{
{\cal L}_1&=\t1 (\t1 - 3 \t2) - (2 \t1 + \t2 ) (2 \t1 + \t2-1) z_1 \cr
{\cal L}_2&=\t2^2(7 \t2-2 \t1)+4 \t2^2(2\t1+\t2-1)z_1-7
\prod_{i=1}^3(2 \t2-\t1-i)\, ,}}
from which the ratio of the topological couplings can
be read off. Fixing, according to~\bki,
the intersection  $K_{J_1J_1J_1}=63$, we get
$K_{J_1J_1J_2}=21$, $K_{J_1J_2J_2}=7$ and
$K_{J_2J_2J_2}=2$. In terms of the complex structure
parameters $z_1,z_2$ the Yukawa couplings are given
as follows
\eqn\yukII{
\eqalign{
K_{111}=&{(63+217 z_1+168 z_1^2+ 16 z_1^3+
63 z_2 (3+7 z_1)(9+7 z_1))\over z_1^3\Delta}\cr
K_{112}=&{z_2 (21-42 z_1-160 z_1^2-32 z_1^3+
7 z_2 (81-189 z_1-539 z_1^2))\over z_1^2\Delta}\cr
K_{122}=&{z_2^2 (7-52 z_1+80 z_1^2+ 64 z_1^3+
7 z_2 (9-28 z_1)(3-14  z_1))\over z_1\Delta}\cr
K_{222}=&{(2-24 z_1+96 z_1^2- 128 z_1^3+
7 z_2 (9-119 z_1+588 z_1^2))\over z_2^3\Delta}\,,}}
where the general discriminant $\Delta$ is
\eqn\disII{\Delta=(1- 4 z_1)^4+27 z_2- 411 z_1 z_2+
           2744 z_1^2 z_2 -38416 z_1^3 z_2-823543 z_1^3 z_2^2\,.}
Using the formulas from \hktyII~one obtains
$\int c_2 J_1= 126$, $\int c_2 J_2=44$. Comparing with \hkty,\hktyII~
we see that these topological numbers indeed coincide with
the ones of the mirror of the Fermat model $X_{14}(1,2,2,2,7)$ after
exchanging the r\^ole of the exceptional divisor and the
one from the ambient space. The Picard-Fuchs equations and the
expression for the Yukawa couplings however do not coincide after
exchanging $z_1$ and $z_2$. The difference is in fact in the
choice of the normalization of the holomorphic three-form and
one of the complex structure coordinates. The other complex
structure coordinate can be identified. If we denote by
$\tilde z_i$ the coordinate of the $X_{14}(1,2,2,2,7)$ model
associated to $l^{(1)}=(-7,0,1,1,1,-3,7)$ (in the notation of \hkty),
we have the
following simple relation between the mirror maps in both models:
$\tilde z_1(q_1,q_2)=z_2(q_2,q_1)$. The
resulting topological invariants, modulo the interchange of
the divisors, are exactly the ones calculated in \hkty,\hktyII~ for
$X_{14}(1,2,2,2,7)$. We will further discuss the relation between
the models in section five.
The first few are listed below in Table~3.3. Note that also in this
case, as well as in all other cases we consider here, the exponent
of the holomorphic ambiguity at the general discriminant turns out to be
$r_0=-1/6$.

{\vbox{\ninepoint{
$$
\vbox{\offinterlineskip\tabskip=0pt
\halign{\strut\vrule#
&\hfil~$#$
&\vrule#&~
\hfil ~$#$~
&\hfil ~$#$~
&\hfil $#$~
&\hfil $#$~
&\hfil $#$~
&\hfil $#$~
&\hfil $#$~
&\vrule#\cr
\noalign{\hrule}
&n_{i,j}&& j=0& j=1 &  j=2  &  j=3  &  j=4 & j=5 & j=6& \cr
\noalign{\hrule}
&n^r_{j,0} && 0& 28&  0&\forall j>1&&&&\cr
&n^e_{j,0} && 0&\forall j&&&&&&\cr
\noalign{\hrule}
&n^r_{j,1}&& 3&-56&378&14427&14427&378 &-56&\cr
&n^e_{j,1} && 0&\forall j&&&&&&\cr
\noalign{\hrule}
&n^r_{j,2}&&-6&140&-1512&9828&-69804&500724&29683962&\cr
&n^e_{j,2}&& 0&  0&    0&   0&     0&   378&6496&\cr
\noalign{\hrule}
&n^r_{j,3}&& 27&-896&13426& -122472 & 837900&-5083092 &27877878 &\cr
&n^e_{j,3}&&-10&  252& -3024   &  22932&   -122850& 489888&-1474200&\cr
\noalign{\hrule}}
\hrule}$$
\vskip-7pt
\noindent
{\bf Table 2.3} The invariants of rational and elliptic curves of degree
$(i,j)$, $n^r_{i,j}$ and $n_{i,j}^e$ respectively, for $X_7(1,1,1,1,3)$.}
\vskip7pt}}

\subsec{An Exotic Example}
As the simplest example, in terms of the number of K\"ahler deformations,
for which the naive application of~\bh\ fails we now consider
a hypersurface of degree $13$
in $\IP^4(1,2,3,3,4)$ that has Euler number $-114$. The second
Betti number is $5$. In this case we do not expect to find
a subsector of the algebraic deformations from a LG-potential
to be related to the complexified K\"ahler structure deformation
of $X_{13}(1,2,3,3,4)$. Instead we should learn everything
from the dual polyhedron $\Delta^*$, whose corners are the
following points
$$\eqalign{\matrix{
\nu^{*(1)}=(-2,-3,-3,-4),&
\nu^{*(2)}=(\-1,\-0,\-0,\-0),&
\nu^{*(3)}=(\-0,\-1,\-0,\-0),\cr
\nu^{*(4)}=(\-0,\-0,\-1,\-0),&
\nu^{*(5)}=(\-0,\-0,\-0,\-1),&
\nu^{*(7)}=(-1,-1,-1,-2),\cr
\nu^{*(8)}=(-1,-2,-2,-2),&
\nu^{*(9)}=(-1,-2,-2,-3),&}}$$
and which contains in addition the point $\nu^{*(6)}=(0,-1,-1,-1)$
on the edge $(\nu^{*(2)}-\nu^{*(8)})$ apart from the interior
point $\nu^{*(0)}=(0,0,0,0)$. In order to find the generators
of the Mori cone, we have to specify a particular triangulation; more precisely
a star subdivision of $\Delta^*$ from the point $\nu^{*(0)}$.
This triangulation is not unique, in contrast to the previous examples.
First note that $\Delta^*$, which has volume $22$
is bounded by thirteen three-dimensional hyperplanes,
on which the natural bilinear form $\langle \nu^{(i)},.\rangle$
$i=1,\ldots, 13$ takes the value $-1$ for $\nu^{(i)}$ the corners of $\Delta$
$$\eqalign{\matrix{
\nu^{(1)}=(-1,-1,-1,2),&
\nu^{(2)}=(-1,-1,-1,-1),&
\nu^{(3)}=(\-5,-1,-1,-1),\cr
\nu^{(4)}=(\-0,\-0,-1,\-1),&
\nu^{(5)}=(\-4,\-0,-1,-1),&
\nu^{(6)}=(-1,\-2,-1,\-0),\cr
\nu^{(7)}=(\-1,\-2,-1,-1),&
\nu^{(8)}=(-1,\-3,-1,-1)&
\nu^{(9)}=(\-0,-1,\-0,\-1),\cr
\nu^{(10)}=(\-4,-1,\-0,-1),&
\nu^{(11)}=(-1,-1,\-2,\-0),&
\nu^{(12)}=(\-1,-1,\-2,-1),\cr
\nu^{(13)}=(-1,-1,\-3,-1).&&}}$$
There exists six different subdivisions which admit a K\"ahler
resolution~\foot{In this respect the model is similar to the much celebrated
example of topology change, studied in~\agm.
We will however refrain from considering
this aspect
since our motivation is different; although an interesting problem, the
outcome would not imply anything new in terms of the phase
structure~\ref\phases{E. Witten, \npb403(1993)159, hep-th/9301042.}.}.
One of them is given by \foot{This subdivision and the other K\"ahler
subdivisions were found from among the 305 subdivisions (phases)
produced by the computer
program PUNTOS written by J.~De~Loera \ref\jdl{J.~De~Loera, {\sl
Triangulations of Polytopes and Computational Algebra}, Ph.D.\ thesis,
Cornell University, in preparation.}.}

$$\eqalign{\matrix{
s_1=(3, 4, 7, 9),&
s_2=(3, 4, 5, 7),&
s_3=(2, 3, 4, 5),&
s_4=(2, 3, 5, 6),\cr
s_5=(2, 4, 5, 6),&
s_6=(2, 3, 4, 9),&
s_7=(4, 5, 6, 8),&
s_8=(3, 5, 6, 8),\cr
s_9=(3, 5, 1, 7),&
s_{10}=(1, 4, 5, 7),&
s_{11}=(1, 4, 7, 9),&
s_{12}=(1, 3, 7, 9),\cr
s_{13}=(1, 3, 5, 8),&
s_{14}=(1, 4, 5, 8),&
s_{15}=(1, 2, 4, 6),&
s_{16}=(1, 2, 3, 6),\cr
s_{17}=(1, 3, 6, 8),&
s_{18}=(1, 4, 6, 8),&
s_{19}=(1, 2, 4, 9),&
s_{20}=(1, 2, 3, 9).\cr
}}$$

We now apply the algorithm in section~2; step {\sl iv)} is
straightforward but tedious. The generators of the Mori cone
are  \foot{There is an interesting point here.  The Mori cone calculated here
is the Mori cone of the toric variety determined by our triangulation of
$\Delta^*$.  This cone differs from the Mori cone of the Calabi--Yau
hypersurface.  To our knowledge, this is the first known example of this
phenomenon; see also Appendix~A.3.  We return to this point in section~4.3.}
\eqn\mcexotic{
\eqalign{
l_1&=(-1,\-1,\-0,\-0,\-0,\-1,\-1,\-0,-2,\-0)\cr
l_2&=(\-0,\-0,\-1,-1,-1,\-0,\-0,\-3,\-0,-2)\cr
l_3&=(-1,\-1,\-0,\-1,\-1,\-0,\-0,-2,\-0,\-0)\cr
l_4&=(\-0,\-0,\-1,\-0,\-0,\-0,-2,\-0,\-1,\-0)\cr
l_5&=(\-0,-1,-1,\-0,\-0,\-0,\-1,\-0,\-0,\-1)\cr
}}

{}From the dual of the Mori cone (which is the K\"ahler cone of the toric
variety in which $X_{13}(1,2,3,3,4)$ is a hypersurface) we get the classical
intersection numbers, see Table~B.1.  As a check, the identical
intersection numbers have been calculated using \ref\schubert
{S.~Katz and S.A.~Str\o mme, {\sl Schubert: a Maple package for
intersection theory}.  Available by anonymous ftp from ftp.math.okstate.edu
or linus.mi.uib.no, cd pub/schubert.}.

To each of the above generators we associate a coordinate in the usual way,
\ie\
$$\z1=-{\a1 \a5 \a6\over \a8^2 \a0},\quad\z2={\a2 \a7^3\over \a3 \a4 \a9^2},
\quad\z3=-{\a1 \a3 \a4\over \a0 \a7^2},\quad
\z4={\a2 \a8\over \a6^2},\quad\z5={\a6 \a9\over \a1\a2}~.
$$

As in the previous example, we use the residue expression for the period
integral to derive the differential equation,
\eqn\periods{\Pi_i(a_0,\ldots,a_p)=
             \int_{\gamma}\int_{\Gamma_j}{\omega\over P\,},
             \quad j=1,\dots , 2(h^{2,1}+1)\,,}
where $P$ is the Laurent polynomial,
$P=\sum_i a_i\phi_i\equiv \sum_i a_i \prod_j Y_j^{\nu^{*(i)}_j}$. Contrary
to the examples earlier in this section we do not have an expression for which
the
mirror is defined as a quotient of a model defined by a transverse
hypersurface in a weighted projective space. However, we can still
write down the differential operators based on the relations between the
$\phi_i$ or equivalently between the $l_i$, the generators of the Mori cone
given in~\mcexotic.
At order two and three we get 6 and 13 differential operators in this
fashion. Since there are five moduli, we would need 10 differential
equations of order two---there are only five non-trivial elements of degree
2, but $5\cdot 6/2=15$ ``monomials'' $\phi_i\phi_j$, with the indices running
over the five moduli, which give ten relations. In most cases (see e.g.
\refs{\hkty,\hktyII})
this problem can be solved by factorization, as was the case
in the previous example. However, one can check (e.g. by using the
code INSTANTON) that this is not the case here. As discussed in~\hkty\
one can circumvent this obstacle by explicitly using the ideal based on
the defining polynomial by which the mirror is defined. Since we lack a
monomial representation of the complex structure deformations in terms of
a defining equation\foot{Alternatively, we could have used the Laurent
polynomial and the relations from the ideal as explained in section~2.
However, we find it easier to use method outlined below.}, we will make use
of the construction of a ring structure
due to Batyrev and Cox~\bc\ as discussed in section~2.
One then gets the following polynomial
\eqn\batcox{
\eqalign{
W=& \a1 x_1 x_2^{13} x_3 x_8 x_{13}+
 \a2 x_3^6x_4x_5^5x_7^2x_9x_{10}^5x_{12}^2 +
  \a3 x_4 x_5 x_6^3 x_7^3 x_8^4 +
  \a4 x_9 x_{10} x_{11}^3 x_{12}^3 x_{13}^4 +\cr
  &\a5 x_1^3 x_4^2 x_6 x_9^2 x_{11}+
  \a6 x_1 x_2^4 x_3^4 x_4 x_5^3 x_7 x_9 x_{10}^3 x_{12}+
  \a7 x_2^6 x_6 x_7 x_8^2 x_{11} x_{12} x_{13}^2+\cr
  &\a8 x_1^2 x_2^8 x_3^2 x_4 x_5 x_9 x_{10} +
  \a9 x_2^9 x_3^3 x_5^2 x_7 x_8 x_{10}^2 x_{12} x_{13} +
  \a{10} x_1 \ldots x_{13}
}}

The idea is now to use the traditional reduction scheme \`a la
Dwork--Griffiths--Katz~\refs{\dwork,\Katz,\griffiths}.
It is rather straightforward to go through the
second order monomials, $\phi_i\phi_j$ and to show that
the only non-trivial relation is\foot{Strictly speaking there is one further
relation involving a second order monomial which cannot be written as
$\phi_i\phi_j$, a phenomenon which was treated in~\hkty. However, as it
turns out the third order relations (differential operators) which are
obtained in this way are encompassed by the  second order operators
found below.}
\eqn\bcrel{
\a1\f9^2+\a0\f1\f7+3\a5\f0\f6+\a6\f1\f9+2\a8\f1\f5~=~0.
}
Note that unlike the generic case of the reduction method there
is no lower order piece from the partial integration.

This is still three short of the ten operators which we need, \ie\
at the current level the ring is not adequate for describing the
moduli space of interest. Rather than insisting on using all of the
thirteen $x_i$, let us now restrict to
$\x1,\x2,\x5,\x8,\x{11}$, \ie\ formally set the remaining $x_i=1$. Renaming
these $y_i,i=1,\ldots,5$ the Batyrev-Cox potential is then reduced to
\eqn\bcreduced{
\eqalign{
\tilde W=&\a1\y1\y2^{13}\y4 +\a2\y3^5 +\a3\y3\y4^4+\a4\y5^3+\a5\y1^3\y5+\cr
&\a0\y1\y2\y3\y4\y5+\a6\y1\y2^4\y3^3+\a7\y2^6\y4^2\y5+\a8\y1^2\y2^8\y3+
\a9\y2^9\y3^2\y4~.}
}
Note that the first five terms can be thought of as obtained from a
transposition of a degenerate potential for the original model, degenerate
because we need to add extra terms in order to make it transversal; this
was indeed the reason why the above $x_i$ were chosen. Since the $x_i$
correspond to points in $\Delta$, choosing the five $x_i$ corresponds to
refraining from resolving the singularities of $X$, the existence of which the
remaining $x_i$ are based on.

Finally, let us then consider the following three monomials
\eqn\opfinal{
\f6\f8,\quad\f7^2,\quad\f0\f7~,
}
which are the among the fifteen $\phi_i\phi_j,i,j=0,6,...,9$ which
do not appear in any of the previous relations at second order among the
$\phi_i\phi_j$.
By vigorous application of the ideal, $\del \tilde W$,
 it is possible in all three cases
to arrive at non-trivial relations of second order involving only monomials
of the form $\phi_i\phi_j$. This completes the story; together with the
 operators obtained from the Mori cone,~\mcexotic, and~\bcrel\
it can be shown that
the triple intersection numbers on the original toric variety
$\IP^4(1,2,3,3,4)$ are reproduced, see Table~B.1.
Note that they differ from those we would compute from the K\"ahler cone of
the manifold $X_{14}(1,2,3,3,4)$, see the discussion in section~4.3.
In Table~B.2 we record
the lowest order
instanton corrections, some of which  are verified in the following section.

\newsec{Verifications.}
In this section we geometrically explain some of the instanton numbers
computed in the earlier sections.

\subsec{$\IP^4(1,1,1,2,2)$}
The first step is to desingularize $\IP^4(1,1,1,2,2)$.  This has already
been accomplished implicitly in \cornersdeltadual\ by the presence
of the vector $\nu^{*(6)}$, since the other 5~vectors in \cornersdeltadual\
determine the toric fan for $\IP^4(1,1,1,2,2)$.  However, we prefer to
use an explicit calculation.

$\IP^4(1,1,1,2,2)$ is singular along the $\IP^1$ defined by $x_1=x_2=x_3=0$.
We desingularize by using an auxiliary $\IP^2$ with coordinates
$(y_1,y_2,y_3)$ and define
$$\tilde{\IP^4}\subset\IP^4(1,1,1,2,2)\times\IP^2$$
by the equations
\eqn\desing{
x_iy_j=x_jy_i,\qquad i,j=1,2,3.
}
The exceptional divisor is just $\IP^1\times\IP^2$, where the two projective
spaces have coordinates $(x_4,x_5)$ and $(y_1,y_2,y_3)$ respectively.
The proper transform of a general degree~7 hypersurface $X$ is seen to
intersect the exceptional divisor in a surface defined by a polynomial
$f(x_4,x_5,y_1,y_2,y_3)$ which is cubic in the $x$'s and linear in the
$y$'s.  This can be seen by direct calculation, the essential point being
the presence of monomials cubic in $x_4,x_5$ and linear in $x_1,x_2,x_3$
in an equation for $X$.  The fibers of the projection of this surface
to $\IP^1$ are lines; thus the desingularized Calabi--Yau manifold $\X$
contains a ruled surface with $\IP^1$ as a base.

This fact was noticed and pointed out to us by D.R.~Morrison.
Other examples of Calabi--Yau
manifolds containing ruled surfaces
parameterized by curves of higher genus have been studied in \hktyII\ and
in~\cdfkm.  The presence of a ruled surface with $\IP^1$ as base yields
different and interesting geometry; this has in part motivated us
to calculate the instanton numbers for this example.

To understand this ruled surface, we first think of $\IP^1\times\IP^2$
as the projective bundle $\IP({\cal O}^3)$ on $\IP^1$ (see
\ref\eishar{D.~Eisenbud and J.~Harris, Proc.\ Symp.\ Pure Math.\
{\bf 46} Part I (1987) 3.}
for notation).  Writing $f=f_1(x)y_1+f_2(x)y_2+f_3(x)y_3$, we see that
$f$ is determined by the three $f_i$; these can be identified with a
map ${\cal O}(-3)\to {\cal O}^3$; the cokernel of a generic such map
is ${\cal O}(1)\oplus{\cal O}(2)$.  So the exceptional divisor of $X$
is just $\IP({\cal O}(1)\oplus{\cal O}(2))$.  Rational ruled surfaces
such as this are very well understood \eishar.  We quickly review the relevant
points.

Let $B$ denote the rank~2 bundle ${\cal O}(a)\oplus{\cal O}(b)$ on $\IP^1$
for some integers $a\le b$.  The abstract rational ruled surface $\IP(B)$ is
isomorphic to the Hirzebruch surface $F_{b-a}$; this surface is characterized
by the minimum value of $C^2$ for $C$ a section of the ruled surface, the
minimum being $a-b$.  The section achieving this minimum is even unique
if $a<b$.

$H^2(\IP(B))$ is generated by two classes---a hyperplane class $H$ and the
fiber $f$.  The cohomology $H^0(mH+nf)$ is calculated for $m\ge 0$ as
$H^0(\IP^1,{\rm Sym}^m(B)\otimes{\cal O}(n))$.  The space of curves in the
class $mH+nf$ is parameterized by the projectivization of this vector space.
We also note the intersection numbers $H^2=a+b,\ H\cdot f=1,\ f^2=0$.
In our case, this gives
\eqn\intersec{
H^2=3,\ H\cdot f=1,\ f^2=0.
}
We also note that the curve of minimum self intersection is in the
class $H-2f$.

In particular, if any families of rational curves are of this type,
and are parameterized by $\IP^r$, then its contribution to
the Gromov-Witten invariant for this
type of curve is $(-1)^r(r+1)$ by \cdfkm.  Later, we will also need that if
a family of rational curves is parameterized by a smooth curve of genus~g,
then its contribution to the Gromov-Witten invariant is $2g-2$.

The classes $J_1$ and $J_2$ are interpreted as the classes given by the
zeros of quadratic and linear polynomials, respectively.  Since all linear
polynomials vanish along $x_1=x_2=x_3=0$, and therefore along the
exceptional divisor $E$ after the blowup, we conclude that
$J_1=2J_2+E$.

To compute $n^r_{i,j}$, we note that if $i<2j$ and if $C\cdot J_1=i,\
C\cdot J_2=j$, then $C\cdot E=i-2j<0$, which implies that a component of
$C$ must be contained in $E$.  So in this case we are often reduced to
understanding curves in $E$.  Next, we need to observe that
$J_1|_E\simeq f$ and $J_2|_E\simeq H$.

In the case of $n^r_{0,j}$ for $j\ge 1$, we see inductively that all
of these curves lie in $E$.  For these curves $C$, we have that
(thought of as a curve in the ruled surface $E$) $C\cdot f=0$ and
$C\cdot H=j$.  The first equation and~\intersec\ say that $C$ must be a
fiber; in that case, necessarily $C\cdot H=1$, so $j$=1.  If $j=1$, then
we have
$$H^0(\IP(B),f)\simeq H^0(\IP^1,{\cal O}(1))\simeq\IC^2$$
so the curves
are parameterized by $\IP^1$ and $n^r_{0,1}=-2$.  We have also shown that
$n^r_{0,j}=0$ for $j\ge 2$.

For $n^r_{1,j}$ with $j\ge 2$,
these curves $C$ are again contained in $E$ (at this point,
it is clear that a component is contained in $E$; but we will
check later that all such curves are entirely contained in $E$).

We have $C\cdot f=1$ and $C\cdot H=j$, which implies that $C$ is of
type $H+(j-3)f$.  In addition, $C\cdot f=1$ implies that $C$ is a
section or the union of a section and fibers, hence is rational.
Finally, $H^0(\IP(B),H+(j-3)f)$ is isomorphic to
$$H^0(\IP^1,B\otimes {\cal O}(j-3))=H^0(\IP^1,{\cal O}(j-2)\oplus
{\cal O}(j-1)),$$
which has dimension $2j-1$.  So the curves are parameterized by a
projective space of dimension $2j-2$.  This gives $n_{1,j}=2j-1$ for
$j\ge 2$.

Note in passing that we have also shown that $n^e_{j,k}=0$ for
$j=0,1$, since only rational curves arise as sections or fibers.
In fact, for all of our two parameter models we note that for exactly the same
reason as given above, a curve $C$ of type $(0,1)$ necessarily
lies in the exceptional divisor $E$.  In each case, $E$ is
an explicitly given rational surface, and by consideration of the
possible curves on $E$, the condition $C\cdot J_2=1$ is either not
possible or forces $C$ to be rational.  Either way, we get $n^e_{0,1}=0$,
justifying the procedure given in section~3.1 for determining $r_0$.

Consider $n^e_{2,5}$.  These curves are contained in $E$ by the
now-familiar argument. Since $C\cdot f=2$ and $C\cdot H=5$, we get
that $C$ is in the class $2H-f$.  This is a family of elliptic curves,
and is parameterized by a $\IP^8$.  This gives $n^e_{2,5}=9$.
The general curve in the class $2H+bf$ is elliptic only for $b=-1$.  This
leads quickly to the conclusion that $n^e_{2,j}=0$ for $j\neq 5$.

We can make further verifications (and finish an earlier argument) by
using $J_2$.  The three linear forms which are the sections of $J_2$ give
a map to $\IP^2$.  Actually, we study this by means of the map
$g:\tilde{\IP^4}\to \IP^2$ given by projection onto $(y_1,y_2,y_3)$;
the map on the Calabi--Yau
hypersurface $\X$ is given by restriction.  The fibers of $g$ are isomorphic
to $\IP^2$; in fact $\tilde{\IP^4}$ can be constructed as the projective
bundle $\IP({\cal O}^2\oplus{\cal O}(2))$ on $\IP^2$.  The ${\cal O}^2$
corresponds to the coordinates $x_4,x_5$; the ${\cal O}(2)$ arises because
$y_1,y_2,y_3$ can be rescaled in $\tilde{\IP^4}$ independent of $x_4,x_5$.
Like all projective bundles, this one comes with a tautological rank~1
quotient, the bundle of linear forms in the fibers.  In this case,
it turns out to be $J_1$.  That is, there is a map
$g^*({\cal O}^2\oplus{\cal O}(2))\to J_1$.  The hypersurface
$\X$ is in the class $3J_1+J_2$.  This says that fibers of $g$, restricted to
the hypersurface of interest, are plane cubic curves.  The fibers are readily
seen to be of type $(3,0)$.

We are now equipped to compute $n^r_{1,0}$.  If $C\cdot J_2=0$, then
$C$ is contained in a fiber of $g$.  The condition $C\cdot J_1=1$ says
that $C$ is a line.  Hence we must enumerate points of $\IP^2$ with
the property that the cubic fiber factors into a line and a conic.
In particular, there is a 1-1 correspondence between lines and conics;
hence $n^r_{1,0}=n^r_{2,0}$.

The technique is standard
\ref\ak{A.~Altman and S.~Kleiman, Comp.\ Math.\ 34 (1977) 3.}.
We form the Grassmann bundle
$G={\rm Gr}^2({\cal O}^2\oplus{\cal O}(2))$ of lines in the fibers of $g$.
Note that ${\rm dim}(G)=4$.  There is a projection map $\phi:G\to\IP^2$.
On $G$, there is a canonical rank~2 bundle $Q$, the bundle of linear forms on
these lines. The equation of $\X$ induces a section of the rank~4 bundle
${\rm Sym}^3(Q)\otimes\phi^*{\cal O}(1)$.  The line is contained in
$\X$ if and only if this section vanishes at the corresponding point
of $G$.  So we must calculate $c_4({\rm Sym}^3(Q)\otimes\phi^*{\cal O}(1))$.
This is immediately calculated to be 177 using Schubert
\schubert.

Turning to $n^r_{1,1}$, we note that $C\cdot J_1=1$ and $C\cdot J_2=1$
implies that a component of $C$ is contained in $E$.  There are
two cases: either $C\subset E$, or $C$ is a union of a curve of type
$(1,0)$ and a curve of type $(0,1)$ (we have already seen that the latter
curve is a fiber of $E$).

In the former case, $C\cdot f=1$
and $C\cdot H=1$ implies that $C$ is in the class $H-2f$.
Since $H^0(\IP(B),H-2f)$ is 1~dimensional, there is only one curve of
this type, contributing 1 to $n^r_{0,1}$.  Alternatively, note that
$(H-2f)^2=-1$, the minimum value; hence $C$ is unique.

If on the other hand $C=C'\cup C''$ with $C'$ of type $(1,0)$, then
$C'\cdot E=C'\cdot(J_1-2J_2)=1$.  For $C$ to be connected, $C''$ must
be the unique fiber $f$ containing the unique point of intersection on
$C'$ and $E$.  So each curve of type $(1,0)$ yields a degenerate
instanton (see appendix to~\bcov) of type $(1,1)$.

Putting these two cases together, we see that $n^r_{1,1}=
n^r_{1,0}+1=178$.

Returning to the final details of $n^r_{1,j}$ for $j\ge 2$, we see that
if $C$ were not contained in $E$, then it would have to be a union
$$C=C'\cup C_1\cup\ldots\cup C_j$$
where $C'$ has type $(1,0)$ and the $C_i$ have type $(0,1)$.  But $C'$
meets $E$ in just one point, so there is no way to add on $j$ fibers
and obtain a connected curve.  So this case does not occur, as asserted
earlier.

Next, $n^e_{3,0}$ is easy; these are the elliptic cubic curves in the
fibers of $g$.  But the general fibers of $g$ {\it are\/} elliptic
cubic curves.  So these curves are parameterized by $\IP^2$; hence
$n^e_{3,0}=3$,

The calculation of $n^r_{3,0}$ is more intricate.  Here we must study
the rational cubic curves in the fibers of $g$.  A cubic curve is rational
only if it is singular.  So we investigate the locus of singular cubic
curves.  This is seen to be a plane curve of degree~36.
But this curve is singular.  In fact, the curve has nodes at each of the
177~cubics that we found earlier that factor into a line and a conic.
This is because such curves have 2~singularities, which can be smoothed
independently, yielding 2~distinct tangent directions in the space of
singular curves.  In addition, there are 216~cusps.\foot{The statements about
the degree and number of cusps can for instance be checked by the standard
technique of considering pairs $(C,p)$ where $C$ is one of the cubic curves
and $p\in C$, then projecting onto $C$.  The Schubert code for these
calculations is available upon request.}

It will be illuminating to generalize this situation before continuing
the calculation.  We suppose that we have a family of nodal elliptic
curves with parameter space of curve $B$ of arithmetic genus $p_a$,
containing $\delta$ nodes and $\kappa$ cusps.  The nodes parameterize
2-nodal elliptic curves (which necessarily split up into 2 smooth rational
curves, meeting twice), and the cusps parameterize cuspidal curves.
In our example, $p_a=(36-1)(35-1)/2=595$, while $\delta=177$ and
$\kappa=216$.

Resolving the singularities of $B$, we get a curve $\tilde{B}$ of
genus $p_a-\delta-\kappa$; and the computation from \cdfkm\ shows that
we must make a correction for the cusps, finally obtaining the
Gromov-Witten invariant $c_1(\Omega^1(\tilde{B}))-\kappa=
2(p_a-\delta-\kappa)-2-\kappa$, which simplifies
to $2p_a-2\delta-3\kappa-2$.  In our example, this gives $n^r_{3,0}=186$.

This is not coincidentally the negative of the Euler characteristic of
the target space $\tilde{X}$.  We calculate the Euler characteristic
by decomposing $\tilde{X}$.  Again we generalize, assuming that there
is a map $f:\tilde{X}\to S$,
where $S$ is a smooth complex surface. We suppose that the general fiber
of $f$ is a smooth elliptic
curve, while over some curve $B$ contained in the surface, the fiber
degenerates in the manner described above.  We keep the notation $p_a,\delta,
\kappa$ as before.

We divide $S$ into four pieces: $S-B$, the complement of the nodes and cusps
in $B$, the nodes, and the cusps.  We can then divide $\tilde{X}$ into
four corresponding pieces, the inverse images of these four pieces via $f$.
We will calculate the Euler characteristic of these four pieces.

We start the calculation by a preliminary calculation on $B$.
We obtain a node by pinching a 1-cycle
to a point, and a cusp by pinching two 1-cycles.  If we then remove the
resulting singularities, we compute that the Euler characteristic of the
complement of the nodes and cusps in $B$ is $2-2p_a+\delta+2\kappa-(\delta+
\kappa)$, which simplifies to $2-2p_a+\kappa$.

We also need to observe that the Euler characteristic of a smooth elliptic
curve, the nodal curves, the 2-nodal curves, and the cuspidal curves
have respective Euler characteristics $0,\ 1,\ 2,\ 2$.

We can obtain the Euler characteristic of each piece by multiplying
the Euler characteristics of the base and the fiber, then summing
over all pieces.  The result is
$0+(2-2p_a+\kappa)(1)+\delta(2)+\kappa(2)$, which simplifies
to $2-2p_a+2\delta+3\kappa$.  This is plainly the negative of the
Gromov-Witten invariant obtained above.  This phenomenon occurs in
several of the the examples discussed in \refs{\hkty,\hktyII\cdfkm}.

For $n^e_{3,1}$, we claim that a curve $C$ with $C\cdot J_1=3$ and
$C\cdot J_2=1$ is reducible.  Suppose it were irreducible.  Then $J_2$
restricts to a degree~1 bundle on $C$, which has a 1~dimensional space of
global sections.  Since $J_2$ has a 3~dimensional space of global sections
(spanned by $y_1,y_2,y_3$), the kernel of the restriction map is at least
2~dimensional.  This constrains $C$ to lie in a fiber of $g$.  But we have
already noted that the fibers of $g$ are of type $(3,0)$, so this is
impossible.  Looking at our prior discussion of curves of type
$(a,b)$ with $a\le 3$ and $b\le 1$, we see that the only possibility is
$C=C_1\cup C_2$, where the $C_i$ are
of type $(3,0)$ and $(0,1)$ respectively.
Such curves are parameterized by $E$---take the fiber $f$
through the point $p\in E$ union the fiber of $g$ through
the image of $p$ in $\IP^2$.  Since $c_2(E)=4$, this gives
$n^e_{3,1}=4$.

Finally, we can easily see that $n_{4,0}=n_{5,0}=177$.  For each of
the 177~line-conic pairs, we have two families of degenerate instantons.
These are analyzed by the method in the appendix to \bcov; since our situation
is simpler, we will content ourselves with sketching the construction.
The first consists of maps from $\IP^1$ to the conic, with lines bubbling
off at each of the two nodes.  These are degenerate instantons
of type $(4,0)$.  The other family comes from maps from $\IP^1$ to the line,
with conics bubbling off at each of the nodes.  These are degenerate
instantons of type $(5,0)$.

One can in fact verify that there is a non-zero contribution to $n_{i,0}$
for all $i$ by using stable maps \ref\kont{M.~Kontsevich, {\sl Enumeration
of Rational Curves via Torus Actions}, alg-geom/9405035.} as the method
to compactify the space of instantons.  All stable maps needed will restrict
to degree~1 maps on each irreducible component.

If $i=3k$, we can take a tree
$C=C_1\cup\cdots\cup C_k$ of rational curves, and map the intersection
points of the components to the nodes of any of the nodal cubic curves
described above, with the rule that if $C_{j-1}$ maps to one branch of a nodal
cubic near $p=C_{j-1}\cap C_j$, then $C_j$ maps to the other branch near
$p$.

If $i=3k+1$, we take $C=C_1\cup\cdots\cup C_{2k+1}$, mapping $C_{2j-1}$
to one of the lines found above, and mapping $C_{2j}$ to the intersecting
conic.  Each $C_{l-1}\cap C_l$ maps to one of the two intersection points of
a line and a conic.  If $l<2k+1$, then $C_l\cap C_{l+1}$ maps to the other
intersection point.

The case $i=3k+2$ is similar, this time mapping $C_{2j-1}$ to conics and
$C_{2j}$ to lines.

\subsec{$\IP^4(1,1,1,1,3)$}
We desingularize the weighted projective space
in this case by using an auxiliary $\IP^3$ with coordinates
$(y_1,y_2,y_3,y_4)$ and define
$$\tilde{\IP^4}\subset\IP^4(1,1,1,1,3)\times\IP^3$$
by the equations
$$x_iy_j=x_jy_i,\qquad i=1,\ldots,4.$$
The exceptional locus is clearly a copy of $\IP^3$.  The proper transform of
a general degree 7 hypersurface is seen to intersect this locus in a linear
hypersurface, \ie\ a projective plane.
This can be seen by direct calculation, the essential point
being the presence of the monomials $x_ix_5^2$ for $i=1,\ldots,4$ which
are linear in the variables $x_1,\ldots,x_4$.

Now $J_1$ and $J_2$ are respectively the classes of degree~3 and degree~1
polynomials.  This leads to the equality $J_1=3J_2+E$, where $E$ is the
class of the exceptional $\IP^2$.  So if a curve $C$ with $C\cdot J_i=a_i$
and $a_1<3a_2$, then a component of $C$ necessarily lies on $E$.

To compute $n^r_{0,j}$, we see that these curves lie on $E$, and have degree
$j$ as a curve on $E$.  Since lines are parameterized by $\IP^2$, we get
$n^r_{0,1}=3$.  Since conics are parameterized by $\IP^5$, we get
$n^r_{0,2}=-6$.  Elliptic curves are similar; the lowest degree possible
is cubic, so $n^e_{0,j}=0$ for $j=1,2$, while $n^e_{0,3}=-10$ since plane
cubics are parameterized by a $\IP^9$.

To compute $n^r_{1,0}$, we note that a curve $C$ with $C\cdot J_2=0$ maps
to a point under the projection to $\IP^3$, since the projection map is
defined by the sections of $J_2$.  We write the equation of the proper
transform of the hypersurface
as $f_7(y)+f_4(y)x_5+f_1(y)x_5^2$, where the $f_j(y)$
are homogeneous polynomials of degree $j$ in $y_1,\ldots,y_4$.  The 28~values
of $y$ for which $f_7(y)=f_4(y)=f_1(y)=0$ yield a $\IP^1$ in the
hypersurface (since $x_5$ may be arbitrary); this curve is easily seen
to have $J_1$-degree~1.  This gives $n^r_{1,0}=28$.  This geometry also
implies that $n^r_{i,0}=0$ for $i>1$.
Each such curve $C$
satisfies $C\cdot E$=$C\cdot J_1-3C\cdot J_2=1$, so that $C\cap E$
is a point $p$.  We can get reducible
curves with $C\cdot J_1=1$ and $C\cdot J_2=1\ ({\rm resp.\ }2)$ by taking
the union of $C$ with a line (resp.\ a conic) in $E$ passing through $p$.
Since lines (resp.\ conics) passing through $p$ are parameterized by
$\IP^1$ (resp.\ $\IP^4$), we get $n^r_{1,1}=28(-2)=-56$ and
$n^r_{1,2}=28(5)=140$.
Also, we can construct rational curves of type $(2,1)$ by taking a pair
of curves of type $(1,0)$ and taking their union with the unique line
in $E$ passing through the respective intersection points of these two
curves with $E$.  Thus $n^r_{2,1}={28\choose 2}=378$.

Irreducible elliptic curves $C$ of type $(5,2)$ satisfy $C\cdot E=-1$,
so these lie on $E$.  Such curves do not exist, since we know that
the plane curves in $E$ have type $(0,j)$ for some $j$.  The only possible
reducible curves are obtained as follows.  Consider one of the rational
curves $C_1$ of type $(5,1)$.  These meet $E$ in 2~points, which may be joined
by the unique line $C_2$ (of type $(0,1)$) meeting these points.  The curve
$C=C_1\cup C_2$ is of type $(5,2)$, and is elliptic.  This verifies
$n^e_{5,2}=n^r_{5,1}$.

Considering elliptic curves of type $(i,3)$ for $i<9$, we see that such
curves have a component contained in $E$.  We see that if $i<5$,  this
component
must be one of the $(0,3)$ elliptic curves considered above, and there are $i$
additional components, each rational curves of type $(1,0)$.  This
gives $n^e_{i,3}={28\choose i}\cdot(-1)^{i-1}(10-i)$, since plane cubics
passing through $i$ fixed points are parameterized by a $\IP^{9-i}$.
These are in agreement with the invariants produced from the instanton
expansion.  This does not hold for $5\le i<9$, due to the presence of
reducible elliptic curves containing a rational component of type
$(5,1)$.

\subsec{$\IP^4(1,2,3,3,4)$}
We start by describing the geometry of the toric variety
$\IP_{\Delta^*}$, especially its K\"ahler cone.
It will be convenient in the sequel to consult the following table.

\eqn\Moritable{\matrix{
\nu^{*(1)} & (-2,-3,-3,-4) & 1 & 0  & 1  & 0  & -1\cr
\nu^{*(2)} & (1,0,0,0)     & 0 & 1  & 0  & 1  & -1 \cr
\nu^{*(3)} & (0,1,0,0)     & 0 & -1 & 1  & 0  & 0 \cr
\nu^{*(4)} & (0,0,1,0)     & 0 & -1 & 1  & 0  & 0 \cr
\nu^{*(5)} & (0,0,0,1)     & 1 & 0  & 0  & 0  & 0 \cr
\nu^{*(6)} & (0,-1,-1,-1)  & 1 & 0  & 0  & -2 & 1 \cr
\nu^{*(7)} & (-1,-1,-1,-2) & 0 & 3  & -2 & 0  & 0 \cr
\nu^{*(8)} & (-1,-2,-2,-2) & -2& 0  & 0  & 1  & 0 \cr
\nu^{*(9)} & (-1,-2,-2,-3) & 0 & -2 & 0  & 0  & 1}}

Here, the five columns on the right come from the representation of the
Mori cone found in~\mcexotic\
(the first coordinate of the earlier representation, corresponding to
the origin, is not needed in the current context).  This
representation gives a basis for the divisor class group of
the toric variety; the coordinates of the classes of the divisors $D_i$
given by the equations $x_i=0$
(corresponding to the edges $\nu^{*(i)}$) can be read off horizontally in
the above.

Now we find the edges of the K\"ahler cone by taking the dual basis
(recall that the K\"ahler cone is dual to the Mori cone).  We find linear
combinations of the generators $D_1,\ldots,D_9$ such that in the coordinates
given by the corresponding rows, we get the five standard basis vectors of
$\Z^5$.  There are many ways of expressing these in terms of the $D_i$;
here is one choice.

$$\eqalign{
J_1 &= D_5 \cr
J_2 &= D_2+D_5+D_6+D_8    \cr
J_3 &= D_2+D_3+D_5+D_6+D_8 \cr
J_4 &= 2D_5+D_8    \cr
J_5 &= 3D_5+D_6+2D_8    \cr
}$$

Toric varieties can also be described as quotients of torus actions
\lref\audin{M.~Audin,
Prog.\ Math.\ {\bf 93}, Birkh\"auser, Basel 1991.}
\lref\cox{D.~Cox, {\sl The Homogeneous Coordinate
Ring of a Toric Variety}~\refs{\audin,\cox},
Preprint, Amherst College, 1993, alg-geom/9210008.}.
We see that the toric variety $\IP_{\Delta^*}$
is naturally identified with $(\IC^9-F)/{\IC^*}^5$,\foot{
This and other calculations in this section were performed using
an extensive collection of Maple procedures for toric varieties
which has been written jointly by
the second author and S.A.~Str\o mme; a modified version of these will
appear in the next release of Schubert.}
 where
$$F=\{x_2=x_7=0\}\cup \{x_2=x_8=0\} \cup \{x_5=x_9=0\} \cup \{x_6=x_7=0\}
\cup \{x_6=x_9=0\}$$
$$\cup \{x_7=x_8=0\} \cup \{x_8=x_9=0\} \cup
\{x_1=x_2=x_5=0\} \cup \{x_1=x_3=x_4=0\}$$
$$\cup \{x_3=x_4=x_6=0\} \cup
\{x_3=x_4=x_8=0\} \cup \{x_1=x_5=x_6=0\}.$$

The only singular point of the toric variety $\IP_{\Delta^*}$ is the point
$x_2=x_3=x_4=x_9=0$
corresponding to the determinant~3 cone spanned by $\nu^{*(2)},\nu^{*(3)},
\nu^{*(4)},\nu^{*(9)}$.  Since the general Calabi--Yau hypersurface $X$
misses this point, $X$ is smooth.

We now identify the classes of curves $C_1\ldots,C_5$ which generate the
Mori cone.  These curves by definition are dual to the divisors $J_i$ via
the intersection pairing.
The intersection numbers $C_i\cdot D_j$ can be read from the
last five columns of \Moritable, at least up to a scalar multiple of
each column.  If the intersection number is negative,
then necessarily $C_i$ is contained in $D_j$.  So we see that $C_1\ldots
C_5$ are respectively contained in the divisor $D_8$, the empty set
$D_3\cap D_4\cap D_9$, the divisor $D_7$, the divisor $D_6$, and the
curve $D_1\cap D_2$.

We immediately observe that the Mori cone of $X$ differs from the Mori
cone of the toric variety, since $C_2$ is not present on $X$.  For comparison,
the locus $x_3=x_4=x_9=0$ is a curve on the toric variety $\IP_{\Delta^*}$,
whose intersection
numbers with the $D_i$ are given by $1/3$ of the corresponding column of
\Moritable.  The fractional intersection numbers are not surprising, since
$C_2$ passes through the singular point of $\IP_{\Delta^*}$.
The fact that $D_3\cap D_4\cap D_9$ has dimension less than~1 is to be expected
on dimensional grounds; but can be verified conclusively by explicitly writing
down the general equation for $X$, substituting $x_3=x_4=x_9=0$, and
noting that vanishing of the remaining term would imply that the point
is contained in the disallowed set $F$.
So this set is in fact empty.

We next set out to identify the Mori cone of $X$.  Consider the class of
a curve of type $C=\sum_ia_iC_i$.  In the sequel, we will sometimes call
this a curve of type $(a_1,\ldots,a_5)$.
Suppose that this is the class of an
effective curve
on $X$.  We claim that
\eqn\Moriineq{a_1\ge0,a_3\ge a_2\ge0, a_4\ge 0, a_5\ge0.}
The inequalities $C\cdot J_i\ge0$ hold because $C$ is an effective curve on
$\IP_{\Delta^*}$ and the $J_i$ are in the K\"ahler cone.  But $C\cdot J_i=a_i$.
Now suppose that $a_2>a_3$.  Without loss of generality, we may assume
that $C$ is irreducible.  Then $C\cdot D_3=C\cdot D_4=a_3-a_2<0$.  Thus
$C\subset D_3\cap D_4$.  We have already seen that $D_3\cap D_4\cap D_9$
is empty.  The intersection of $D_3\cap D_4$ with any of $D_1,\ D_6,\
{\rm or\ }D_8$ is empty as well, by consideration of the locus $F$.  So
we only have to show that $C\cdot D_i<0$ for at least one of $i=1,6,8,9$.
But the system of inequalities in the $a_i$ deduced from the inequalities
$C\cdot D_i\ge 0$ for $i=1,6,8,9$ together with $a_i\ge 0$ is readily seen
to be inconsistent, justifying \Moriineq.

To show that the system \Moriineq\ indeed defines the Mori cone, we first
note that it is a simplicial cone spanned by $C_1,\ C_2+C_3,\ C_3, C_4,\ C_5$.
So it suffices to show that all of these are classes of a curve on $X$.
We denote these classes by $\tilde{C_1},\ldots,\tilde{C_5}$.

Let $\tilde{J_i},\ i=1,\ldots,5$ be the dual basis of divisors (which
will turn out to be the generators of the K\"ahler cone of $X$).    This
gives explicitly $\tilde{J_i}=J_i$ for $i=1,2,4,5$, and $\tilde{J_3}=
J_3-J_2=D_3$.

We have already seen that if the respective curves $\tilde{C_1},\
\tilde{C_3},\ \tilde{C_4}$ exist,
they lie in the divisors $D_8,\ D_7,\ D_6$ and $\tilde{C_5}$ lies in the curve
$D_1\cap D_2$.  By the same reasoning, we see that $\tilde{C_2}$ lies
in the divisor $D_9$.

Let $\hat{{\cal J}_i}$ denote the wall of the K\"ahler cone obtained by
taking non-negative linear combinations of all of the $\tilde{J_k}$ except
$\tilde{J_i}$.  According to \ref\Wilson{P.M.H.~Wilson, Inv.\ Math. 107
(1992) 561.},
these walls consist of divisor classes which contract $\tilde{C_i}$ (\ie\
the divisors intersect each curve in the locus trivially).
Let us find these loci.  First, we identify all curves
$x_i=x_j=0$ which meet the divisors in
$\hat{{\cal J}_i}$ trivially.  We may ignore the curves which are components
of $F$. Direct calculation gives

$$\matrix{
1 & (x_3=x_8=0)\cup (x_4=x_8=0)\cr
2 & (x_3=x_9=0)\cup (x_4=x_9=0)\cr
3 & (x_5=x_7=0)\cup (x_7=x_9=0) \cr
4 & (x_1=x_6=0) \cup (x_3=x_6=0)\cup (x_4=x_6=0) \cup (x_5=x_6=0)\cr
5 & (x_1=x_2=0)
}$$

The above table identifies interesting curves, which leads to the following
results.

The divisors of $\hat{{\cal J}_1}$ blow down the fibers of a base point free
pencil on $x_8=0$ generated by $x_3,x_4$ (base point free since
$x_3=x_4=x_8=0$ is a component of $F$).  The fibers are calculated to be
rational curves of type $(1,0,0,0,0)$.  This verifies
$n_{1,0,0,0,0}=-2$ and $n_{j,0,0,0,0}=0$ for $j>1$.

The divisors of $\hat{{\cal J}_2}$ blow down the fibers of
a base point free pencil on $x_9=0$ generated by
$(x_3,x_4)$ (while $x_3=x_4=x_9=0$ is non-empty on $\IP_{\Delta^*}$
since it is not a component of $F$, it is in fact empty on $X$).
The fibers are calculated to be
rational curves of type $(0,1,1,0,0)$.  This verifies
$n_{0,1,1,0,0}=-2$ and $n_{0,j,j,0,0}=0$ for $j>1$.

Similarly, the divisors of $\hat{{\cal J}_3}$
blow down the fibers of a base point free
pencil on $x_7=0$ generated by $x_5,x_9$.
Note that while $(x_5,x_9)$ scale differently,
the description of $x_7=0$ as a toric variety shows that
$x_5=0$ and $x_9=0$ are linearly equivalent on $x_7=0$.
The fibers are calculated to be
rational curves of type $(0,0,1,0,0)$.  This verifies
$n_{0,0,1,0,0}=-2$ and $n_{0,0,j,0,0}=0$ for $j>1$.

The divisors of $\hat{{\cal J}_4}$
blow down the fibers of the base point free
pencil on $x_6=0$ generated by $(x_3,x_4)$.
The fibers are calculated to be
rational curves of type $(0,0,0,1,0)$.  This verifies
$n_{0,0,0,1,0}=-2$ and $n_{0,0,0,j,0}=0$ for $j>1$.
Note that the curves $x_1=x_6=0$ and $x_5=x_6=0$ are actually disconnected
unions of 3 and 4 fibers respectively, so contribute nothing new.

The divisors of $\hat{{\cal J}_5}$ blow down the curve
$x_1=x_2=0$, which is easily seen to have three irreducible
connected components, each of which is of type $(0,0,0,0,1)$.
This verifies $n_{0,0,0,0,1}=3$ and $n_{0,0,0,0,j}=0$ for $j>1$.

We can also verify several other numbers.  For example, degenerate
instantons can be formed by taking the unions of transversely intersecting
irreducible rational curves which have no closed loops (hence still have
genus~0).  Here is a short list.

Each of the 3~curves of type $(0,0,0,0,1)$ meets each of $x_6=0$ and $x_9=0$
once. So there is a unique fiber of either or both of these surfaces that
can be joined to each of these 3~curves to get a rational curve (a glance
at $F$ shows that $x_6=0$ and $x_9=0$ are disjoint).  This verifies
$n_{0,0,0,1,1}=n_{0,1,1,0,1}=n_{0,1,1,1,1}=3$.

Now consider the curve $C$ given by $x_7=x_9=0$.  This is one of the
$\hat{{\cal J}_3}$-exceptional curves already considered, so is of type
$(0,0,1,0,0)$.  The curve $C$ meets each of the exceptional
fibers of type $(0,1,1,0,0)$
on $x_9=0$ in one point, so can be joined to each of these fibers. We
again get a family of curves
parameterized by $\IP^1$.  This verifies $n_{0,1,2,0,0}=-2$.  Now each of
the 3~curves of type $(0,0,0,0,1)$ can be joined to a unique element of this
family; and then again a unique fiber of $x_6=0$ can be joined afterwards.
This verifies $n_{0,1,2,0,1}=n_{0,1,2,1,1}=3$.

Note also that the fibrations on $x_6=0$ and $x_8=0$ are defined by the
same pencil $(x_3,x_4)$, and that the corresponding fibers meet in one
point.  Taking unions again gives a family of curves parameterized by
$\IP^1$.  Then each of the 3~curves of type $(0,0,0,0,1)$ can be joined
to precisely one of these, yielding a rational curve (each such curve meets
$x_6=0$ once but is disjoint from $x_8=0$).  This verifies
$n_{1,0,0,1,0}=-2$ and $n_{1,0,0,1,1}=3$.

We next observe that curves of type $(0,2,3,0,0)$ are restrictions to
$x_9=0$ of the divisors linearly equivalent to $J_2$.  These restrictions
are parameterized by $\IP^3$ (explicitly, the divisors are defined by linear
combinations of $x_3^2x_7,x_3x_4x_7,x_4^2x_7,x_2x_5x_6x_8$).  This gives
$n_{0,2,3,0,0}=-4$.  Each of the 3~curves of type $(0,0,0,0,1)$ meets
$x_9=0$ in one point; the divisors of $J_2$ passing through this point
(the only ones that can be joined to get a connected curve) are parameterized
by $\IP^2$.  This gives $n_{0,2,3,0,1}=3\cdot 3=9$.  Similarly we can add
on two curves (which restricts divisors of $J_2$ to a $\IP^1$) or all
three (uniquely specifying the divisor of $J_2$).  This verifies
$n_{0,2,3,0,2}=3\cdot(-2)=-6$ and $n_{0,2,3,0,3}=1$, while
$n_{0,2,3,0,j}=0$ for $j>3$.

Finally we observe that curves of type $(0,3,4,0,0)$ are restrictions to
$x_9=0$ of the divisors linearly equivalent to $J_2+J_3$.  These restrictions
are parameterized by $\IP^5$ (explicitly, the divisors are defined by linear
combinations of $f_3(x_3,x_4)x_7,x_2x_3x_5x_6x_8,x_2x_4x_5x_6x_8$, and
$x_1x_5^2x_6x_8^2$, where $f_3(x_3,x_4)$ is a homogeneous cubic; note that
when restricted to $X\cap(x_9=0)$, the last monomial becomes a linear
combination of the others).  This gives $n_{0,3,4,0,0}=-6$, and proceeding as
before, we also get $n_{0,3,4,0,1}=3\cdot 5=15,\ n_{0,3,4,0,2}=3\cdot (-4)=
-12$, and $n_{0,3,4,0,3}=3$.

The calculation of the Mori cone of $X$ shows that the true large complex
structure coordinates are $q_1,q_2q_3,q_3,q_4,q_5$.  In other words, the
boundary divisor $q_3=0$ in the moduli space constructed from toric geometry
gets blown down, slightly changing the geometry of the large complex
structure limit.  The instanton expansions remain valid, the only change
being that we get the more precise result that the instanton expansions
must be power series in $q_1,q_2q_3,q_3,q_4,q_5$, agreeing with our
calculations.

\newsec{Applications and Generalizations}

\subsec{Non-Landau--Ginzburg cases}

The classification of the three dimensional polyhedra
appears as the natural generalization of the work of~\ks.
We will not solve this complicated combinatorial problem here,
but rather give the first examples of reflexive pairs of polyhedra,
which do not come from the list of \ks\ nor from moddings of the examples
in that list \ref\moddings{
M. Kreuzer and H. Starke, \npb405(1993)305, hep-th/9211047
and  hep-th/9412033;
A. Niemeier, {\sl Klassifizierung von
Calabi--Yau-Stringkompaktifizierungen durch
$N=2$-superkonforme LG-Theorien},
Diploma Thesis, TU-Munich (1993)}.
The important observation in finding such examples is the fact
that one can systematically modify a given polyhedron by moving
one corner to a new position, in such a fashion that the new polyhedron
remains reflexive.
We can then use  the general methods for determining
the Picard-Fuchs equations for reflexive polyhedra, which were
developed in the previous section.

As an example of this phenomenon let us consider some of the
two moduli models of
table~3.1 and  move one of  the vertices to a new position without
destroying reflexivity. In table~5.1 we list models constructed by
moving the point
$\nu^{*(1)}$ such that the resulting
Calabi-Yau spaces have $h_{11}<4$. In $a)-c)$, the original location
for $\nu^{*(1)}$ is $(-k_2,-k_3,-k_4,-k_5)$, see~\vstari.
There are no such natural coordinates for the model in table
5.1.d); a choice of coordinates has been made here.\foot{We are grateful to
Maximilian Kreuzer for checking that the models marked with a star in
table~5.1 do  not appear as a result of various moddings
\moddings\ of the models in \ks.}

{\vbox{\ninepoint{
$$a)\quad\vbox{
{\offinterlineskip\tabskip=0pt
\halign{
\strut\vrule#&~#~&\hskip-6pt\vrule#&\hfil$#$~&\vrule#&\hfil$#$~&\vrule#
&\hfil$#$~&\vrule#\cr
\noalign{\hrule}
&$\nu^{*(1)}$
&& \chi
&&h_{11}
&&h_{21}& \cr
\noalign{\hrule}
& $(-1,-1,-2,-2)$
&&-186
&& 2 (0)
&& 95(0)& \cr
& $(-1,-1,-1,-2)^*$
&& -180
&& 2(0)
&& 92(0)&\cr
&  $(-1,-1,-1,-1)$
&& -168
&& 2(0)
&& 86(0) & \cr
&  $(-1,-1,0,0)^*$
&& -162
&&  2(0)
&&  83(0) &\cr
\noalign{\hrule}
&  $(-1,-1,-3,3)$
&& -324
&& 3(0)
&& 165(0)  &\cr
&  $(-1,-1,3,3)^*$
&&  -216
&&  3(0)
&& 111(0)&\cr
& $(-2,-1,-2,0)^*$
&& -156
&&  3(0)
&&  81(0)  &\cr
&  $(-1,-2,0,0)$
&& -144
&&   3(0)
&&  75(1) & \cr
& $(-3,-2,-2,2)^\diamond$
&& -120
&&  3(0)
&& 63(1) & \cr
\noalign{\hrule}}}}
\qquad b)\quad
\vbox{
{\offinterlineskip\tabskip=0pt
\halign{
\strut\vrule#&~#~&\hskip-6pt\vrule#&\hfil$#$~&\vrule#&\hfil$#$~&\vrule#
&\hfil$#$~&\vrule#\cr
\noalign{\hrule}
&$\nu^{*(1)}$
&& \chi
&&h_{11}
&&h_{21}& \cr
\noalign{\hrule}
& $(-1,-2,-2,-6)$
&& -252
&& 2(0)
&& 128(0)&\cr
& $(-1,-1,-1,-3)$
&&-240
&& 2 (0)
&& 122(0)& \cr
&  $(-1,-1,-1,-2)^*$
&& -200
&& 2(0)
&& 102(0) & \cr
&  $(-1,-1,-2,-3)$
&& -208
&&  2(0)
&&  106(0) &\cr
&  $(-1,-1,-1,-1)\diamond$
&& -176
&& 2(0)
&& 90(0)  &\cr
&  $(-1,-1,-1,0)$
&&  -168
&&  2(0)
&& 86(0)&\cr
\noalign{\hrule}
&  $(-1,-1,-3,-6)$
&& -324
&&   3(0)
&&  165(1) & \cr
& $(-1,-2,-2,-4)^*$
&& -192
&&  3(0)
&&  99(0)  &\cr
& $(-1,-2,-2,-2)^*$
&& -156
&&  3(0)
&&  81(0)  &\cr
& $(-1,-2,-2,0)$
&& -144
&&  3(0)
&& 75(2) & \cr
\noalign{\hrule}}}
}
$$
\vskip-20pt

$$
c)\quad\vbox{
{\offinterlineskip\tabskip=0pt
\halign{
\strut\vrule#&~#~&\hskip-6pt\vrule#&\hfil$#$~&\vrule#&\hfil$#$~&\vrule#
&\hfil$#$~&\vrule#\cr
\noalign{\hrule}
&$\nu^{*(1)}$
&& \chi
&&h_{11}
&&h_{21}& \cr
\noalign{\hrule}
& $(0,0,-1,-1)$
&& -168
&& 2(0)
&& 86(0)&\cr
\noalign{\hrule}
& $(-1,0,-2,-2)$
&&-168
&& 3 (0)
&& 87(0)& \cr
&  $(-1,0,-2,0)^*$
&& -172
&& 3(0)
&& 89(0) & \cr
&  $(-1,-1,-2,-3)$
&& -132
&&  3(0)
&&  69(3) &\cr
\noalign{\hrule}}}}
\qquad d)\quad
\vbox{
{\offinterlineskip\tabskip=0pt
\halign{
\strut\vrule#&~#~&\hskip-6pt\vrule#&\hfil$#$~&\vrule#&\hfil$#$~&\vrule#
&\hfil$#$~&\vrule#\cr
\noalign{\hrule}
&$\nu^{*(1)}$
&& \chi
&&h_{11}
&&h_{21}& \cr
\noalign{\hrule}
& $(-1,-1,-1,-1)^*$
&& -152
&& 3(0)
&& 79(1)&\cr
& $(-1,-1,0,0)$
&&-144
&& 3 (0)
&& 75(1)& \cr
&  $(-1,-1,-1,0)$
&& -132
&& 3(0)
&& 69(3) & \cr
&  $(-1,-1,-2,-3)^*$
&& -124
&&  3(0)
&&  65(3) &\cr
& $(-1,-1,0,0)\diamond$
&&-112
&& 3 (0)
&& 59(6)& \cr
&  $(-2,-1,-1,1)\diamond$
&& -108
&& 3(0)
&& 57(4) & \cr
\noalign{\hrule}}}
}
$$
\vskip-10pt

\noindent{\bf Table 5.1:} Calabi--Yau hypersurfaces, $a)-d)$
 with $h_{11}=2,3$
in toric varieties, which derive from the
polyhedron of  $X_7 (1,1,1,2,2)$, $X_7(1,1,1,1,3)$,
$X_{12}(1,1,3,3,4)$ and $X_{28}(2,2,3,7,14)$
by moving the point $\nu^{*(1)}$. Examples which do not appear
in the list of \ks\ are marked
with a star. Models for which the Hodge numbers agree with that of a
Landau-Ginzburg model with more than five fields are marked
with a $\diamond$, although it is not clear that this makes
the theories the same.}
\vskip10pt}}

The generators of the Mori cone for the second case in the table~5.1.a),
which is a new  two moduli example, are
\eqn\genmorinp{l^{(1)}=(-3,0,0,0,1,1, 1),\quad
               l^{(2)}=( 0,1,1,1,-1,0,-2)}
In this case the third and second order Picard-Fuchs operators
follow in a very simple manner by factorization
\eqn\pfnp{
\eqalign{
{\cal L}_1&=(\t2-\t1)(\t1 - 2 \t2) -3 (3 \t1 - 2 ) (3 \t1 -1) z_1 \cr
{\cal L}_2&=\t2^3-(2\t2-\t1-1)(2\t2-\t1-2)(\t1-\t2+1) z_2\, ,}}
The discriminant and the Yukawa couplings are calculated as
$$\Delta=(1- 27 z_1)^3+
27 z_1 z_2(1- 540 z_1 + 5832 z_1^2+11664 z_1^2 z_2)$$
\eqn\yuknp{
\eqalign{
K_{111}&={3(7-54 z_1-216 z_1 z_2)
\over z_1^3\Delta},\quad K_{112}={9 z_2  (1-27 z_1+36 z_1-32 z_1 z_2)
\over z_1^2\Delta}\cr
K_{122}=&{3 ((1-27 z_1)^2 + 54 z_1 z_2+2916 z_1^2 z_2)
\over z_1 z_2^2\Delta}\, \quad K_{222}={81 z_1(1+216 z_1)\over z_2^2\Delta~,}}}
Instead of \normwcp, which applies only to hypersurfaces
in weighted projective spaces, we use the formulas of \hktyII
\eqn\myform{\eqalign{
\int_{X} c_2 \wedge J_p=&{1\over 2}\sum_{m,n}
\left(\sum_{i} l_i^{(m)} l_i^{(n)}\right) K_{J_p\, J_m\, J_n}, \cr
\chi=\int_{X} c_3 =&{1\over 3}\sum_{j,m,n}
\left(\sum_{i} l_i^{(j)} l_i^{(m)} l_i^{(n)}\right) \, K_{J_j\, J_m\, J_n},}}
which gives a relation between the evaluation of the second Chern
class on $H^{(1,1)}$ as well as the Euler number
with the intersection numbers. Formulas \myform~apply to
canonical resolved hypersurfaces and also to canonical resolved
complete intersections in general toric varieties. Knowing that
$\chi=-168$ we get from~\myform~the following normalization of the couplings
$K_{J_1J_1J_1}=21$, $K_{J_1J_1J_2}=9$, $K_{J_1J_2J_2}=3$, $K_{J_2J_2J_2}=0$.
{}From the first relation in \myform~follows $\int c_2 J_1= 78$ and $\int c_2
J_2=36$, which fixes $s_1=-{15\over 2}$,
$s_2=-4$. The fact that $n^e_{1,0}$ vanishes enforces $r_0=-1/6$,
and leads to predictions for the instantons, as given in Table~5.2.

\vskip7pt
{\vbox{\ninepoint{
$$
\vbox{\offinterlineskip\tabskip=0pt
\halign{\strut\vrule#
&\hfil~$#$
&\vrule#&~
\hfil ~$#$~
&\hfil ~$#$~
&\hfil $#$~
&\hfil $#$~
&\hfil $#$~
&\hfil $#$~
&\hfil $#$~
&\vrule#\cr
\noalign{\hrule}
&n_{i,j}&& j=0& j=1 &  j=2  &  j=3  &  j=4 & j=5 & j=6& \cr
\noalign{\hrule}
&n^r_{0,j} && 0&\forall j&&&&&&\cr
&n^e_{0,j} && 0&\forall j&&&&&&\cr
\noalign{\hrule}
&n^r_{1,j}&& 180&27&0&\forall j>1&&&&\cr
&n^e_{1,j} && 0&\forall j&&&&&&\cr
\noalign{\hrule}
&n^r_{2,j}&&180&6804&-54&0& \forall j>2&&&\cr
&n^e_{2,j}&& 0& \forall j &&&&&&\cr
\noalign{\hrule}
&n^r_{3,j}&& 180&138510& 4860& 243 & 0&\forall j>3 & &\cr
&n^e_{3,j}&& 3&  0&0 &-4&0&0&0&\cr
\noalign{\hrule}
&n^r_{4,j}&& 180&1478520& 5103972& -29520&-1728 &0&0 &\cr
&n^e_{4,j}&& 0&  -54&    6804&  540&135&0&0&\cr
\noalign{\hrule}}
\hrule}$$
}}
\vskip-10pt
\noindent
{\bf Table 5.2} The invariants of rational and elliptic curves of degree
$(i,j)$, $n^r_{i,j}$ and $n_{i,j}^e$ respectively, for the non-LG two moduli
case
described above.}
\vskip10pt

\subsec{Mirror nesting of the moduli spaces}

In~\refs{\cdfkm,\hkty}\ it was realized that certain models when
restricted to a specific codimension one surface of the K\"ahler
moduli space are birationally equivalent to a different Calabi--Yau
manifold. The models in question have $h^{1,1}=2$ and a
$Z_2$ curve singularity,
$X_{8}(2,2,1,1,1)_{-168}^{2,86}$,
$X_{12}(1,1,2,2,6)_{-252}^{2,128}$,
$X_{12}(1,2,2,3,4)_{-144}^{2,74}$,
$X_{14}(1,2,2,2,7)_{-240}^{2,122}$
There exists a one-dimensional subspace in the K\"ahler moduli space, which
corresponds to the following one K\"ahler moduli Calabi--Yau spaces
$X_{4,2}(1,1,1,1,1,1)_{-176}^{1,89}$,
$X_{6,2}(1,1,1,1,1,3)_{-256}^{1,129}$,
$X_{6,4}(1,1,1,2,2,3)_{-156}^{1,79}$ and
$X_{8}  (1,1,1,1,4  )_{-296}^{1,149}$ respectively. In particular the
 relation
manifests itself in the following relation between the
topological invariants, $\sum_{d_D} n_{d_J,d_D}=n_{d_{J}}$~\cdfkm.

In the last case the Calabi--Yau manifold with the one dimensional
K\"ahler deformation space is itself defined as a hypersurface of a toric
variety given by a reflexive polyhedron
This gives the following very simple interpretation\foot{This
observation was also made by Shinobu Hosono, whom we thank for an email
correspondence on this point.} of the situation, namely that the polyhedra
are nested into each other, \ie\ $\Delta^*_{X_{8}(1,1,1,1,4)}\subset
\Delta^*_{X_{14}(1,2,2,2,7)}$.

Let us now look in more detail at this nesting phenomenon and its
implication on the moduli spaces and proceed with the model
$X_{14}(1,2,2,2,7)$.
In the case at hand $\Delta^*_{X_{14}(1,2,2,2,7)}$
is the convex hull of the following points
\eqn\dualpoints{\eqalign{\matrix{
\nu^{*(0)}=(\-0,\-0,\-0,\-0)^4,&
\nu^{*(1)}=(-2,-2,-2,-7)^0,&
\nu^{*(2)}=(\-1,\-0,\-0,\-0)^0,\cr
\nu^{*(3)}=(\-0,\-1,\-0,\-0)^0,&
\nu^{*(4)}=(\-0,\-0,\-1,\-0)^0,&
\nu^{*(5)}=(\-0,\-0,\-0,\-1)^0,\cr
\nu^{*(6)}=(-1,-1,-1,-3)^1,&
\nu^{*(7)}=(-1,-1,-1,-4)^3,&
\nu^{*(8)}=(\-0,\-0,\-0,-1)^3},}}
where we indicated the dimension of the lowest
dimensional face the points lie on as upper
index on the points.
Since we include the codimension~1 points $\nu^{*(7)},\nu^{*(8)}$ in
the Laurent polynomial $P=\sum_{i=0}^8 a_i \phi_i$ the two dimensional moduli
space is redundantly parameterized. Part of the redundancy is
removed by choosing representatives of the ring of
deformations\foot{The rest of the redundancy
is removed by Euler type homogeneity conditions
given by the generators of the Mori cone as usual.}.
This can be done due to two relations of type
\trivialidentityII~but at level one: $a_0 \phi_8+2 a_5 \phi_0 + a_6
\phi_1={d_5\over X_4}$ and
$a_0 \phi_7+2 a_5 \phi_5+a_6\phi_1={d_5\over X_1X_2X_3 X_4^4}$ with
$d_5=(1-\Theta_1-\Theta_2-\Theta_3+\Theta_4)$, which allows us either to
set $a_7=a_8=0$, $a_1=a_8=0$ or $a_1=a_7=0$. In all
cases the convex hull of the remaining Newton polyhedron is reflexive
and as we argue below $(\Delta,\Delta^*)$ describes birationally
equivalent manifolds in the three possible cases,
although the polyhedra are different combinatorial objects.
Note that the third possibility corresponds to the two
moduli case $X_{7}(1,1,1,1,3)$ treated in section two.

Another possibility is to restrict oneself to a subspace
of the moduli space, by forcing an additional
coordinate $a_i$ to vanish. The only possible way to do this
for the case at hand in such a way that the remaining Newton polyhedron
stays reflexive is to set $a_1=a_6=a_8=0$, which
leads to the Newton polyhedron describing
the one dimensional moduli space of
$X_{8}  (1,1,1,1,4  )_{-296}^{1,149}$ and explains
the relation between the topological numbers of the two models.
Let us summarize this situation in the table~5.3.

{\vbox{\ninepoint{
$$\vbox{
{\offinterlineskip\tabskip=0pt
\halign{
\strut\vrule#&~#~&\hskip-6pt\vrule#&
\hfil$#$~&\vrule#&\hfil$#$~&\vrule#
&\hfil$#$~&\vrule#&\hfil$#$~&\vrule#\cr
\noalign{\hrule}
&$(\Delta,\Delta^*)$
&&(\Delta_I,\Delta^*_I)
&&(\Delta_{II},\Delta_{II}^*)
&&(\Delta_{III},\Delta_{III}^*)
&&(\Delta_{IV},\Delta_{IV}^*) &\cr
\noalign{\hrule}
\noalign{\hrule}
&$P$
&&a_1=a_6=a_8=0
&&a_1=a_8=0
&&a_1=a_7=0
&&a_7=a_8=0&\cr
\noalign{\hrule}
&$\Delta^*$
&&{\rm conv}(7,2,3,4,5)
&&{\rm conv}(6,7,2,3,4,5)
&&{\rm conv}(6,8,2,3,4,5)
&&{\rm conv}(1,2,3,4,5)&\cr
\noalign{\hrule}
&$ \Delta$
&&  {\rm conv}(1,2,3,4,5)
&&{\rm conv}(1,2,6,7,8,9,10,11)
&&{\rm conv}(1,2,6,7,8,12,13,14)
&&{\rm conv}(1,2,6,7,8)&\cr
\noalign{\hrule}
&$LG$
&& X_{8}(1,1,1,1,4)
&&---
&&X_{7}(1,1,1,1,3)
&&  X_{14}(1,2,2,2,7)&\cr
\noalign{\hrule}
&$h^{1,1}$
&&1(0)
&&2(0)
&&2(0)
&&2(0)&\cr
\noalign{\hrule}
&$h^{2,1}$
&&149(0)
&&122(0)
&&122(0)
&&122(15)&\cr
\noalign{\hrule}
&$z_1$
&&{a_5^4 a_2 a_3 a_4 a_7\over a_0^8}
&&{a_2 a_3 a_4 a_6^4\over a_0^4 a_7^3}
&&-{a_2 a_3 a_4 a_6\over a_0 a_8^3}
&&-{a_2 a_3 a_4 a_6^7\over a_0^7 a_1^3} &\cr
&$z_2$
&&
&&-{a_5 a_7\over a_0 a_6}
&& {a_6 a_8\over a_0^2}
&& {a_1 a_5\over a_6^2}
&\cr
\noalign{\hrule}}}}
$$
}}
\vskip-10pt
\noindent{\bf Table 5.3:} {\ninepoint{
The reflexive polyhedra inside  $\Delta^*_{IV}=
\Delta^*_{X_{14}(1,2,2,2,7)}$ as well as their
duals. The polyhedra are specified as convex hulls of the points given
in~\dualpoints\ and~(5.2), with the row below indicating the corresponding
LG-configuration if it exits.
Note that $\Delta^*_I\subset\Delta^*_{II}\subset \Delta^*_{IV}$ hence
$\Delta_{IV}\subset\Delta_{II}\subset \Delta_I$.
Furthermore we list the number of K\"ahler and complex structure deformations
of $X_\Delta$ (the number of non-algebraic
deformations is indicated in parentheses) as well
as the vanishing coefficient in the
Laurent polynomial and the canonical large complex structure
coordinates of $X^*_\Delta$,
which are determined by the Mori cone. For each case the
latter is generated by positive linear combinations of
the generators of the Mori cone of $\IP_{\Delta_1^*}$ which have
vanishing entries at the places of the
corresponding Laurent monomials.
}}}
\vskip10pt

The polyhedra $\Delta^*$, $\Delta$ in table 5.3 are specified as the convex
hull of the points
given in \dualpoints~ and the following points inside of
$\Delta_{X_8(1,1,1,1,4)}$
\eqn\points{\eqalign{\matrix{
\nu^{(1)}=(-1,-1,-1,-1)^0,&
\nu^{(2)}=(-1,-1,-1,\-1)^0,&
\nu^{(3)}=(-1,-1,\-7,-1)^0,&\cr
\nu^{(4)}=(-1,\-7,-1,-1)^0,&
\nu^{(5)}=(\-7,-1,-1,-1)^0,&
\nu^{(6)}=(-1,-1,\-6,-1)^1,&\cr
\nu^{(7)}=(-1,\-6,-1,-1)^1,&
\nu^{(8)}=(\-6,-1,-1,-1)^1,&
\nu^{(9)}=(-1,-1,\-3,\-0)^1,&\cr
\nu^{(10)}=(-1,\-3,-1,\-0)^1,&
\nu^{(11)}=(\-3,-1,-1,\-0)^1,&
\nu^{(12)}=(-1,-1,\-0,\-1),&\cr
\nu^{(13)}=(-1,\-0,-1,\-1)&
\nu^{(14)}=(\-0,-1,-1,\-1)&&\cr.}}}

This nesting phenomenon, which is ubiquitous among the reflexive
polyhedra, has the following simple implications\foot{We
conjecture that ${\cal M}_{2,1}(X_d)\subset {\cal M}_{2,1}(X_{d_1,d_2})$
is true for the first three examples given at the beginning of
this section.} for the moduli
spaces of K\"ahler ${\cal M}_{1,1}(X_{\Delta_{1,2}})$ and complex structure
deformation ${\cal M}_{2,1}(X_{\Delta_{1,2}})$ of the manifolds
$X_{\Delta_{1,2}}$

\eqn\emb{\eqalign{\Delta_1^*\subset \Delta^*_2
&\Longleftrightarrow \Delta_2\subset \Delta_1\Longrightarrow \cr
{\cal M}_{1,1}(X_{\Delta_1})\subseteq {\cal M}_{1,1}(X_{\Delta_2}),
&\qquad
{\cal M}_{2,1}(X_{\Delta_2}) \subseteq {\cal M}_{2,1}(X_{\Delta_1})}~~.}
If also ${\rm dim} ({\cal M}_{1,1}(X_{\Delta_1}))
   ={\rm dim}({\cal M}_{1,1}(X_{\Delta_2}))$ and
${\rm dim}( {\cal M}_{2,1}(X_{\Delta_1}))
   ={\rm dim}({\cal M}_{2,1}(X_{\Delta_2}))$
the manifolds will be birationally equivalent to each other.
This is in fact the case for the last three examples
in table~5.3.

Let us return to the example discussed in the beginning of this section
and compare the manifolds associated to
$(\Delta_I,\Delta_I^*)$ and $(\Delta_{II},\Delta^*_{II})$, which from
now on will be denoted the  I model and II model.
The representation $(\Delta_{II},\Delta^*_{II})$ of table 5.3
is the most suitable for our purposes, because the restriction
to the moduli space of the I model is simply given by $a_6=0$.

\figinsert\figone{Schematic Picture of the mirror nesting of the
moduli spaces for the I model and the II model }{3 in}{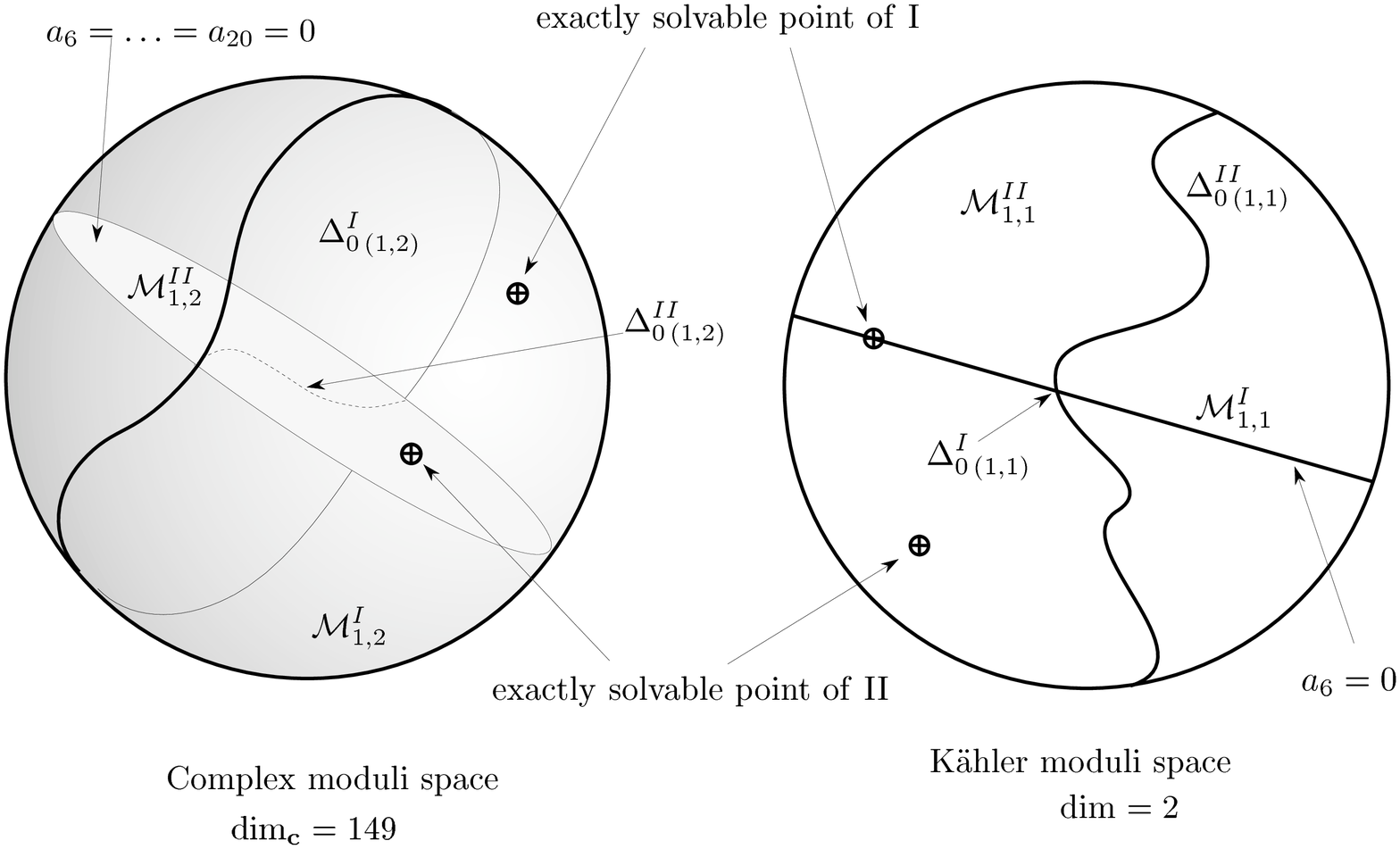}

In the large K\"ahler structure variables $z_1,z_2$  the Picard-Fuchs system
consists of the following two differential operators
of second and third order
\eqn\pfnewII{\eqalign{{\cal L}_1&=\t1^2 (7 \t1 - 2 \t2) + 7 z_1
\Bigl\{\bigl[(3\t1-\t2+2)-2 z_2 (12\t1 +\t2+6)\bigr]
(3 \t1-\t2+1)(3\t1-\t2)\cr &
+\bigl[(4 z_2^2 (8\t1+2 \t2+3)(3 \t1-\t2)
           -8 z_2^3(4 \t1 +\t2+1)(4 \t1 +\t2+3)\bigr](8 \t1+5)\Bigr\}\cr
&- 2 z_2 \t1^2 (4 \t1 -\t2) \cr
{\cal L}_2&=\t2 (3 \t1+\t2) + z_2 (4 \t1-\t2) (4 \t1 + \t2+1)\, ,}}
The differential operator for the I-model in the large K\"ahler structure
variable is
\eqn\pfnewI{{\cal L} =
\theta^4 - 64 z (8\theta+7)(8\theta+5)(8\theta+3)(8\theta+1)}

Let us then transform to the new variables $(a,b)$
\eqn\newvar{a^8={a_2a_3a_4a_5^4a_7\over a_0^8}=z_1 z_2^4=z,\qquad \qquad
            b^8={a_2a_3a_4a_6^8\over a_5^4a_7}={z_1\over z_2^4}}
in which the limit $b\rightarrow 0$ corresponds to restricting the
K\"ahler moduli space of the II-model  to the K\"ahler moduli space of the
I-model. The general discriminants of I and II read in this parametrisation
\eqn\gendis{\eqalign{\Delta^{II}_{0(1,1)}=&
a^3(1-2^{16} a^8) +
a^2 b ( 3 - 7\cdot 2^{15} a^8) +
a b^2(3 -7\cdot 41\cdot 2^{10} a^8) +
b^3 (1 -5\cdot 7 \cdot 2^{9} a^8) \cr &
+ a^4 b^4 (
7\cdot 23 a^2 b
- 7\cdot 3^2 a b^2
- 2 \cdot 3\cdot 7\cdot 11\cdot 67 a^3
+ 3^3 b^3
+ 7^7 a^7 b^4)
\cr
\Delta^I_{0,(1,1)}=&\lim_{b\rightarrow 0}v\Delta^{II}_{0(1,1)}=
a^3(1-2^{16} a^8).}}

The (unnormalized) Yukawa couplings of the II-model are

\eqn\yuknewII{
\eqalign{
K_{111}=&{1\over a^3 \Delta^{II}_{0(1,1)}   }
         \Bigr(2^6 a^2 (2^4 a {+}3^2 \cdot 5 b){+}
         7^4 a^4 b^5 ( 2^6 a^2
         {-}2^2 5 a b {+}3^2 b^2)
         {+} b^2 ( 2^2 661 a{+}19\cdot 41 b))\cr
K_{112}=&{1\over a^2 \Delta^{II}_{0(1,1)}  }
          \Bigr(2^6 a^2 + 2^2 37 a b+ 3 \cdot 31 b^2+
           7^3 a^4 b^3 (2^{10} a^3
           + 2^6 3 a^2 b
           - 2^2 11 a b^2
           + 3^2 b^3)\Bigl)\cr
K_{122}=&{1 \over a \Delta^{II}_{0(1,1)}  }
         \Bigr((2^2 a - 5 b)+
               7^2 a^4 b ( 2^{14} a^4
               + 2^{10} 11 a^3 b
               + 2^7 3 a^2 b^2
               + 3^2 b^4)\Bigl)\cr
K_{222}=&{7 a \over b \Delta^{II}_{0(1,1)} }
       \Bigr(
          2^2(1- 2^{16} a^8) +
          a^3 b( 2^14 a^4 +
          20^{10}3\cdot 29 a^3 b - 2^7 5 a^2 b^2 -
          2^2 23 a b^3 + 3^2 b^4)\Bigl)\,.}}

We note that as $b\to 0$, both $K_{112}$ and $K_{122}$ stay finite (and in
particular non-zero) while $K_{222}$ blows up as $1/b$. Comparison
with~\ref\cfkm{P. Candelas, A. Font, S. Katz and D. Morrison,
\npb429(1994)626, hep-th/9403187.} would indicate that
$b=0$ corresponds to a conifold point; this will indeed be shown below. Also,
$\lim_{b\rightarrow 0} K_{111}=2^{10}/\Delta^{I}_{0(1,1)}$ which
agrees with the result obtained by a direct computation on the I
model~\onemod.

Let us comment further on the limit $b\to 0$
of the II model and the correspondence with model I.
{}From table~5.3 we read off that the II model has
homogeneous coordinates $(y_2,y_3,\ldots,y_7)$ which are identified by
two $\IC^*$ actions with exponents $(1,1,1,0,4,-3)$ and
$(0,0,0,1,-1,1)$, corresponding to the generators of the Mori cone,
$l^{(1)}=(-4,1,1,1,0,4,-3)$ and $l^{(2)}=(-1,0,0,0,1,-1,1)$.
The values of the $y_i$ are restricted so that
we cannot have $y_2=y_3=y_4=y_5=0$ or $y_5=y_7=0$ (the combinatorics leading
to this conclusion is identical to that which
leads to the calculation of the numerators of the entries for $z_1$ and
$z_2$ in table~5.3).  Consider the vector function
\eqn\blowdown{(y_2,\ldots,y_7)\mapsto(y_2,y_3,y_4,y_6y_7,y_5y_6).}
Consideration of the $\IC^*$ actions shows that this gives a well
defined map from the II model to $\IP(1,1,1,1,4)$, which blows down
the divisor $y_6=0$ to a codimension~2 subset.  This map is precisely the
one that contracts the second generator of the Mori cone, as
indicated by the $-1$ in the corresponding coordinate of $l^{(2)}$.
The hyperplane class of $\IP(1,1,1,1,4)$ pulls back to $J_1$.
So we are in a similar situation to that in the first paper of
\cdfkm, obtaining the relation
\eqn\gwsum{\sum_jn_{i,j}(II)=n_i(\IP(1,1,1,1,4)),}
where the spaces in parentheses indicate the space for which the Gromov-Witten
invariants are calculated.

It is also interesting to analyze the singularities resulting from the blow
down \blowdown.  In the coordinates $(x_1,\ldots,x_5)$ of
$\IP^4(1,1,1,1,4)$, the form of our map shows that the image of the II model
is a degree~8 hypersurface containing the surface $x_4=x_5=0$.
The situation is now entirely analogous to that studied by Greene,
Morrison and Strominger
\lref\gms{B.R. Greene, D.R. Morrison and A. Strominger,
{\sl Black Hole Condensation and the Unification of String Vacua},
hep-th/9504145.}
\gms.  The octic equation
has the form $x_4f+x_5g=0$, where $f$ and $g$ have degrees 4 and 7.  This
Calabi-Yau is singular at the 28 points $x_4=x_5=f=g=0$, having conifold
singularities (nodes).  There are
correspondingly 28 vanishing cycles on a nearby smooth octic hypersurface,
which, when compactifying the type IIB string on the Calabi-Yau in
question, give rise to 28 hypermultiplets
which become light as we approach our hypersurface.  The 28 vanishing
cycles sum to zero, as they are bounded by the complement of 28 balls
in the surface $x_4=x_5=0$.  By the argument in \gms, we check that this is
consistent with $h^{2,1}$ decreasing by 27 from 149 to 122, while
$h^{1,1}$ increases by 1 from 1 to 2 as we perform this conifold transition.

\subsec{Construction of algebraic realizations of the deformation ring}
It is a well-known fact that in certain situations not all of the
K\"ahlermoduli space
can be described by toric divisors of $X$. Similarly, the deformation of
the defining equation
of a hypersurface $X^*$ in a toric variety $\IP^4_{\Delta^*}$ is in
general not sufficient to describe the complex structure moduli space of
$X$.
 More precisely, there are non-toric divisors in $X$ when~\batyrevI
\eqn\twistone{
\sum_{{\rm codimension}\,\theta^*=2,\,
\theta^*\in \Delta^*}l'(\theta^*)l'(\theta)}
is non-zero. In a similar fashion, when
\eqn\twisttwo{
\sum_{{\rm codimension}\,\theta=2,\,\theta^*\in \Delta}l'(\theta)l'(\theta^*)}
is non-zero there are non-algebraic deformations of the complex structure
of $X$. $l'(\theta)$ and $l'(\theta^*)$ is the number of interior
points in a face $\theta$ and $\theta^*$ respectively, where $\theta$ and
$\theta^*$ are dual faces; if $\theta\in\Delta$ is an
$n$-dimensional face given by the
set of vertices $\{\nu^{i_1},\ldots,\nu^{i_k}\}$ then the dual face,
$\theta^*\in\Delta^*$ is
an $3-n$ dimensional face  given by,
\eqn\dualface{
\theta^*=\{y\in\Delta^*\,|\,(y,\nu^{i_1})=\ldots,(y\nu^{i_k})=-1\}~,}
where $(\theta^*)^*=\theta^*$.
We would like to propose a method by which we under certain circumstances
can circumvent this problem such that either all of the non-toric divisors
in~\twistone\ or the non-algebraic deformations of the complex structure
in~\twisttwo, can be treated by the methods described in section~2.

Let us assume that there exists a set of points $\{\tilde\nu^{j_1},\ldots
\tilde\nu^{j_r}\}$ in codimension 1 faces of $\Delta$; we will consider
the situation in~\twistone. In a number of examples which we have studied
we have found that it was then possible to promote $\nu^{j_i}\in\theta$, given
by~\twistone, to vertices, while at the same time removing the previous
vertices of $\theta$. This also implies that the $\tilde\nu^{j_i}$ become
points in dimension~2 faces. After all this procedure will result in
new polyhedra $\tilde \Delta^*$ and $\tilde \Delta$ such that
\eqn\newpoly{
\tilde\Delta^*\supset\Delta^*\,,\quad\tilde\Delta\subset\Delta~.}
Hence, as we remove points coming from~\twistone\ we at the
same time increase the contribution from~\twisttwo, \ie\ there are more points
in codimension 2 faces of $\tilde\Delta$ and so the number of non-algebraic
deformations of the complex structure has increased.
Repeated application of this procedure will then have completely removed the
contribution from~\twistone. The same can be applied to~\twisttwo\ with the
obvious changes in the above discussion.
Unfortunately, we do not know at this point how general the above idea is.

Let us now consider an example, $X_{12}(1,1,1,3,6)$,
 in which a non-algebraic sector occurs
in the K\"ahler structure moduli space.
Batyrev's original construction leads to
the A-representation of the model, where
$\Delta_A$ is the convex hull of
\eqn\pointsA{\eqalign{\matrix{
\nu^{(1)}=(-1,-1,-1,-1)^0,&
\nu^{(2)}=(-1,-1,-1,\-1)^0,&
\nu^{(3)}=(-1,-1,\-3,-1)^0,&\cr
\nu^{(4)}=(-1,\-11,-1,-1)^0,&
\nu^{(5)}=(\-11,-1,-1,-1)^0.&\cr}}}
The non-algebraic state is related to the fact that
the point $\nu^{*(6)}$ in the dual polyhedra $\Delta^*_A$
\eqn\dualpointsA{\eqalign{\matrix{
\nu^{*(0)}=(\-0,\-0,\-0,\-0)^4,&
\nu^{*(1)}=(-1,-1,-3,-6)^0,&
\nu^{*(2)}=(\-1,\-0,\-0,\-0)^0,\cr
\nu^{*(3)}=(\-0,\-1,\-0,\-0)^0,&
\nu^{*(4)}=(\-0,\-0,\-1,\-0)^0,&
\nu^{*(7)}=(\-0,\-0,\-0,\-1)^0,\cr
\nu^{*(6)}=(\-0,\-0,-1,-2)^2,&
\nu^{*(11)}=(\-0,\-0,\-0,-1)^3},}}
lies on a face ${\rm conv}(\nu^{*(1)},
\nu^{*(2)},\nu^{*(3)})$, whose dual
${\rm conv}(\nu^{(2)},\nu^{(3)})$ is not empty,
but contains the point $\tilde \nu^{(3)}=(-1,-1,1,0)$.
The dimension of the cohomologies are $h^{2,1}=165(0)$
and $h^{1,1}=3(1)$, \ie\ in the A-representation,
we have an algebraic description of
${\cal M}_{2,1}$ but not of ${\cal M}_{1,1}$.
We hence cut $\Delta_A$ in a way, which
turns $\nu^{(3)}$ into a corner, \ie\ by replacing
$\nu^{(3)}$ with ${\tilde \nu}^{(3)}$. This leads to the
B-representation of the model in which $\Delta_B$ is the convex
hull of
\eqn\pointsB{\eqalign{\matrix{
\nu^{(1)}=(-1,-1,-1,-1)^0,&
\nu^{(2)}=(-1,-1,-1,\-1)^0,\cr
\nu^{(3)}=(-1,-1,\-1,0)^0,&
\nu^{(4)}=(-1,\-11,-1,-1)^0,&
\nu^{(5)}=(\-11,-1,-1,-1)^0.}}}
and the enlarged $\Delta^*_B$ contains
\eqn\dualpointsB{\eqalign{\matrix{
\nu^{*(0)}=(\-0,\-0,\-0,\-0)^4,&
\nu^{*(1)}=(-1,-1,-3,-6)^0,&
\nu^{*(2)}=(\-1,\-0,\-0,\-0),\cr
\nu^{*(3)}=(\-0,\-1,\-0,\-0)^0,&
\nu^{*(4)}=(\-0,\-0,\-1,\-0)^0,&
\nu^{*(5)}=(\-0,\-0,-1,\-2)^0,\cr
\nu^{*(6)}=(\-0,\-0,-1,-2)^2,&
\nu^{*(7)}=(\-0,\-0,\-0,\-1)^1,&
\nu^{*(8)}=(\-0,\-0,-1,-1)^3,\cr
\nu^{*(9)}=(\-0,\-0,-1,\-0)^3,&
\nu^{*(10)}=(\-0,\-0,-1,\-1)^3,&
\nu^{*(11)}=(\-0,\-0,\-0,-1)^3.}}}
In the B-representation the data of the cohomologies for the new
pair $(\Delta_B,\Delta^*_B)$ are $h^{2,1}=165(55)$ and
$h^{1,1}=3(0)$, \ie\ we can pick the Laurent
monomials of $\nu^{*(0)},\nu^{*(6)},\nu^{*(7)}$
as representatives of the ring of K\"ahler structure deformations.
To summarize we can describe the complete complex structure
deformations algebraically in the A-representation and
the full K\"ahler structure deformations in the B-representation.
We will use the B-representation to solve the K\"ahler deformations
part of the $X_{12}(1,1,1,3,6)$ model including the  "twisted" sector
and investigate how the "untwisted" deformation space of the
A-representation\hkty~is
embedded in the full deformation space.
The generators of the Mori cone for \dualpointsB~are
\eqn\morigenB{\eqalign{
l_B^{(1)}=&(\-0,\-0,\-0,\-0,\-1,\-1,\-0,-2),\quad
l_B^{(2)}=(\-0,\-1,\-1,\-1,\-0,\-0,-3,\-0)\cr
l_B^{(3)}=&( -4,\-0,\-0,\-0,\-0,-1,\-1,\-4).}}
The subspace of the Mori cone, which corresponds to the
A-representation, is simply obtained by picking the smallest positive
linear combination of the generators $l_B^{(i)}$, s.t. the
components of the new point $\nu^{*(5)}\in \Delta^*_B$ are zero, \ie\
\eqn\morigenA{\eqalign{
l_A^{(1)}=l_B^{(1)}+l_B^{(3)}=&(-4,\-0,\-0,\-0,\-1,\-0,\-1,\-2),\cr
l_A^{(2)}=&(\-0,\-1,\-1,\-1,\-0,\-0,-3,\-0).}}

The principal parts of the complete set of the Picard-Fuchs
equations, which governs the B-representation read
\eqn\princpart{\eqalign{
{\cal L}_1&=\t1(\t1-\t3)+{\cal O}(z),\quad\,\,\,\,
{\cal L}_2=(\t3-3 \t2)(\t1-\t3)+{\cal O}(z),\cr
{\cal L}_3&=\t3(\t3-3 \t2)+{\cal O}(z),\quad
{\cal L}_4=\t2^3+{\cal O}(z),\cr
{\cal L}_5&=\t2^2(2\t2-\t1)+{\cal O}(z).}}
Using the normalization from \myform~we have the
following intersection numbers
\eqn\intersectionsB{\eqalign{
K_{J_1,J_1,J_1}&=18,\quad
K_{J_1,J_1,J_2}=6,\quad
K_{J_1,J_1,J_8}=18,\quad
K_{J_1,J_2,J_2}=2,\cr
K_{J_1,J_3,J_3}&=18,\quad
K_{J_1,J_2,J_3}=6,\quad
K_{J_2,J_2,J_3}=1,\phantom{8}\quad
K_{J_2,J_3,J_3}=3,\cr
K_{J_3,J_3,J_3}&=9,\quad\,\,\,
\int_X c_2 J_1=96,\quad\,\,\,
\int_X c_2 J_2=36,\quad\,\,\,\,
\int_X c_2 J_3=102.}}
Comparison with the untwisted sector of the A-representation
for which one has
\eqn\intersectionsA{
K_{I_1,I_1,I_1}=18,\quad
K_{I_1,I_1,I_2}=6, \quad
K_{I_1,I_2,I_2}=2, \quad
\int_X c_2 I_1=96,\quad
\int_X c_2 I_2=36}
shows that one  has to identify $J_1$ with $I_1$ and
$J_2$ with $I_2$. For the topological invariants
we have the following relations
\eqn\relinvariants{n_{ d_{I_1},d_{I_2} }=
\sum_{d_{J_3}} n_{d_{J_1},d_{J_2},d_{J_3}},\qquad
n_{d_{J_1},d_{J_2},d_{J_3}}= n_{d_{J_1},d_{J_2},(3 d_{J_2}+d_{J_1}-d_{J_3})}
.}
This is the analog of the situation for the embeddings of the
one moduli cases in the two moduli ones, mentioned at the
beginning of this section.

{\vbox{\ninepoint{
$$
\vbox{\offinterlineskip\tabskip=0pt
\halign{\strut\vrule#
&\hfil~$#$
&\vrule#&~
\hfil ~$#$~
&\hfil ~$#$~
&\hfil $#$~
&\hfil $#$~
&\hfil $#$~
&\hfil $#$~
&\hfil $#$~
&\hfil $#$~
&\hfil $#$~
&\hfil $#$~
&\vrule#
&\hfil $#$~
&\vrule#\cr
\noalign{\hrule}
&(i,j)&&n_{ij0}& n_{ij1}&n_{ij2}  & n_{ij3}  & n_{ij4}& n_{ij5}& n_{ij6}&
n_{ij7}&n_{ij8}&n_{ij9}&& n_{ij}&\cr
\noalign{\hrule}
&(0,1) &&     3&       0&        0&         3&       0&      0&       0&
0&      0&      0&&6      &\cr
&(0,2) &&    -6&       0&        0&         0&       0&      0&     -6&
0&      0&      0&&-12      &\cr
&(0,3) &&    27&       0&        0&         0&       0&      0&       0&
0&      0&     27&&54      &\cr
\noalign{\hrule}
&(1,0) &&   108&     108&        0&        0&       0&       0&       0&
0&      0&      0&&216      &\cr
&(2,0) &&     0&     324&        0&        0&       0&       0&       0&
0&      0&      0&&324      &\cr
&(3,0) &&     0&     108&      108&        0&       0&       0&       0&
0&      0&      0&&216      &\cr
&(4,0) &&     0&       0&      324&        0&       0&       0&       0&
0&      0&      0&&324      &\cr
&(5,0) &&     0&       0&      108&     108&       0&        0&       0&
0&      0&      0&&216      &\cr
\noalign{\hrule}
&(1,1) &&     0&    -216&        0&    -216&       0&       0&      0&       0&
     0&       0&&-432      &\cr
&(1,2) &&     0&     540&        0&        0&      0&       0&    540&       0&
     0&       0&&1080      &\cr
&(1,3) &&     0&   -3456&        0&        0&      0&       0&      0&       0&
     0&-3456  &&-6912     &\cr
\noalign{\hrule}
&(2,1) &&     0&    -648&     5778&    5778&   -648&       0&     0&     0&
 0&      0&&10260     &\cr
&(2,2) &&     0&    1620&   -23112&       0&   1296&       0&-23112&  1620&
 0&      0&&-41688&\cr
&(2,3) &&     0&  -10368&   202986&       0&  -3240&       0&     0&  -3240&
 0&202986 &&378756&\cr
\noalign{\hrule}}
\hrule}$$
\vskip-10pt
\noindent
{\bf Table 5.4} The invariants of rational  curves of degree
$(i,j,k)$ for $X_{12}(1,1,1,3,6)$ in the B-representation including
the "twisted state". The last column contains for comparison the
invariants which correspond to the "untwisted sector" in the
A-representation.}
\vskip10pt
}}
Finally, it is very instructive to investigate the intersection numbers in the
basis, which is suggested by the Laurent-monomials divisor relation in
the A-representation. Denoting this divisors with the same labels
as the corresponding points in
$\Delta_A^*$ we have after the identification
\eqn\basechangeA{I_0=I_1,\quad I_6=I_1-3 I_2,}
the from \intersectionsA~the following topological data
\eqn\intersectionsAMD{
K_{I_0,I_0,I_0}=18, \quad K_{I_6,I_6, I_6}=18, \quad
\int_X c_2 I_0=96,\quad
\int_X c_2 I_6=-12.}
The divisor $I_6$ with multiplicity two on
the triangle ${\rm conv} (\nu^{*(1)},\nu^{*(2)},\nu^{*(3)})$
has to be split symmetrically between $I'_6$ and $I''_6$. In this
way we arrive at the intersection numbers~\hkty
\eqn\intersectionsAMD{
K_{I_0,I_0,I_0}=18, \,\,
K_{I'_6,I'_6, I'_6}=9, \,\,
K_{I''_6,I''_6, I''_6}=9,\,\,
\int_X c_2 I_0=96,\,\,
\int_X c_2 I'_6=-6,\,\,
\int_X c_2 I''_6=-6.}
Now in order to make the consistency check, that the A- and the
B-representation are homotopy equivalent we must find
linear transformation, by which \intersectionsB~transform into
\intersectionsAMD. As the reader may check the simple transformation
\eqn\basechangeB{I_0=J_1,\quad I_6'=J_3-3 J_2,\quad I_6''=J_1-J_3}
has precisely this property. Note that this is also in agreement with the
second relation in~\relinvariants.

We will now mention an example for the situation when there are
non-algebraic complex
structure deformations. This also serves as an important technical
application of the discussion in the previous
section. Let us
use the $(\Delta_{II},\Delta_{II}^*)$- and the
$(\Delta_{III},\Delta_{III}^*)$ representation of the moduli space
${\cal M}_{2,1}(X_{14}(1,2,2,2,7))$ as representations in which all
deformations are algebraic, see section 5.2. The possibility of utilizing
$X_{7}(1,1,1,1,3)$ in this way was already pointed out, however
in general we need non-Landau-Ginzburg models, as the second
one, to obtain an algebraic representation. We can understand the
modification of $\Delta_{IV}^*$ as follows. The point $\nu^{*(6)}$ on
the edge ${\rm conv}(\nu^{*(1)},\nu^{*(5)})$ gives rise to the
non-algebraic states of the deformation space for ${\cal M}_{2,1}$
in the representation $(\Delta_{IV},\Delta_{IV}^*)$, because its dual
face ${\rm conv}(\nu^{(6)},\nu^{(7)},\nu^{(8)})$ contains 15 inner
points. Removing the points $\nu^{*(1)},\nu^{*(8)}$
($\nu^{*(1)},\nu^{*(7)}$) promotes the point $\nu^{*(6)}$ into an
edge. The polyhedra $\Delta_{II}$($\Delta_{III}$) contains 15(18) points
more then $\Delta_{IV}$ (3(3) on codim 4, 9(3) on codim 3,
3(9) on codim 2 and 0(3) on codim 1 faces), which correspond (partly)
to the missing algebraic deformations.

Finally, let us show, in cases where there are no points in codimension
1 faces, how we can circumvent the failure of removing the non-algebraic
complex structure deformation.
Consider $X_{8}(1,1,2,2,2)^{2,86}$. Beside the five corners
the simplicial polyhedron $\Delta^*$ contains the interior
point $\nu^*_0=(0,0,0,0)$ as well as one other point
$\nu^*_6(0,-1,-1,-1)$ on a codimension three edge
$\theta^*$ spanned by $\nu_1^*=(1,0,0,0)$ and $\nu_5=(-1,-2,-2,-2)$.
The dual face $\theta$ of codimension two in $\Delta$ is
spanned by $(3,-1,-1,-1),(-1,3,-1,-1),(-1,-1,3,-1)$ and
contains three points $(1,0,0,-1),(0,1,0,-1),(0,0,1,-1)$.
By \twistone\ we hence count three non-algebraic $(1,2)$-forms.

As there are no points on codimension one faces in $\Delta^*$,
cutting one of the corners $\nu^*_1$ or $\nu^*_5$ will
not lead to a reflexive pair $(\Delta_B,\Delta^*_B)$,
which could serve as an algebraic description for the complex
structure deformations.
Instead use $|H|=(x_1^2,x_1x_2,x_2^2,x_3,x_4,x_5)$ to map
to $\IP^5$; the image is contained in the singular quadric $y_1y_3=y_2^2$.
The singularity $(y_1=y_2=y_3=0)$ gets blown up to produce the extra
exceptional toric divisor.
The degree 8 equation becomes degree 4 in the $y_i$'s.  So
we get a degenerate blown up $\IP^5[2,4]$.

Now vary the quadric from rank 3 to rank 4 (the prototype, which is unique up
to coordinate change, is $y_1y_2=y_3y_4$).  The quadric is singular at
the smaller locus $y_1=y_2=y_3=y_4=0$, so can still be blown up to produce
the extra $(1,1)$-form.  But the space of rank
3 quadrics has codimension 3 inside
the space of rank 4 quadrics --- so by allowing rank 4 quadrics, we
have introduced 3 more parameters.

\newsec{Discussion}

The purpose of this article has been two-fold.
On one hand we have shown that the
original ideas of Candelas et al~\cdgp, studying the moduli space of complex
structure and Kahler structure deformations by means of special geometry and
mirror symmetry extends to general Calabi--Yau hypersurfaces in toric
varieties.
Although we do not claim to have shown this for {\it all} Calabi--Yau
manifolds of this type
we believe that the non-trivial examples considered is evidence that a
description along the lines outlined below will serve the purpose.

Starting from $X$, its mirror, $X^*$ is constructed via the dual polyhedron. In
resolving the singularities, subdividing $\Delta^*$, one has in general
several subdivisions which lend themselves to a K\"ahler resolution,
corresponding to a different geometric interpretation in the phase picture.
For each of these subdivisions, we construct the generators of the Mori
cone for $P_{\Delta^*}$.
Note that in general it is not true that the Mori cones of the
toric variety and $X$ are the same.
Using linear relations between the generators we find all second
and third order differential operators. Application of INSTANTON~\hktyII\
 gives
information on whether this is a sufficient set of operators. In general
that is not the case.

The next step is to find the Batyrev-Cox ring and then restrict to a subset
of the coordinates, for which the original polyhedron $\Delta$ is
unresolved. The explicit use of the monomial representation, and in particular
the ideal allows us to compute all the second order relations. Although
in the example studied in section~3 this turned out to be sufficient, we
believe that it will not be so for the general situation. Rather, we
need to consider order three monomials and then reapply the reduction
method. Collecting all of the order two and order three operators
constructed in this way, one will be able to first show that the
classical intersection numbers agree with the above computation. Finally,
the instanton contributions can (in principle) be computed to arbitrary
multidegree of the rational curve.

There is an alternative approach to the method described here which is under
investigation.  It appears that
the principal parts of the Picard-Fuchs operators
can be deduced from the topological couplings directly.  This would
be a consequence of mirror symmetry if mirror symmetry could be established
for Batyrev's construction via dual polyhedra.
It is more likely that the result about the principal parts can be proven
directly within a  purely mathematical framework.  It can be checked that
the principal parts of all
the second and third order operators
in this paper may be derived much more easily by the method under
investigation. At present however, this method is less powerful since its
correctness rests on the validity of at least one conjecture.

Using these methods we have presented some  new ways of exploring
the moduli space. Given a toric variety represented in terms of a reflexive
polyhedron, $\Delta$, we may alter $\Delta$ with the requirement that the
new polytope, $\tilde\Delta$ is reflexive. In such a fashion new theories can
be constructed as hypersurfaces in toric varieties based on~$\tilde\Delta$,
taking a first step towards a classification of all reflexive polyhedra,
and hence further enlarge the class of $N=2$ SCFT. (Recall that the toric
description is equivalent to the $N=2$ supersymmetric gauged linear
sigma model~\refs{\phases,\agm}, where the latter is believed to have
non-trivial fixed point in the IR-limit.) A less severe alteration of $\Delta$
such that the Hodge numbers do not change for the hypersurfaces in question
can still result in interesting phenomena. In some cases it is possible to
find a $\tilde \Delta (\tilde \Delta^*)$ such that all the K\"ahler (complex)
structure deformations can be described algebraically. Thus, we can study the
moduli space for all deformations including the so called twisted deformations.
At this point it is not clear how general this phenomenon is; we know of
twisted complex structure deformations which seems to allude the above
prescription. Finally, given a $\Delta$ we can choose to embed the
hypersurface in a variety of ways, and in particular restricting to (singular)
hypersurfaces which however in their own right correspond to smooth (possibly
after some desingularization) Calabi--Yau manifolds. It may be that it is
possible to understand all hypersurfaces in toric varieties as special points
in a moduli space. Naively, although the process in going from one
theory to another
is singular this may provide us with a better understanding of
the theories at hand. However,
in light of the recent developments
in understanding conifold transitions in type IIB string
theories in terms of condensation of black
holes\lref\strom{A.~Strominger, {\sl Massless Black Holes and
Conifolds in String Theory}, hep-th/9504090.}
{}~\refs{\strom,\gms}; though, rather than being singular the transition is
smooth when the effect of the massless black holes are properly taken
into account. Indeed, we find that the transition between model I and
II in section~5.2 can be explained in perfect analogy with~\gms.

Related to this is the recent evidence of a spacetime duality between
type II string theory compactified on Calabi-Yau manifolds and that of
compactification of the heterotic string on $K3\times
T^2$~\ref\sqm{S. Ferrara, J. A. Harvey, A. Strominger and C. Vafa,
{\sl Second-Quantized Mirror Symmetry}, hep-th/9505162.}. It would be
interesting to see if it is possible to find the heterotic dual
corresponding to the type of Calabi-Yau compactifications discussed in
this paper.

On the other hand, with the improved knowledge of the moduli space of
heterotic string vacua one could try to make some first modest steps in
probing the phenomenological implications. Apart from the well-known Yukawa
couplings the moduli dependence of the threshold corrections to
the gauge couplings in string theory (as well as in the effective quantum
field theory)~\kl\ have come  to be understood. We have computed the
corrections for the two parameter
models studied in this article (as was done for some Fermat-type models
in~\refs{\cdfkm,\hkty}), and since the techniques at hand
 allows us to find the Picard-Fuchs equation (and its solutions) for any model,
similarly the threshold corrections can be computed for any Calabi--Yau
hypersurface for which our method applies. Finally, it is interesting to note
that so
many quantities only depend on the massless part of the theory, or
can be deduced from the massless sector. One may wonder what other non-trivial
properties can be computed in a similar fashion.

\vskip20pt
\noindent
{\bf Acknowledgments}:
It is a pleasure to thank D.R.~Morrison and M.R.~Plesser for useful
discussions. P.B.\ was supported by the DOE grant DE-FG02-90ER40542 and
acknowledges the hospitality of the Theory Division, CERN and Aspen Center
for Physics where this work was initiated.  S.K.\ was supported by NSF
grant DMS-9311386.  A.K.\ thanks Maximilian Kreuzer, Wolfgang Lerche,
Bong Lian, Stefan Theisen, Shing-Tung Yau and especially Shinobu
Hosono for discussions and comments.

\vfill\eject

\appendix{A}{Further examples}

\subsec{The case $X_8(1,1,1,2,3)_{-208}^{2,106}$}
\eqn\genmoriIII{l^{(1)}=(-2,0,0,0,0,1,1),\quad
                l^{(2)}=(-2,1,1,1,2,0,-3)}
\eqn\pfIII{
\eqalign{
{\cal L}_1&=\t1 (\t1 - 3 \t2) - (2 \t1 + \t2 ) (2 \t1 + \t2-1) z_1 \cr
{\cal L}_2&=
(1-4\z1)^2\Big[\t2^2(\t1-4\t2)-2\z1\t2^2(2\t1+2\t2+1)+4\z2\prod_{i=0}^2
(\t1-3\t2-i)\Big]\cr
&+16(1-4\z1)\z1\z2(\t1-3\t2)(\t1-3\t2-1)((\t1-3\t2-2)+2(2\t1+2\t2+1))\cr &+
128\z1^2\z2(\t1-3\t2)(2\t1+2\t2+1)+16\z1\z2(\t1-3\t2)(\t1-3\t2-1)}}

\eqn\yukIII{
\eqalign{
K_{111}=&4{(1-4 \z1 )(9+64 \z1 +96 \z1^2+16 \z1^3)+
(243+432 \z1+4608 \z1^2 +4096 \z1^3)\z2\over \z1^3\Delta}\cr
K_{112}=&4{(1-4 \z1)^2(3+ 12 \z1 + 2 \z1^2)-(81-432\z1-512 \z1^2 +4096
\z1^3)\z2\over \z1^2 \z2 \Delta}\cr
K_{122}=&4{(1-4 \z1)^3(1+\z1) +(27-336 \z1 +1536 \z1^2 -4096 \z1^3)\z2\over \z1
\z2^2 \Delta}\cr
K_{222}=&{(1-4 \z1)^4+4(9-176 \z1+1536 \z1^2+4096 \z1^3)\z2
\over\z2^3 \Delta}\, ,}}
\eqn\disIII{\Delta=(1-4 \z1)^5 +\z2(27-576 \z1+5120 \z1^2-
180224 \z1^3-131072\z1^4-4194304 \z1^3 \z2)}
\eqn\topIII{\eqalign{
K_{J_1J_1J_1}=36,\,\,
K_{J_1J_1J_2}&=12,\,\,
K_{J_1J_2J_2}=4,\,\,K_{J_2J_2J_2}=1\cr
\int c_2 J_1&= 96,\quad \int c_2 J_2=34}}
$r_0=-{1\over 6},\,\,s_1=-9,\,\,s_2=-{23\over 6},\,\, n_{8i-j,i}=n_{j,i}$

{\vbox{\ninepoint{
$$
\vbox{\offinterlineskip\tabskip=0pt
\halign{\strut\vrule#
&\hfil~$#$
&\vrule#&~
\hfil ~$#$~
&\hfil ~$#$~
&\hfil $#$~
&\hfil $#$~
&\hfil $#$~
&\hfil $#$~
&\hfil $#$~
&\vrule#\cr
\noalign{\hrule}
&n_{i,j}&& j=0& j=1 &  j=2  &  j=3  &  j=4 & j=5 & j=6& \cr
\noalign{\hrule}
&n^r_{j,0} && 0& 40 & 0 &\forall j>1&&&&\cr
&n^e_{j,0} && 0&\forall j&&&&&&\cr
\noalign{\hrule}
&n^r_{j,1}&& 3 &-80&780&54192&121410& 54192&780&\cr
&n^e_{j,1} && 0&  0&  0&   40&   200&    40&  0&\cr
\noalign{\hrule}
&n^r_{j,2}&&-6&200&-3120&29640&-425600&5297640&558340176&\cr
&n^e_{j,2}&&0& 0& 0  & 0& -400& 62032& 6426648&\cr
\noalign{\hrule}
&n^r_{j,3}&& 27&-1280&27580&-365040& 3953900 & -41185408&371614680&\cr
&n^e_{j,3}&& -10& 360&   -6240&   69160&     -547340&3041832   & -8415720&\cr
\noalign{\hrule}}
\hrule}$$
\vskip-7pt
\noindent
{\bf Table A.1} The invariants of rational and elliptic curves of degree
$(i,j)$, $n^r_{i,j}$ and $n_{i,j}^e$ respectively, for $X_8(1,1,1,2,3)$.}
\vskip7pt}}

$J_1$ is the class of cubic polynomials; $J_2$ is the class of linear
polynomials.  The exceptional divisor $E$ is $\IP^2$, and $J_1=3J_2+E$.
{}From this, we see that the numbers $n_{0,j}$ are identical to those for
$\IP^4(1,1,1,1,3)$.

\subsec{The case $X_9(1,1,2,2,3)^{2,86}_{-168}$}
\eqn\genmoriIII{l^{(1)}=(-3,-1,-1,1,1,0,3),\quad
                l^{(2)}=(-1,1,1,0,0,1,-2)}
\eqn\pfIV{
\eqalign{
{\cal L}_1&=\t2 (\t2 - \t1)^2 - (3 \t1 + \t2 ) (3 \t1 -2 \t2+2)
(3 \t1-2 \t2 +1) z_2 \cr
{\cal L}_2&=(\t2{-}\t1)^2{-}(\t2{-}\t1)(3\t1{-}\t2\t2) {+}
4\t1(3\t1{-}2\t2){+}3\z1(3\t1{-}2\t2)(3\t1{-}2\t2{-}1)\cr
&{-}48\z1\z2(3\t1{+}\t2{+}1)(3\t1{+}\t2{+}2)
{-}48\z1\z2(3\t1{+}\t2{+}1)(3\t1{-}2\t2) {-} 16\z1(\t2{-}\t1)^2 \, ,}}
\eqn\yukIV{
\eqalign{
K_{111}=&{1\over \z1^3 \Delta}\Bigr(
6 - 5z_1 - 48z_2 + 269z_1z_2 - 768z_1^2z_2 +
96z_2^2 - 1677z_1z_2^2 \cr &
+ 6144z_1^2z_2^2 + 2916z_1z_2^3 - 11664z_1^2z_2^3\Bigl)
\cr
K_{112}=&{1\over \z1^2 \z2 \Delta} \Bigr(9 - 9z_1 -
 72z_2 - 400z_1z_2 + 256z_1^2z_2 +
  144z_2^2 + 4005z_1z_2^2 -\cr & \qquad
  11520z_1^2z_2^2 -
  8748z_1z_2^3 + 34992z_1^2z_2^3 \Bigl)\cr
K_{122}=&{1 \over \z1\z2^2\Delta}\Bigr
(13 - 13z_1 - 106z_2 - 1525z_1z_2 +
1280z_1^2z_2 + 216z_2^2 + 14634z_1z_2^2+
\cr &\quad   13824z_1^2z_2^2 - 32805z_1z_2^3 -
104976z_1^2z_2^3 \Bigl) \cr
K_{222}=&{1\over \z2^3\Delta}
\Bigr(17 - 17z_1 - 153z_2 - 2610z_1z_2 +
  2304z_1^2z_2 + 324z_2^2 +\cr & \qquad 32886z_1z_2^2 +
  20736z_1^2z_2^2 - 78732z_1z_2^3 +
  314928z_1^2z_2^3\Bigl)\,,}}
\eqn\disIV{\eqalign{\Delta=&
  1 - z_1 - 8z_2 - 531z_1z_2 + 512z_1^2z_2 + 16z_2^2 +
  4335z_1z_2^2 + 74496z_1^2z_2^2 -\cr &
  65536z_1^3z_2^2 - 8748z_1z_2^3 -
  1539648z_1^2z_2^3 + 8503056z_1^2z_2^4 -
  14348907z_1^2z_2^5}}
$$
K_{J_1J_1J_1}=6,\,\,
K_{J_1J_1J_2}=9,\,\,
K_{J_1J_2J_2}=13,\,\,K_{J_2J_2J_2}=17,\,\,\int c_2 J_1= 48,\,\,
\int c_2 J_2=74$$

$r_0=-{1\over 6},\,\,s_1=-5,\,\,s_2=-{43\over 6}$

{\vbox{\ninepoint{
$$
\vbox{\offinterlineskip\tabskip=0pt
\halign{\strut\vrule#
&\hfil~$#$
&\vrule#&~
\hfil ~$#$~
&\hfil ~$#$~
&\hfil $#$~
&\hfil $#$~
&\hfil $#$~
&\hfil $#$~
&\hfil $#$~
&\vrule#\cr
\noalign{\hrule}
&n_{i,j}&& j=0& j=1 &  j=2  &  j=3  &  j=4 & j=5 & j=6& \cr
\noalign{\hrule}
&n^r_{0,j} && 0& -2&  0&\forall j>1&&&&\cr
&n^e_{0,j} && 0&\forall j&&&&&&\cr
\noalign{\hrule}
&n^r_{1,j}&& 1&640&641&4&5&7 &9 &\cr
&n^e_{1,j} && 0&\forall j&&&&&&\cr
\noalign{\hrule}
&n^r_{2,j}&&0&0&10032&208126&8734&-2596&-3900&\cr
&n^e_{2,j}&& 0& 0& 0&   640& 12& 24&36&\cr
\noalign{\hrule}
&n^r_{3,j}&& 0&0&0&288384& 23177356 & 23347504&798855&\cr
&n^e_{3,j}&&0&  0& 0  & -1280&  158140& 164290& 17152&\cr
\noalign{\hrule}}
\hrule}$$
\vskip-7pt
\noindent
{\bf Table A.2} The invariants of rational and elliptic curves of degree
$(i,j)$, $n^r_{i,j}$ and $n_{i,j}^e$ respectively, for $X_9(1,1,2,2,3)$.}
\vskip7pt}}

$J_1$ is the class of quadratic polynomials; $J_2$ is the class of
cubic polynomials.  We have $3J_1=2J_2+E$, where the exceptional divisor
$E$ is a rational ruled surface, with map to $\IP^1$ defined by $(x_3,x_4)$.
This shows as in the case of $\IP^4(1,1,1,2,2)$ that $n^r_{0,1}=-2$.
Similarly, $n^r_{1,j}=2j-3$ for $j\ge 4$, corresponding to the sections
of the ruled surface.

\subsec{The case $X_{14}(1,1,2,3,7)_{-260}^{2,132}$}
\eqn\genmoriIII{l^{(1)}=(0,-2,-2,-4,1,0,7),\quad
                l^{(2)}=(-2,1,1,2,0,1,-3)}
\eqn\pfV{
\eqalign{
{\cal L}_1&=\t1 (7 \t1-3 \t2)-4 z_1 z_2^2(2 \t2+1)(2 \t2+3) \cr
{\cal L}_2&= 2 (1{-}64\z1\z2^2)^2 \Big[ (\t2{-}2\t1)^3 {-}
\z2 (7\t1 {-} 3\t2)(7\t1 {-} 3\t2 {-}1)(7\t1 {-} 3\t2 {-}2)\Big] \cr
&{-}14336 \z1^2\z2^5(2\t2{+}1)(7\t1{-}3\t2) {-}
128\z1\z2^3(7\t1{-}3\t2)(7\t1{-}3\t2{-}1)\cr
&{-}32\z1\z2^3(1{-}64\z1\z2^2)(7\t1{-}3\t2)(7\t1{-}3\t2{-}1)\Big[
7(2\t2{+}1){+}4(7\t1{-}3\t2{-}2)\Big]\, ,}}
\eqn\yukV{
\eqalign{
K_{111}=&{1\over \z1^3 \Delta}\Bigr(9(1+ 27 \z2) -
          16 \z1\z2^2    (11 + 378 \z2)-
          256 \z1^2\z2^4 (88 + 2205 \z2)-\cr
        &
         \qquad 8192 \z1^3\z2^6 (24 +  343 \z2)-
          2809856 \z1^3 \z2^7\Bigr)\cr
K_{112}=&{1\over \z1^2 \z2 \Delta} \Bigr(
              21(1{+}27 \z2){-}336\z1\z2^2(2{+}63 \z2){-}
              256\z1^2 \z2^4 (160{+}3773 \z2){-}
              131072 \z1^3\z2^6\Bigl)\cr
K_{122}=&{1 \over \z1\z2^2\Delta}\Bigr(
          49 (1+27 \z2)-
          64 \z1 \z2^2 (34 +1029\z2)-
         512 \z1^2 \z2^4 (120+2401 \z2)\Bigl)
           \cr
K_{222}=&{1\over \z2^3\Delta} \Bigr(
          114 + 1029 \z2-80\z1 \z2^2 (80 +2401 \z2)-
          573444 \z1^2 \z2^4 \Bigl)\,,}}

\eqn\disV{\eqalign{\Delta=&1+27 \z2 - 16 \z1 \z2^2(16-441 \z2)+
                    2048 \z1^2\z2^4 (12+343 \z2)-\cr
                    &4096\z1^3\z2^6(256+38416\z2+823543\z2^2)+
                    16777216 \z1^4\z2^7}}
$$
K_{J_1J_1J_1}=9,\,\,
K_{J_1J_1J_2}=21,\,\,
K_{J_1J_2J_2}=49,\,\,
K_{J_2J_2J_2}=114,\,\,
\int c_2 J_1= 66,\,\,
\int c_2 J_2=156$$
$r_0=-{1\over 6},\,\, s_1=-{13\over 2},\,\, s_2=-14$

{\vbox{\ninepoint{
$$
\vbox{\offinterlineskip\tabskip=0pt
\halign{\strut\vrule#
&\hfil~$#$
&\vrule#&~
\hfil ~$#$~
&\hfil ~$#$~
&\hfil $#$~
&\hfil $#$~
&\hfil $#$~
&\vrule#\cr
\noalign{\hrule}
&n_{i,j}&& j=0& j=1 &  j=2  &  j=3  &  j=4 & \cr
\noalign{\hrule}
&n^r_{j,0} && 0&\forall j&&&&\cr
&n^e_{j,0} && 0&\forall j&&&&\cr
\noalign{\hrule}
&n^r_{j,1}&& 3&0&\forall j>0&&&\cr
&n^e_{j,1}&& 0&\forall j&&&&\cr
\noalign{\hrule}
&n^r_{j,2}&&-6&220&0&\forall j>1&&\cr
&n^e_{j,2}&& 0&  1&    0&   \forall j>1&&\cr
\noalign{\hrule}
&n^r_{j,3}&& 27&-440&0&\forall j>1&  & \cr
&n^e_{j,3}&&-10&  -2& 0  & \forall j>1&   & \cr
\noalign{\hrule}
&n^r_{j,4}&& -192&  1100& 260& 0&\forall j>2& \cr
 &n^e_{j,4}&& 231& 5& 2& 0&\forall j>2 & \cr
\noalign{\hrule}}
\hrule}$$
\vskip-7pt
\noindent
{\bf Table A.3} The invariants of rational and elliptic curves of degree
$(i,j)$, $n^r_{i,j}$ and $n_{i,j}^e$ respectively, for $X_{14}(1,1,2,3,7)$.}
\vskip7pt}}

$J_1$ is the class of cubic polynomials and $J_2$ is the class of degree
seven polynomials.  The exceptional divisor $E$ is a $\IP^2$, and
$7J_1=3J_2+E$.
{}From this, we see that the numbers $n_{0,j}$ are identical to those for
$\IP^4(1,1,1,1,3)$.

Note that the K\"ahler cone of the toric variety differs from the K\"ahler cone
of the Calabi--Yau hypersurface.  The Mori cone of $X_{14}(1,1,2,3,7)$ is
generated by $l^{(1)}+2l^{(2)},l^{(2)}$, \ie\ by the curves of
type $(1,2)$ and $(0,1)$.  A curve of type $(a,b)$ with $b<2a$ would be
contained in the locus $x_1=x_2=x_3=0$, which is a curve on the toric
variety but a point on the Calabi--Yau hypersurface.
\vfill\eject

\appendix{B}{Pichard-Fuchs operators, intersection and instanton numbers
for $\IP^4(1,2,3,3,4)$}

$$
\eqalign{
{\cal L}_1=&(\t2+\t4-\t5)(-2\t1+\t4) +{\cal O}(z_i)\cr
{\cal L}_2=&(\t1-2\t4+\t5)(-2\t2+\t5) +{\cal O}(z_i)\cr
{\cal L}_3=&(-2\t1+\t4)(-2\t2+\t5) +{\cal O}(z_i)\cr
{\cal L}_4=&(3\t2-2\t3)(-2\t1+\t4)+{\cal O}(z_i)\cr
{\cal L}_5=&\t1(-2\t2+\t5)+{\cal O}(z_i)\cr
{\cal L}_6=&(\t1-2\t4+\t5)(3\t2-2\t3)  +{\cal O}(z_i)\cr
{\cal L}_7=&\t1(\t1+\t3-\t5)(\t1-2\t4+\t5) +{\cal O}(z_i)\cr
{\cal L}_8=&(\t2{+}\t4{-}\t5)(\t1{+}\t3{-}\t5)(3\t2{-}2\t3) +{\cal O}(z_i)\cr
{\cal L}_9=&(\t3-\t2)^2(\t1+\t3-\t5)+{\cal O}(z_i)\cr
{\cal L}_{10}=&\t1(\t2+\t4-\t5)(\t1+\t3-\t5)+{\cal O}(z_i)\cr
{\cal L}_{11}=&(-2\t1{+}\t4)(-2\t2{+}\t5)(-2\t2{+}\t5-1) +{\cal O}(z_i)\cr
{\cal L}_{12}=&(\t1+\t4-\t5)(3\t2-2\t3)(-2\t1+\t4)+{\cal O}(z_i)\cr
{\cal L}_{13}=&(\t1{-}2\t4{+}\t5)(\t1{-}2\t4{+}\t5{-}1)(3\t2{-}2\t3)
+{\cal O}(z_i)\cr
{\cal L}_{14}=&\t1(\t2+\t4-\t5)(-2\t2+\t5) +{\cal O}(z_i)\cr
{\cal L}_{15}=&\t1(-2\t2+\t5)(-2\t2+\t5-1)+{\cal O}(z_i)\cr
{\cal L}_{16}=&\t1(\t1-2\t4+\t5)(3\t2-2\t3)+{\cal O}(z_i)\cr
{\cal L}_{17}=&(\t3-\t2)^2(-2\t1+\t4)+{\cal O}(z_i)\cr
{\cal L}_{18}=&\t1(\t2+\t3-\t5)(3\t2-2\t3)+{\cal O}(z_i)\cr
{\cal L}_{19}=&\t1(3\t2-2\t3)(-2\t2+\t5)+{\cal O}(z_i)\cr
{\cal L}_{20}=&(\t2+\t4-\t5)(3\t2-2\t3)+{\cal O}(z_i)\cr
{\cal L}_{21}=&\t1(\t1-2\t4+\t5)-4(\t3-\t2) (\t1-2\t4+\t5)+{\cal O}(z_i)\cr
{\cal L}_{22}=& -\t1(\t1+\t3-\t5)+\t1(-2\t1+\t4) -4(\t3-\t2)(-2\t1+\t4)  \cr
&+(\t3-\t2)(\t1+\t2+\t3-\t5) -
(\t1+\t3-\t5)(3\t2-2\t3) \cr
&+2 (3\t2-2\t3)(-2\t2+\t5) +{\cal O}(z_i)\cr
{\cal L}_{23}=& -39 (-2\t1+\t4)\t3+ 13 (-2\t1+\t4)(-2\t2+\t5)\cr
&+13 (-2\t1+\t4)(\t1+\t3-\t5)-13 \t1 (\t1+\t3-\t5)+
\t4(\t3-\t2)- 6\t4 (3\t2-2\t3)\cr
&-8 (\t3-\t2)(-2\t1+\t4)-
4 (\t1-2\t4+\t5)(\t3-\t2) + {\cal O}(z_i)\cr
 }$$

\noindent
The first 19 operators are obtained from the Mori cone~\mcexotic, while
the last four we got from the relation~\bcrel\ and via the
Dwork-Katz-Griffith's
reduction scheme applied to~\opfinal.
\vskip7pt
\vbox{{\ninepoint{
$$
\vbox{\offinterlineskip\tabskip=0pt
\halign{\strut\vrule#
&~$#$~\hfil\quad
&~$#$~\hfil\quad
&~$#$~\hfil\quad
&~$#$~\hfil\quad
&~$#$~\hfil
&\vrule#\cr
\noalign{\hrule}
&K_{J_1J_1J_1}=8&K_{J_1J_1J_2}=16&K_{J_1J_2J_2}=26& K_{J_2J_2J_2}=38&
K_{J_1J_1J_3}=24&\cr
&K_{J_1J_2J_3}=39&K_{J_2J_2J_3}=57&K_{J_1J_3J_3}=58&K_{J_2J_3J_3}=85&
K_{J_3J_3J_3}=125&\cr
&K_{J_1J_1J_4}=20&K_{J_1J_2J_4}=34&K_{J_2J_2J_4}=52&K_{J_1J_3J_4}=51&
K_{J_2J_3J_4}=78&\cr
&K_{J_3J_3J_4}=116&K_{J_1J_4J_4}=42&K_{J_2J_4J_4}=68&K_{J_3J_4J_4}=102&
K_{J_4J_4J_4}=84&\cr
&K_{J_1J_1J_5}=32&K_{J_1J_2J_5}=52&K_{J_2J_2J_5}=78&K_{J_1J_3J_5}=78&
K_{J_2J_3J_5}=117&\cr
&K_{J_3J_3J_5}=174&K_{J_1J_4J_5}=68&K_{J_2J_4J_5}=104&K_{J_3J_4J_5}=156&
K_{J_4J_4J_5}=136&\cr
&K_{J_1J_5J_5}=104&K_{J_2J_5J_5}=156&K_{J_3J_5J_5}=234&K_{J_4J_5J_5}=208&
K_{J_5J_5J_5}=312&\cr
\noalign{\hrule}}
\hrule}$$
\vskip-7pt
\noindent
{\bf Table B.1}: Intersection numbers for $\IP^4(1,2,3,3,4)$.}
\vskip7pt}}

\vskip7pt
\vbox{
{\ninepoint{
$$
\vbox{\offinterlineskip\tabskip=0pt
\halign{\strut\vrule#
&~$#$~\hfil
&~\hfil$#$\quad\quad
&~$#$~\hfil
&\hfil~$#$\quad\quad
&~$#$~\hfil
&\hfil~$#$\quad\quad
&~$#$~\hfil
&\hfil~$#$\quad\quad
&\vrule#\cr
\noalign{\hrule}
&[0, 0, 0, 1, 0] & -2&[0, 0, 1, 0, 0] & -2&[1, 0, 0, 1 , 0] &\ph -2&
[1, 0, 0, 0, 0] & -2&\cr
&[0, 0, 0, 0, 1] &\ph 3&[0, 0, 0, 1, 1] &\ph 3&[1, 0, 0, 1, 1]&\ph 3&
[1, 1, 1, 1, 1] &\ph 3&\cr
&[0, 1, 1, 0, 0] &-2&[0, 1, 1, 0, 1] &\ph 3&[0, 1, 1, 1, 1]&\ph 3&
[0, 1, 1, 1, 2] &\ph 7&\cr
&[0, 1, 2, 0, 0]  &-2&[0, 1, 2, 0, 1] &\ph 3&[0, 1, 2, 1, 1]&\ph 3&
[0, 1, 2, 1, 2] &\ph 7&\cr
&[0, 2, 3, 0, 0] & -4&[0, 2, 3, 0, 1] &\ph 9&[0, 2, 3, 0, 2]& -6&
[0, 2, 3, 0, 3] &\ph 1&\cr
&[0, 2, 3, 1, 1]&\ph 9&[0, 2, 3, 1, 2] &-18&[0, 2, 3, 1, 3]&\ph 9&&&\cr
&[0, 2, 3, 2, 2] & -6&[0, 2, 3, 2, 3] &\ph 9&&&&&\cr
&[0, 3, 4, 0, 0] & -6&[0, 3, 4, 0, 1] &\ph 15&[0, 3, 4, 0, 2]& -12&
[0, 3, 4, 0, 3] &\ph 3&\cr
&[0, 3, 4, 1, 1]&\ph 15&[0, 3, 4, 1, 2] &-36&&&&&\cr
&[0, 3, 5, 0, 0] &-6&[0, 3, 5, 0, 1] &\ph 15&[0, 3, 5, 0, 2]&-12&
[0, 3, 5, 1, 1] &\ph 15&\cr
&[0, 4, 5, 0, 0]&-8&[0, 4, 5, 0, 1]&\ph 21&[0, 4, 6, 0, 0]&-32&&&\cr
&[1, 1, 1, 1, 2]&\ph 7&[1, 1, 1, 2, 2]& \ph 7&[1, 1, 1, 2, 3]& \ph 7&
[1, 1, 2, 1, 1]& \ph 3&\cr
&[1, 1, 2, 1, 2]& \ph 7&[1, 1, 2, 2, 2]& \ph 7&[1, 1, 2, 2, 3]& \ph 8&
[1, 1, 3, 2, 3]&\ph 3&\cr
&[1, 2, 2, 2, 3]&\ph 7&[1, 2, 3, 1, 1]&\ph 9&[1, 2, 3, 1, 2]& -18&
[1, 2, 3, 1, 3]&\ph 9&\cr
&[1, 2, 3, 2, 2]& -18&[1, 3, 4, 1, 1] &\ph 15&&&&&\cr
\noalign{\hrule}}
\hrule}$$
\vskip-7pt
\noindent
{\bf Table B.2}: A list of the non-zero instanton numbers for rational curves
of degree $[a_1,\ldots,a_5]$ on $\IP^4(1,2,3,3,4)$,
 with the restriction $\sum_{i=1}^5a_i\leq 10$;
the non-existence of the curve of degree $[0,1,0,0,0]$ is explained in
section~4.}}}
\vfill\eject

\listrefs
\bye